\documentclass{aa}  
\usepackage[dvipsnames]{xcolor}
\usepackage{physics}
\usepackage{bm}
\usepackage{orcidlink}
\newcommand\gaia{\textit{Gaia}}

\newcommand\te{$t_{\rm E}$}
\newcommand\pie{$\pi_{\rm E}$}
\newcommand\thetae{$\theta_{\rm E}$}
\usepackage{graphicx}
\usepackage{txfonts}
\usepackage{hyperref}
\hypersetup{
  colorlinks   = true, %Colours links instead of ugly boxes
  urlcolor     = blue, %Colour for external hyperlinks
  linkcolor    = blue, %Colour of internal links
  citecolor   = blue %Colour of citations
}
\begin{document}

   \title{Predictions for astrometric microlensing in \textit{Gaia}}
   
   \author{
          Zofia Kaczmarek\orcidlink{0009-0007-4089-5012}\inst{1}\thanks{e-mail: \url{zofia.kaczmarek@uni-heidelberg.de}}
          \and
          David Sweeney\orcidlink{0000-0002-7528-1463}\inst{2,3}
          \and
          \L{}ukasz Wyrzykowski\orcidlink{0000-0002-9658-6151}\inst{4,5}
          \and
          Ulrich Bastian\orcidlink{0000-0002-8667-1715}\inst{1}  
          }

   \institute{Zentrum für Astronomie der Universität Heidelberg, Astronomisches Rechen-Institut, Mönchhofstr. 12-14, 69120 Heidelberg, Germany
   \and
   Sydney Institute for Astronomy (SIfA), The University of Sydney, Physics Road, Sydney 2050, Australia
   \and
   Donald Bren School of Information and Computer Sciences, University of California, Irvine, CA 92697, USA
   \and
   Astronomical Observatory, University of Warsaw,
   Al. Ujazdowskie 4, 00-478 Warsaw, Poland
   \and
   Astrophysics Division, National Centre for Nuclear Research,
  Pasteura 7, 02-093 Warsaw, Poland}

   \date{{Received 25 March 2026 / Accepted 02 September 2026}}

% \abstract{}{}{}{}{} 
% 5 {} token are mandatory
 
  \abstract
  {\gaia, the European Space Agency's astrometric mission, is uniquely positioned for microlensing studies thanks to its all-sky coverage and simultaneous {astrometric and photometric} observations. In \gaia~Data Release 4 (DR4), astrometric time series will be published for the first time, providing a rich dataset for {isolated} stellar remnant searches.}
   {In anticipation of \gaia~DR4, we {prepared} tools for fitting photometric and one-dimensional astrometric measurements. We {tested} the recovery of true event parameters and population distributions with \gaia-like observations.}
   {We {created} a mock survey of microlensing events mirroring \gaia~DR4's survey design. The events {were} generated using dedicated Galactic simulations to provide realistic expectations on stellar and dark lens yields. For selected events, we {simulated} observations following \gaia's scanning law and precision, using the {\tt astromet} package.
   Such simulations are needed to interpret forthcoming data {(}i.e. {to} update Galactic models to match observed populations{)}; they also enable testing inference tools with known ground truth. We {used} nested sampling to infer full posterior distributions and compare them with the true lens parameters.}
   {Our simulation predicts {322 $\pm$ 70} microlensing events with astrometric signals above \gaia~precision thresholds; {209 $\pm$ 57} of {them} have detectable photometric {signals}. Among those, {82 $\pm$ 9} and {49 $\pm$ 7}, respectively, have remnant lenses. To assess \gaia's capabilities in constraining {remnant} mass distributions, we {modelled} all remnant-lens events with detectable astrometric and photometric signals. We {recovered} the true {Einstein times ($t_{\rm E}$) and radii ($\theta_{\rm E}$)} for most events. {Microlensing parallax ($\pi_{\rm E}$)} measurements are consistent with true values but uncertain, limiting mass determination. We recommend focusing on events with bright sources or anomalous {{five-parameter}} fits to optimize dark lens searches. We make available the mock \gaia~time-series dataset, as well as a {toolkit} repository for working with the \gaia~data format, fitting models, and visualizing results.}
   {}


   \maketitle
%
%-------------------------------------------------------------------

\section{Introduction}
\label{sec:intro}

Gravitational microlensing occurs with the alignment of a massive lens and a background light source, which causes the light to be distorted and amplified in accordance with General Relativity \citep{Einstein1936, Paczynski1986}. As the lens is not required to be luminous, this effect is especially powerful in searching for intrinsically dark objects. Microlensing has been widely applied to search for exoplanets {\citep[e.g.][]{Beaulieu2006, Abe2013, Tsapras2018}, including free-floating planets \citep[e.g.][]{Mroz2018, Sumi2023}, {and} dark stellar remnants \citep[e.g.][]{Wyrzykowski2016, WyrzMandel2020, Kaczmarek2022, Kruszynska2024, Kaczmarek2025}, including the discovery of the first isolated stellar-mass black hole \citep{Sahu2022, Lam2022, Lam2023}. Other applications range from nearby stars \citep[e.g.][]{Paczynski1995, Belokurov2002, Han2008}, where several high-precision lens mass measurements have been made \citep{Sahu2017, Zurlo2018, McGill2023}, to the most central regions of the Galaxy \citep{Paczynski1991, Griest1991}, where microlensing can test} predictions of mass function and Galactic structure models \citep[e.g.][]{Sumi2013, Wyrzykowski2015, Mroz2019, Mroz2024, Specht2020, Koshimoto2021, Kaczmarek2024, Nunota2025}. In summary, thanks to its sensitivity to mass, microlensing is a highly valuable and universal tool for exploring the Milky Way.

Gravitational microlensing is very rare, as the probability of a given source star being lensed is $\sim 10^{-7}$ \citep[e.g.][]{Paczynski1991, EB, Specht2020}. Hence, real-time monitoring of millions of stars is needed. This motivation triggered the first microlensing event searches \citep{Alcock1992, Bennett1993, Udalski1992, Aubourg1993} and continues today with next-generation specialized surveys, such as {the} Optical Gravitational Lensing Experiment \citep[OGLE;][]{OGLEIV}, Microlensing Observations in Astrophysics \citep[MOA;][]{Hearnshaw2006, Sumi2013}, or {the} Korea Microlensing Telescope Network \citep[KMTNet;][]{KMTNet}. Simultaneously, surveys not specifically directed at microlensing, including 
VISTA Variables in {the} Via Lactea \citep[VVV;][]{Minniti2010}, the Zwicky Transient Facility \citep[ZTF;][]{Bellm2019}, or the All-Sky Automated Survey for SuperNovae \citep[ASAS-SN;][]{Shappee2014} have also yielded numerous microlensing event detections \citep[e.g.][]{Navarro2017, Navarro2018Longitude,NavarroForesaken, NavarroFarDisk,NavarroLatitude, Mroz2020ztf, Husseiniova2021, Zhai2023}.

Among the surveys of particularly high interest to the microlensing community is \gaia, the European Space Agency's astrometric mission \citep{thegaiamission}. Between 2014 and 2025, \gaia~was continuously scanning the entire sky, collecting astrometric, photometric and spectroscopic data for {about} two billion sources. While \gaia~was not designed for microlensing {discovery}, it was the first survey to provide all-sky detections of microlensing events over a wide magnitude range, reaching as high in longitude as the Galactic anticentre \citep{DR3lenses}. {The all-sky coverage of \gaia~makes its} microlensing event sample especially valuable for Galactic structure studies and confrontation with models \citep[e.g.][]{Underworld2024}. To maximise the science yield from \gaia's all-sky transients, Gaia Science Alerts \citep{gaia-alerts} were initiated, allowing for fast ground-based follow-up of events of interest and mitigating issues related to \gaia's low observation cadence.

Before the long-awaited Data Release 4\footnote{\url{https://www.cosmos.esa.int/web/gaia/release}} (DR4), \gaia's full power has still not been utilised, as the individual astrometric measurements are not yet publicly available. While photometric data alone do not contain sufficient information to measure lens masses, adding astrometric observations allows for an unambiguous recovery of lens parameters including mass \citep{Hog1995, DominikSahu2000}. In some cases, \gaia~astrometric time series alone could be sufficient to recover lens mass \citep{IKChen2023, Jankovic2025}.

Currently, there are only very few lens mass measurements available{:} for the sample of black hole (BH) candidates selected from MOA/OGLE photometric surveys, including the only known isolated BH so far \citep{Sahu2022, Lam2022, Lam2023}, and for predicted microlensing events \citep{Sahu2017, Zurlo2018, McGill2023}. All of those works used \textit{Hubble Space Telescope} (HST) astrometry{; \citet{Zurlo2018} also used high-contrast imaging with adaptive optics at the Very Large Telescope}.

Capturing the shift in source position caused by gravitational microlensing is a demanding task. Even for the best candidates{, pre-selected for high expected astrometric signal, the apparent shift is of order of 0.1--1\,mas}, which is close to the {HST precision level of 0.3}\,mas expected for well-exposed stars \citep[e.g.][]{Bellini2011}. Coverage during windows of opportunity with sufficient astrometric deviation is needed, while HST observations {generally} need to be scheduled well in advance, with the observing time being highly expensive and oversubscribed.

Conversely, the \gaia~mission routinely {measured} astrometric positions for its {two} billion sources; the precision of those measurements rivals, and on the bright end significantly exceeds, {that of HST} astrometry \citep[e.g.][]{delPino2022, Lindegren2021}. The ability to directly measure masses and the all-sky coverage makes it a powerful instrument to study Galactic populations and structure via the microlensing effect. With all astrometric time-series data from {five and a half} years of operation on track to be released in {December} 2026 in \gaia~DR4, we may soon see the sky through a new lens of {astrometric variability}.

This dataset brings both great promise and unique challenges. Limitations include additional light coming either from the lens or neighbouring stars. Currently, \gaia~only distinguishes separate sources at separations {greater than} 180\,mas \citep{GaiaEDR3} and is highly incomplete up to separations of {about} 1 arcsecond \citep{GaiaEDR3val}. A hard limit is imposed by the {point-spread function} in the {along-scan} direction, which has a median {full width at half-maximum} of 103\,mas \citep{DR1_preprocessing}. All these values significantly exceed angular scales of {standard} microlensing events{, which are at a level of 0.1\,mas;} in {the} case of luminous lenses, astrometric signal also becomes diluted due to the lens light. Furthermore, it is a known effect that \gaia~astrometry is limited in crowded fields due to occasional source confusion \citep{GaiaEDR3val, GaiaverseV, Luna2023, CastroGinard2024}; such fields are also most likely to contain microlensing events. {Besides,} \gaia 's cadence is sparse compared to that of dedicated microlensing ground-based surveys. \gaia~is also limited in highly dust-obscured regions due to observing in the predominantly visible-wavelength $G$ band. Finally, the astrometric data -- in the format of one-dimensional measurements in the time-variable along-scan direction -- is not straightforward to interpret and translate to astrometric shifts caused by lensing.

The potential of a \gaia-like mission for studying astrometric microlensing was already recognised almost two decades before its launch \citep{Hog1995}. This topic then received more attention in the official Gaia Concept and Technology Study \citep{gaiawhitebook}. \citet{Belokurov2002} elaborated in more detail on the unique possibilities and the wealth of microlensing events {that} it would bring. They predicted {that} \gaia~would measure masses of {about} 2500 astrometric events with an error of {less than} $50\%$. Their analysis was built on early predictions for \gaia~mission design {that} were overly optimistic: as we now know, the astrometric along-scan measurement errors are several times higher \citep{Lindegren2021}, and the number of visits per event during the nominal mission also decreased by a factor of {two} \citep{Rybicki2018}. Nonetheless, \citet{Belokurov2002} paved the way for the microlensing discoveries of \gaia. Crucially, they voiced the need to combine event detections with high-cadence ground-based follow-up, which has been realised in Gaia Alerts \citep{gaia-alerts} and has made possible the characterization of \gaia's most prized microlensing events \citep[e.g.][]{Gaia16aye, Gaia19bld, Gaia21blx, Gaia22dkv}.

The advent of the \gaia~era made it possible to predict future events caused by known, nearby lenses \citep[e.g.][]{McGill2018, McGill2020, Bramich2018, Klueter2020}. \citet{Klueter2020} showcased the possibilities of \gaia~astrometry in a systematic study of such predicted events. They concluded {that} it would be possible to detect astrometric deflection in 114 events, including a mass measurement at a {precision of 15\% or better} for 13 events. {In} the regime of predicted events, lenses are bright and proper motions of both lens and source are well-known already before the event{;} contrarily, most lensing events are found serendipitously and more challenging to model.

For dark lenses, the most recent predictions of \gaia 's performance {were} outlined in \citet{Rybicki2018}. They predicted that, over the course of the nominal \gaia~mission, lens masses over 10\,$M_\odot$ could be measured with {accuracies in the range of a few to 15 per cent at the bright event end ($G \lesssim 15.5$\,mag). For events with fainter sources (15.5\,mag < $G$ < 17.5\,mag), \citet{Rybicki2018} stated that lens masses can only be determined if they are over 30\,$M_\odot$.} This was an exploratory analysis, as the predictions were based on a single light curve of an event with a suspected BH lens found in OGLE-III data \citep{Wyrzykowski2016}.

Today, equipped with precise metrics of \gaia's performance based on continuous validation over {the duration of the mission}, close emulators of \gaia~pipelines for the general astrometric solution as well as microlensing event analysis, and state-of-the-art Galactic models, we aim to provide maximally realistic predictions for microlensing in \gaia~DR4 and kickstart the development of data analysis tools for the future \gaia~Data Releases.

The paper is structured as follows. In {Sect.} \ref{sec:data}, we outline the construction of our sample of simulated microlensing events. In {Sect.} \ref{sec:methods}, we introduce the microlensing equations and the data representation employed in \gaia~astrometry; we integrate them into a model of expected observations for a given event. In {Sect.} \ref{sec:results}, we discuss {astrometric and photometric} detectability and the properties of detectable events. We analyse detectability as a function of brightness, lens mass and sky position. For a selected subset of events, we generate realistic mock \gaia~DR4 data. We then fit models to assess how accurately we can recover true event parameters, in particular the lens mass. In {Sect.} \ref{sec:discussion}, we summarise and discuss our work. We suggest {the} best strategies for analysing and mining the upcoming \gaia~DR4 dataset. {The} mock \gaia~data generated in this work, together with code used for its modelling{, are provided in the accompanying GitHub repository\footnote{\url{https://github.com/zofiakaczmarek/gaia-gleams}}}.

\section{Data}
\label{sec:data}

\subsection{Light source population}

For the light source population, we {used} a mock \gaia~Early Data Release 3 catalogue of \citet{Rybizki2020}, hereafter GaiaEDR3mock. The catalogue is based on a modified version of {\tt Galaxia}, a synthetic Milky Way survey generation software \citep{Sharma2011}{, which builds on the Besan\c{c}on model of population synthesis \citep{Robin2003, Robin2004}.}

We {refrained} from the use of the real \gaia~source catalogue due to difficulties with source distances. Precise knowledge of source distance is needed for simulating the event: to determine whether it can happen ({lens distance $D_L$ $<$ source distance $D_S$}) and calculate its essential parameters (\te, \pie, \thetae). \gaia~does not directly report distances, but parallaxes; naively inverting them to obtain distance causes bias \citep{BailerJones2015, Luri2018}. The parallax distribution also has a negative tail, arising as a natural consequence of random noise larger than the true parallax value; while using negative parallaxes as inverse distances gives unphysical results, excluding them would bias the resulting sample to lower distances \citep{Luri2018, GaiaEDR3val}. In consequence, source parallaxes are not sufficient to create a mock catalogue. 

The correct solution for distance determination is to frame it as an inference problem \citep{BailerJones2015}, employing Bayesian statistics and realistic, sky-position dependent Galactic distance priors in combination with the measured parallax and possible auxiliary data. However, while geometric, photogeometric \citep{BailerJones2021} and kinegeometric \citep{BailerJones2023} distance determination have been very successful and found wide applications using this framework, they are still less effective and exhibit some bias in Bulge fields \citep{BailerJones2023}. Our science case is exceptionally impacted by issues in \gaia's distance determination, due to a preference for far-away (i.e. farther than the lenses), Bulge-field (highest probability of microlensing; \citealt{Paczynski1991}) sources. Therefore, we {chose} to use simulated light sources to be maximally careful.

We also {chose} not to use raw simulation (e.g. \texttt{Galaxia}) output, as it would not include sky-variable observational effects, such as \gaia's selection function, extinction and crowding.  GaiaEDR3mock includes a simulation of those effects, while also providing ground truth distance values, making it feasible to construct a realistic population of \gaia~microlensing events. 

\subsection{Lens population}
\label{sec:lenses}
True parameters of the Galactic lens populations, especially dark remnant populations, are not well-constrained. Simulations assuming Galactic models are needed, so that these models can be verified by comparison to the actually observed events. Here, we build on the stellar and dark remnant objects from the \citet{Underworld2022} simulation. Their work {used} a modified version of {\tt Galaxia}, which generates remnants by subjecting stars of sufficient masses to supernova explosions, including natal kicks, at the end of their lifetime. The remnants are then propagated to their present-day positions and kinematics, using {the {\tt StellarMortis} software\footnote{\url{https://github.com/David-Sweeney/StellarMortis}} \citep{StellarMortis} and} the Milky Way potential implemented in {\tt galpy} \citep{Bovy2015}.

{We list the models used for generating remnants from progenitor stars, including initial-final mass relations (IFMRs) and natal kick distributions, in Table~\ref{tab:populations}. `\citet{Igoshev2020}' refers to the kick distribution for young pulsars therein. The `scaled \citet{Igoshev2020}' distribution is rescaled by the remnant mass as $M_{\rm rem} / 1.35 M_{\odot}$ and truncated to 0 km/s for progenitor masses $ \geq 40 M_{\odot}$, which is assumed to be the boundary for direct collapse, following \citet{Underworld2022}. For more information on the underlying simulation we refer to \citet{Underworld2022}.}

We {added} two modifications with respect to \citet{Underworld2022, Underworld2024}. Firstly, the original simulation {returned} mono-mass neutron star (NS) and BH populations, with masses fixed at median values from the literature: $M_{\rm NS} = 1.35\,M_{\odot}$ and $M_{\rm BH} = 7.8\,M_{\odot}$, respectively. However, we expect the remnant {lens} mass distributions to be skewed upwards with respect to the original remnant mass distributions, as more massive objects have larger Einstein radii. For example, \citet{Kaczmarek2025} find average BH lens masses in OGLE-like simulation fields between 9.2 and 12.6\,$M_{\odot}$ (depending on the {IFMR} applied), using the {\tt PopSyCLE} \citep{Lam2020} simulation software. To study this effect, we {applied} the \citet{Spera2015} {IFMR} to obtain the remnant mass $M_{\rm rem}(M_{\rm ZAMS}, Z)$ with the {\tt SPISEA} software \citep{SPISEA}. {We {capped} the metallicity distribution at [Fe/H] = 0.33 to avoid anomalous behaviour of the $M_{\rm rem}(M_{\rm ZAMS}, Z)$ function.}
We {chose to use} the \citet{Spera2015} model as it significantly outperforms other IFMR models in a recent study reproducing OGLE-IV microlensing data \citep{Perkins2025}, as well as for its analytical form making results easily reproducible. BH natal kicks {were} also rescaled (as in \citealt{Underworld2022}) to match the assigned mass before propagating to the present-day position.

Secondly, the original simulation {applied} a lower cut to progenitor mass at 8\,$M_\odot$ and {did} not include white dwarfs (WDs). However, WDs are expected to cause {about} 10\% of Galactic microlensing events, constituting the second largest (after stars) population of lenses \citep{Lam2020}. We {added} a WD population using the same {\tt Galaxia} simulation (now filtering for stars with progenitor masses {lower than} 8\,$M_\odot$ that had exhausted their life cycles) and applying the \citet{Kalirai2008} initial-final mass relation. Following \citet{Hamers2019} and \citet{ElBadry2018}, we also {subjected} the WDs to small natal kicks, {drawn randomly from} a Maxwellian distribution with $\sigma = 0.5$\,{\,km\,s$^{-1}$}. All remnants {were} propagated {from their progenitors} to present-day {Galactic positions and velocities} using the {\tt StellarMortis} software.

We note that our lens population sizes are decreased by the escape of kicked remnants {formed in supernova explosions} from the Galaxy \citep[e.g.][]{Olejak2022, Underworld2022}. {In our samples,} $39\,976$ out of $130\,749$ BHs (31\%) and $114\,746$ out of $304\,131$ NSs (38\%) become unbound from the Milky Way. That corresponds to total stellar-origin, Galaxy-bound populations of $9.1 \times 10^7$ BHs and $1.7 \times 10^8$ NSs, which is in agreement with values predicted by current state-of-the-art population synthesis studies \citep{Sartore2010, Olejak2022}.

\begin{table*}[]
\caption{Summary of the building blocks of the lens population. }
\begin{center}
\begin{tabular}{p{2.5 cm} p{1.8cm} p{1.6 cm} p{3.1 cm} p{3.8cm}}
\hline \hline
Population & Samples & $f_N$ & IFMR & Natal kick distribution \\
\hline
Black holes & $130\,749$  & $10^{-3}$ & \citet{Spera2015} & scaled \citet{Igoshev2020} \\
Neutron stars & $304\,131$  & $10^{-3}$ & \citet{Spera2015} & \citet{Igoshev2020} \\
White dwarfs & {$8\,597\,010$} & {$10^{-3}$} &  \citet{Kalirai2008} & Maxwellian, $\sigma=0.5$ km/s \\
Stars & $7\,050\,937$ & $5 \times 10^{-5}$ & N/A & N/A \\
\hline
\end{tabular}
\end{center}
{\footnotesize \textbf{Notes: }$f_N$ represents the fraction of the total Milky Way population contained in the samples. More information on the lens population can be found in {Sect.}~\ref{sec:lenses}.}
\label{tab:populations}
\end{table*}

\subsection{{Subsampling}}

As evolving the remnants and querying the background source catalogue around lens positions {are} computationally expensive, the populations from {Sect.} \ref{sec:lenses} {were} {subsampled} by a given factor {$f_N$ (fraction of the total Galactic population simulated; listed in Table~\ref{tab:populations}). We compensated for the subsampling of the lens population by matching lenses and sources over $10^3$ mock DR4 surveys. To allow for encountering different background sources, the reference positions of the lenses need to be shifted between mock surveys, but with offsets small enough to preserve properties and densities of their corresponding background source populations. For simplicity, this was realised by extending the total interval over which linear lens--source motions were tracked to $10^3$ times the duration of DR4; as total displacements of the lenses over this interval are only on sub-arcminute levels, {which is} not enough to significantly affect background source populations, the aggregated results are a good approximation of a single DR4 survey with no subsampling. We note that, in this step, we do not aim to reproduce the realistic Galactic orbits of the tracked objects over thousands of years; instead, we simply create $10^3$ linear-motion approximations over very small patches of the sky. The method described here is identical to that used by \citet{Underworld2024} (there termed {``}undersampling{''}); we refer the reader to their study for further details and discussion.} 

{Our final yield of unique events is equal to $f_{\rm sub} = 10^3 f_N$ of that expected in DR4. This factor is equal to 1 for all populations of remnants (BH, NS{,} and WD), but not for stars, which are so numerous in the Galaxy that even under-simulating lensing by $\sim10^{-3}$ of them would be too computationally challenging. Therefore, all stellar lens yields are corrected by $1/f_{\rm sub}$. While imperfect, this simulation design prioritizes remnant lenses as our objects of scientific interest, yet retaining a sufficiently large comparison population of stellar lenses. Additionally, simulating events with stellar lenses is limited by the unpredictable and not fully understood behaviour of \gaia 's instrumentation and astrometric pipeline for very close pairs ({discussed in detail} in {Sect.} \ref{sec:detectable}), and therefore already impacted by severe uncertainties.}

Generating fewer potential lenses and following them for a longer time may make the final distribution of lensing event parameters noisier in cases where one lens causes multiple events. However, it greatly reduces the required computational resources. In the photometrically detectable event sample ({Sect.} \ref{sec:results}), only {small fractions} of events (0.1\%, 0.{3}\%, 3.0\%, 0.0\% respectively for stars, WDs, NSs{,} and BHs) have lenses that are duplicates from previously listed events; only the NS parameter distributions are significantly impacted by repeated events. {Among events with both photometric and astrometric signal,} the sole lens causing repeated events is a NS. This is consistent with the expectation of \citet{Kaczmarek2026} that NSs are likely to be mesolenses{:} single objects that sweep a large on-sky area susceptible to lensing per unit time, and could even cause repeated events.

\subsection{Lens--source pairs}
\label{sec:matching}

We then {used} a procedure similar to \citet{Underworld2024} to match lens -- light source pairs. We first {constructed} a database of potential background sources. To each lens, we {assigned} a maximum angular {separation} $d_{\rm max}$ from the position of the lens at the midpoint of the simulation time interval, where:

\begin{equation}
    d_{\rm max, L} = (\mu_{\rm L} + \mu_{\rm S, max})\frac {t}{2} + u_{\rm max}\theta_{\rm E,max,L} + \pi_{\rm L}\textnormal{,}
    \label{eq:buffer}
\end{equation}

\noindent {which} still makes it possible to approach the lens at the defined impact parameter $u_{\rm max}$ during the simulated survey time. Here, $\mu_{\rm L}$ is the proper motion of the lens and $\mu_{\rm S, max}$ is an approximate upper limit on the expected proper motion of the star. We {assumed} $\mu_{\rm S, max} = 10$\,mas\,yr$^{-1}$, as 95\% of \gaia~stars have lower proper motions (and nearby stars rarely become sources in lensing events). {The interval over which we track our objects, $t$}, is equal to {$10^3$} times the DR4 timespan {-- therefore} the first factor in Eq.~(\ref{eq:buffer}) dominates and is $\sim 10\,''$.  $u_{\rm max}$ is the highest allowed impact parameter; in order to also enable creating mock data for purely astrometric events, we fix $u_{\rm max} = 10$. $\theta_{\rm E,max,L}$ is the maximum physically allowed Einstein radius for a given lens {(}i.e. calculated for a source at infinity {from} Eq.~{(}\ref{eq:thetaE}{)}{)}; $\pi_{\rm L}$ is the lens parallax, equal to the maximum possible deviation from straight line source-lens motion {(}also calculated for a source at infinity {from} Eq.~{(}\ref{eq:thetaE}{)}{)}.

We {queried} the GaiaEDR3mock catalogue for stars where $\exists_{\rm L} \ d(\textnormal{star, L}) < d_{\rm max, L}$, where L refers to objects from the lens population ({Sect.}~\ref{sec:lenses}). We {eliminated} stars that are nearer than their candidate lenses. As the catalogue includes the \gaia~selection function as the probability {\tt d11y} of a source being detected, we also {applied} a random number filter where each star {had} {a} probability of 1 - {\tt d11y} to be eliminated.

We then {simulated} relative star--lens straight line motion and only {selected} the stars that actually {had} a closest approach to the lens at a {separation smaller {than} or equal to} $u_{\rm max}\theta_{\rm E,max,L} + \pi_{\rm L}$ during the simulated survey time. In the last step, we {re-simulated} the relative star-lens motion with the addition of parallax, and {recalculated} $\theta_{\rm E}$ values including the individual star distances {$D_S$}, for the remaining stars. Pairs that {reached a separation below the limit of} $u_{\rm max} \theta_{\rm E}$ {were} then saved into the close approach database to be later analysed in detail ({Sect.}~\ref{sec:results}).

\section{Methods}
\label{sec:methods}

\subsection{Basic microlensing relations}
\label{sec:microlensing_eqs}

A combination of a lens of mass $M_{\rm L}$ situated at distance $D_{\rm L}$ and a source at distance $D_{\rm S}$ defines the Einstein radius $\theta_{\rm E}$:

\begin{equation}
    \theta_{\rm E} = \sqrt{\frac{4GM_{\rm L}}{c^2} \left(\frac{1}{D_{\rm L}} - \frac{1}{D_{\rm S}}\right)}{,}
    \label{eq:thetaE}
\end{equation}

\noindent which describes the angular on-sky scale of the resulting microlensing event.

{In a standard microlensing event,} this scale is very small {(most commonly around 0.1\,mas)}, making it impossible to resolve the generated source images. Instead, the microlensing event is observed as a brightening of the previously invariable source (and, only in rare detections so far, as an astrometric shift of the source's photocentre). Because the light curve information alone is not enough to uniquely determine lens mass, Galactic and stellar models {usually} need to be employed to inject additional information and therefore determine the nature of the lens \citep[e.g.][]{Wyrzykowski2016, Kaczmarek2022, Kaczmarek2025, Kruszynska2024, Bachelet2024, Howil2024, Pylypenko2026}; uncertainties on masses inferred in this way are large.

{For a standard microlensing event, there are} only {two} parameters tied to physical information about the lens {that} can be determined from the light curve. {The first one is} the timescale of the event (called Einstein time and denoted $t_{\rm E}$), which is the time it takes to cross $\theta_{\rm E}$ in relative source-lens motion $\mu_{\rm rel}$: 

\begin{equation}
t_{\rm E} = \frac{\theta_{\rm E}}{\mu_{\rm rel}}.
\end{equation}

\noindent {The second one is the} microlensing parallax ($\vec{\pi_{\rm E}}$), with a value of:

\begin{equation}
    \pi_{\rm E} = \frac{a_\oplus}{\theta_{\rm E}}\left(\frac{1}{D_{\rm L}} - \frac{1}{D_{\rm S}}\right),
    \label{eq:piE}
\end{equation}

\noindent where $a_\oplus = 1$\,au, and the direction of relative lens--source motion. $\vec{\pi_{\rm E}}$ defines the asymmetric patterns in the light curve, occurring with a period of 1\,year and observed as a result of the line of sight slightly changing during orbital motion of the observer. Due to the small amplitude and long period of those changes, as well as multiple parameter degeneracies \citep[][]{Sm03, Gould2004}, $\vec{\pi_{\rm E}}$ is usually weakly constrained from the light curve, and can only be well-determined for long ($t_{\rm E}$ {of order of} 1 year) events and under a sufficiently dense observing cadence.

The position vector of the source with respect to the lens $\bm{u}$ is defined in units of $\theta_{\rm E}$, and is a sum of linear motion and a ({relatively} small) parallax deviation caused by the observer's orbital motion. To define it in objective North--East coordinates, we apply a rotation $\bm{R}$:

\begin{equation}
\bm{u}(t) = \bm{u}_{lin}(t) + \bm{u}_{par}(t) = \bm{R} (\pi - \varphi )\begin{pmatrix} \frac{t-t_0}{t_{\rm E}} \\ u_0 \end{pmatrix} - \pi_{\rm E} \begin{pmatrix} f_{\delta \ }(t) \\ f_{\alpha}(t) \end{pmatrix},
\label{eq:u_general}
\end{equation}

\noindent where $u_0$ is the impact parameter, $\pi_{\rm E}$ is the microlensing parallax, $\varphi$ is the angle of lens--source motion counted from North to East, and $f_{\alpha}, f_{\delta}$ are the parallax factors in the North and East directions, respectively; the parallax factors describe a position on the   unit parallax ellipse (i.e. projected orbit) defined for coordinates ($\alpha$, $\delta$) at time $t$. The minus in the parallax component is due to the source tracing a smaller parallax ellipse than the foreground lens.

The lens--source separation $\bm{u}$ can be directly tied to the corresponding observed position and brightness.
The change in observed source magnitude is:

\begin{equation}
    \Delta m = -2.5 \log{A},
    \newline
    A = \frac{u^2 + 2}{u\sqrt{u^2 + 4}},
    \label{eq:ampl}
\end{equation}

\noindent where $u = |\bm{u}|$ and $A$ is the amplification of the source star flux. The astrometric shift $\Delta \bm{u}$ (in units of $\theta_{\rm E}$) of the photocentre, with respect to the source situated at position $\bm{u}$, is simply:

\begin{equation}
    \Delta \bm{u} = \frac{ \bm{u} }{|\bm{u}|^2+2}
    \label{eq:displacement}
\end{equation}

\noindent \citep{Hog1995, Walker1995, Miyamoto1995}, making $\theta_{\rm E}$ directly measurable from astrometric shift observations. {As the observed source position is obtained by averaging the apparent positions of the two images on either side of the source, the observed astrometric shift must be smaller than the angular scale $\theta_E$; from Eq.~{(}\ref{eq:displacement}{)}, it is strictly no larger than $ \sqrt{2}\theta_E/4$.}

From Eqs.~(\ref{eq:thetaE}) and~(\ref{eq:piE}), lens mass can be expressed as

\begin{equation}
    M_{\rm L} = \frac{\theta_{\rm E}}{\kappa \pi_{\rm E}}\text{,}
    \label{eq:lens_mass}
\end{equation}

\noindent where $\kappa = 4G/c^2$ is a constant. Therefore, combining photometric and astrometric observations leads to a direct mass measurement.

\subsection{Parametrization for \gaia~fits}

\gaia~epoch astrometry, including the upcoming DR4 database, operates in local plane coordinates (LPCs). We refer to \citet{LPC} for more details on the coordinate system and to \citet{Gaia_astrom_core} for a description of the \gaia~astrometric core solution, including the corrections applied; here we provide a short outline.

With each astrometric observation at time $t$, \gaia~{reports} a datapoint containing $(t, w, \sigma_w, \psi, f_w, f_z)$. $w$ is the target measurement of the along-scan position of the source with an error $\sigma_w$, while scan angle $\psi$, and parallax factors in along- and across-scan directions $f_w$, $f_z$ are metadata describing the geometry of the observation. The reported values include corrections for effects of the spacecraft's position and motion, such as aberration, gravitational deflection from Solar System objects or barycentric time correction (R\"{o}mer delay). In particular, the parallax factors account for the orbit of \gaia, so that no spacecraft ephemeris is necessary to analyse time-series astrometry. As such, the LPCs are a maximally simple coordinate system to analyse astrometric signals from an astrophysical source, alleviating the need to consider effects that are not inherent to that source.

The observation strategy containing sets of $(t, \psi, f_w, f_z)$ for each $\alpha, \delta$ position {had} been planned in advance and defined in the so-called \textit{\gaia}~scanning law \citep{LindegrenBastian2010, deBruijne2010, thegaiamission}.

The transformation between an offset from the reference position expressed in local celestial coordinates, ($\Delta \alpha \cos \delta, \Delta \delta$), and in LPCs ($w$, $z$) is uniquely defined by the scan angle $\psi$ \citep{LPC}. When using the LPCs, the parallax component $\bm{u}_{par}$ in Eq.~(\ref{eq:u_general}) can be easily expressed with the $(f_w, f_z)$ values reported by \gaia~for each observation:

\begin{equation}
\bm{u}_{par} = - \pi_{\rm E} \begin{pmatrix} f_w \\ f_z\end{pmatrix}.
\end{equation}

As until now we have expressed all on-sky positions in units of $\theta_{\rm E}$, to get the observable lensing signals $\Delta w$, we need to rescale the along-scan gravitational shift:
\begin{equation}
\Delta w = \theta_{\rm E} \Delta \bm{u}_w =  \theta_{\rm E}  \frac{ \bm{u \cdot e_w}}{|\bm{u}|^2+2},
\label{eq:linear_theta}
\end{equation}

\noindent where the $w$ subscript denotes the along-scan direction, and $\bm{e_w}$ is the unit vector in that direction. The actual observed along-scan position $w$ is then:

\begin{equation}
w^O = w_{\rm 5p,S} + \Delta w\text{,}
\label{eq:overall_astrometry}
\end{equation}

\noindent where $w_{\rm 5p,S}$ is the position of the source in standard {{five-parameter}} motion \citep[e.g.][]{Gaia_astrom_core}, which is defined by the parameters $\alpha_{\rm 0,S}, \delta_{\rm 0,S}, \mu_{\alpha^*,\text{S}}, \mu_{\delta, \text{S}}, \varpi_{\rm S}$ {(where $\alpha^*$ is used as a shorthand notation for $\alpha \cos \delta$)}.

Together with the photometric model of Eq.~(\ref{eq:ampl}), we have defined a complete model of \gaia~time-series of a microlensing event, which we can now apply to the simulated events.

\section{Results}

\begin{figure}
	\includegraphics[width=\columnwidth]{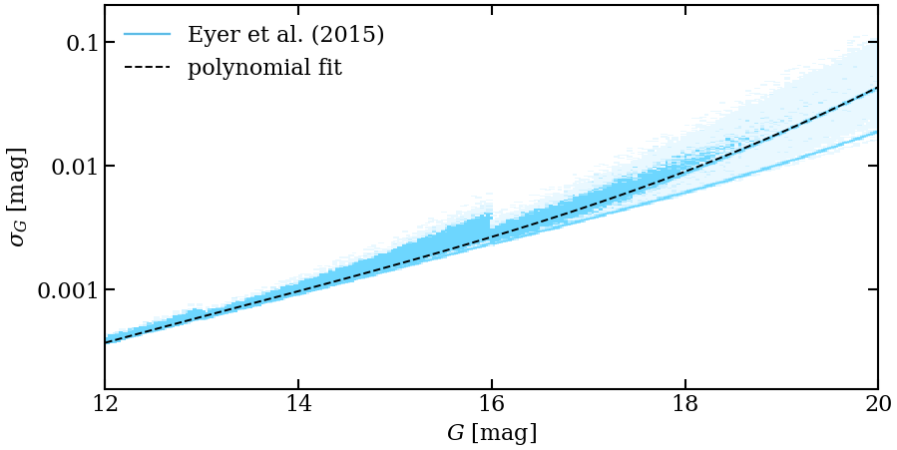}
    \caption{{Models of} per-transit \gaia~$G$-band photometric precision {as a function of magnitude}. The background pixel map is as in {Fig.} 3 {of} \citet{Eyer2015}{, adapted with author permission; blue points represent simulated data and blue solid lines represent model fits (lower: pre-launch expectations, upper: mean in-orbit performance). The polynomial approximation of the mean in-orbit performance by Eq.~(\ref{eq:sigma_g})} is overplotted (black, dashed).}
    \label{fig:sigma_g_fit}
\end{figure}

\label{sec:results}

\subsection{Detectability}
\label{sec:detectable}

\begin{figure*}
	\includegraphics[width=2\columnwidth]{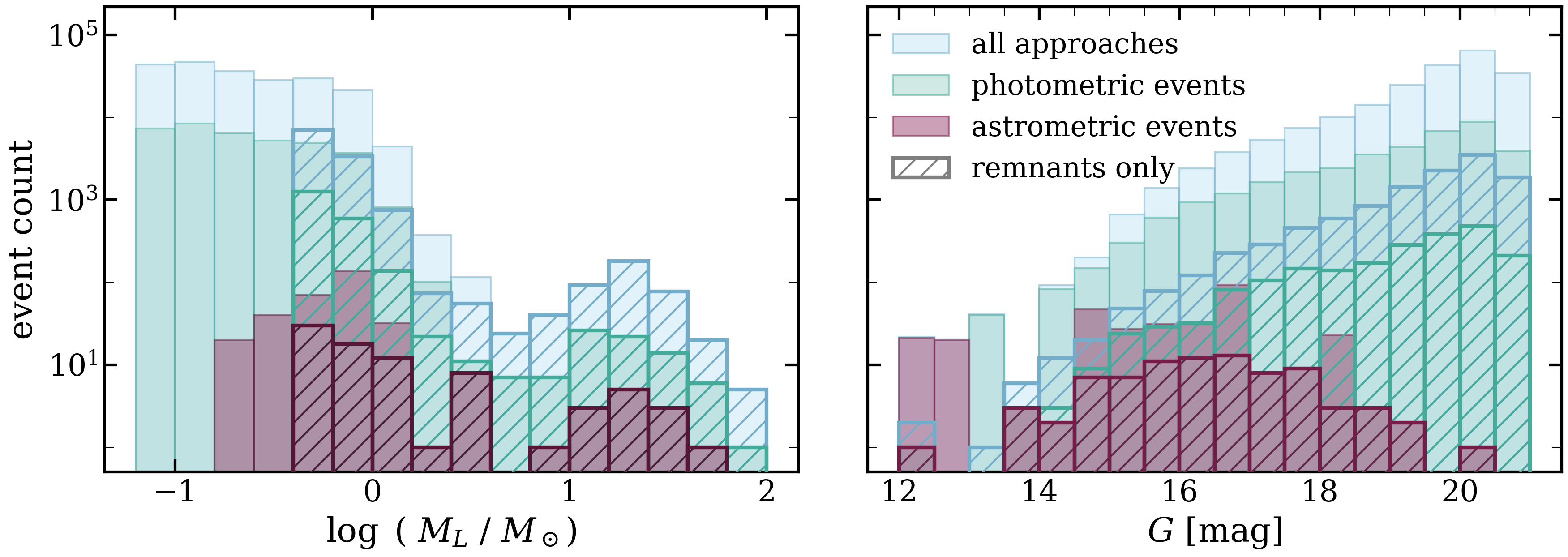}
    \caption{Distributions of all simulated close ($\min(u) \leq 10$) approaches (light blue), events detectable in \gaia~photometry {(green)}, and events detectable in \gaia~astrometry (wine red). {Filled bins represent all events (re-weighted by 1/{$f_{\rm sub}$} for the stellar lens population); hatched bins represent only the remnant lens subset.} \textit{Left:} A histogram of event counts binned by lens mass. \textit{Right:} A histogram of event counts binned by source $G$ magnitude.}
    \label{fig:detectability_hist}
\end{figure*}

{We used} {\tt astromet}\footnote{\url{https://github.com/zpenoyre/astromet.py}} to simulate \gaia~data. {\tt astromet} is a software package for generating astrometric tracks, producing mock \gaia-like observations and emulating \gaia's astrometric fitting pipeline, which has been developed for analysing binary systems \citep{Penoyre2022, Penoyre2022b} and subsequently extended and applied to microlensing events \citep[e.g.][]{Jablonska2022}.

To filter the simulated close approaches and find events whose signal could potentially be detectable by \gaia, we {applied} the criterion of lensing signal being {above the }3$\sigma$ level for per-transit observations (i.e. from near-simultaneous observations from 9 CCDs aggregated as the star passes through \gaia 's focal plane), corresponding to {above the }1$\sigma$ level for per-CCD observations. Precision levels {were} determined conservatively using the baseline source star $G$ magnitude.

For photometry, we {approximated} $\sigma_G$, the per-transit photometric precision, with a polynomial:
\begin{equation}
    \log \sigma_G (G) = 0.001404G^3 - 0.0563G^2 + 0.958G - 9.249.
    \label{eq:sigma_g}
\end{equation}

\noindent {We determined the coefficients with a third-degree polynomial fit} to points from the $G \in$ (12\,{mag}, 21\,{mag}) range in Fig. 3 of \citet{Eyer2015}. {In Fig.~\ref{fig:sigma_g_fit}, we show the original simulated data and models of \citet{Eyer2015}, which include scatter due to different configurations of stray light, as well as our polynomial fit.} For each close approach, we {calculated} $\max_t \Delta G(u(t))$ -- the maximum change in observed source magnitude {from Eq.~(}\ref{eq:ampl}{)} in the \gaia~$G$ band for $t_\text{start, DR4} < t < t_\text{end, DR4}$. Events passing the $\max_t \Delta G(u(t)) > 3 \sigma_G(G)$ cut will hereafter be referred to as {photometric events}.

Similarly, to identify events detectable in astrometry, we {assigned} to each source its maximum possible astrometric shift $\delta_{\max}$. As can be simply derived from Eq.~(\ref{eq:displacement}), this shift reaches {its} maximum at $u=\sqrt{2}$ and then falls monotonically, hence:
\begin{equation}
     \delta_{\max}(u_0, \theta_{\rm E}) = \Delta u (\max(\sqrt{2}, u_0)) \theta_{\rm E}.
\end{equation}

\noindent We {filtered} by $\delta_{\max}(u_0, \theta_{\rm E}) > \sigma_{\rm AL}(G)$, where $\sigma_{\rm AL}$ is the per-CCD precision of \gaia~along-scan direction observations (i.e. {criteria equivalent to those} used for photometric events) as derived {by} \citet{Lindegren2021} from robust scatter estimation and implemented in {\tt astromet}. Events passing this cut will hereafter be referred to as {astrometric events}.

Based on these criteria, our simulated DR4 survey contains {322 $\pm$ 70} astrometric events with detectable signal, out of which {209 $\pm$ 57} have both photometric and astrometric signal. These estimates include the re-weighting by 1/{$f_{\rm sub}$} for stellar lenses (Table~\ref{tab:populations}){. All} reported uncertainties only encompass Poisson noise and not the uncertainty of the underlying parameters of the simulation. {82 $\pm$ 9} and {49 $\pm$ 7} of these {events}, respectively, are caused by remnants. Some events in DR4 will be detectable in astrometry but not in photometry -- following from the yields listed above, we expect 113 $\pm$ 40 such events including 33 $\pm$ 6 with remnant lenses ({about one third} of all stellar-lens and {two fifths} of all remnant-lens astrometric events). While in this work we focus on events having detectable signals in both channels, astrometry-only events could still be modelled for lens parameters and, in some cases, even yield precise lens mass measurements{. For details on fitting astrometry-only microlensing events in \gaia, including recoverability of their parameters, we refer to \citet{Jankovic2025} and the {\tt GAME-Filter} repository\footnote{\url{https://github.com/tajjankovic/GAME-Filter/}}.}

This designation as {`photometric'} or {`astrometric'} events does not guarantee {discovery}, as the \gaia~cadence is sparse and the time when signal is detectable may be missed. This will be apparent in {Sect.}~\ref{sec:mock_obs}--\ref{sec:models}, where some events will not have indications of anomalous signal. We intentionally {modelled} these events as well in order to assess how gaps in the \gaia~cadence affect lens characterisation.

\begin{figure}
	\includegraphics[width=\columnwidth]{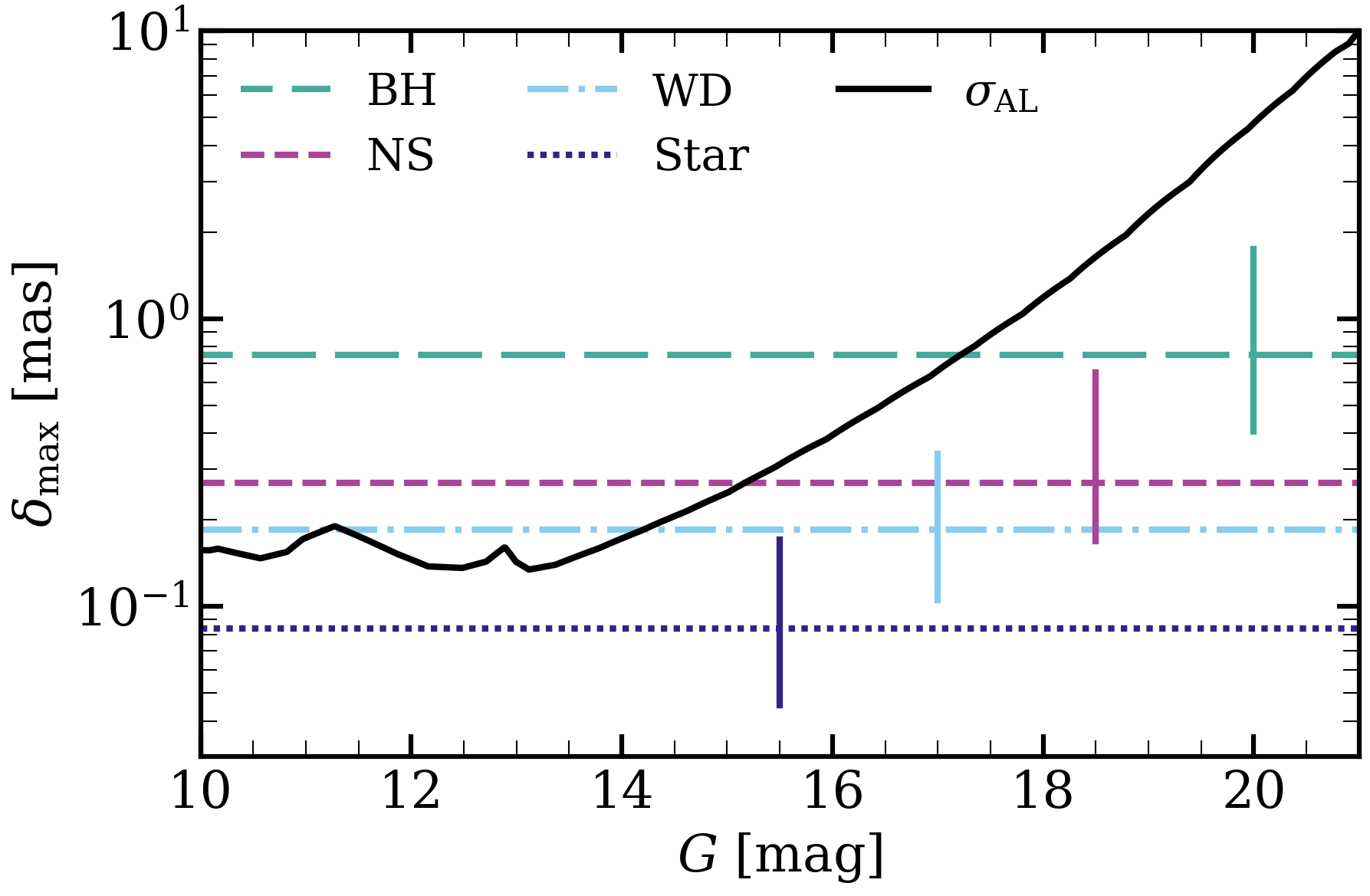}
    \caption{Typical astrometric signal for photometric microlensing events from each lens class (from top: black hole - green long dash, neutron star - purple short dash, white dwarf - light blue dash-dotted, star - dark blue dotted). For each class, the horizontal line is the median value of $\delta_{\rm max}$ in the sample of all photometric events, while the error bar represents {16-84 percentile ranges} (the $x$-axis positions of the error bars encode no information, and a horizontal offset of 1.5 mag is applied only for figure clarity). Overplotted in solid black is the per-CCD precision of \gaia~{along-scan} observations, $\sigma_{\rm AL}$.}
    \label{fig:thresholds_AL}
\end{figure}

We {did} not include the effects of blending with neighbour stars or with the lens in these detectability criteria. We checked that including the effect of blending with a luminous stellar lens on the light curve would only impact classifications of $\lesssim 1\%$ (16 removed, 1 added) of simulated photometric events with star lenses. However, modelling the effect of blending on astrometry is difficult. There is no clear analytic mapping from combinations of flux ratios and angular {separations} of close pairs to the treatment of the scan result in the \gaia~pipeline, and factors impacting this treatment are complex \citep[e.g.][]{Torra2021, GaiaEDR3val, Holl2023, Cifuentes2025}. Possible scenarios include \gaia~reporting two resolved sources, one source situated at the photocentre, one source situated at the position of either component, or none (spurious detections, rejected transits, failed crossmatches, outliers not to be included in the astrometric solution); {\tt astromet} applies a simplified treatment, always situating data points at the photocentre. We leave a more careful handling of astrometric events with luminous lenses and their detectability to future work.

\begin{figure}
	\includegraphics[width=\columnwidth]{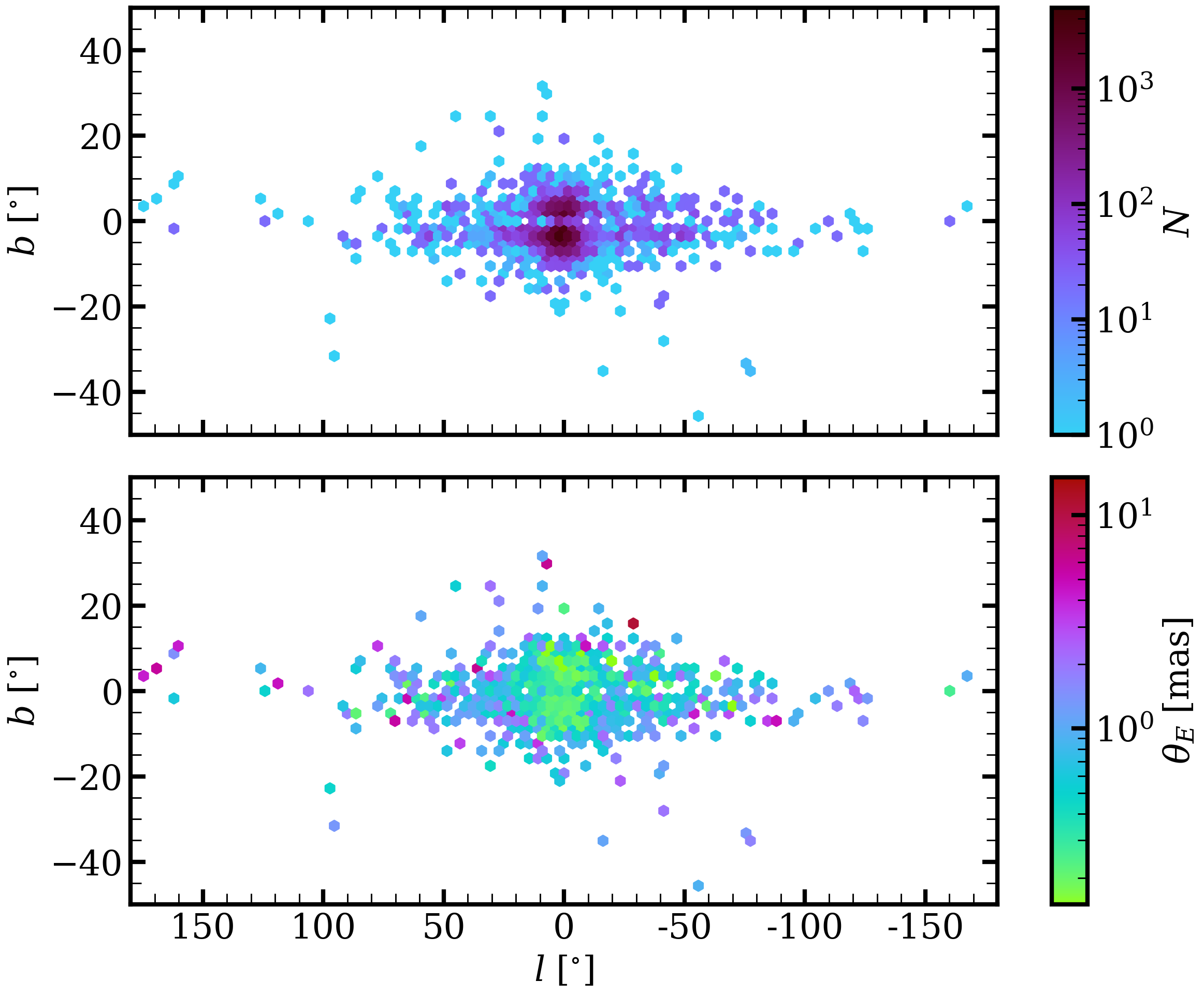}
    \caption{On-sky distributions of all photometric events in the simulated event catalogue. Events with stellar lenses have been upsampled accordingly by 1/{$f_{\rm sub}$}. \textit{Top:} A 2D hexagonal binning of simulated events over Galactic longitude and latitude, coloured by the number of microlensing events per bin. \textit{Bottom:} Like top panel, but with bins coloured by median $\theta_{\rm E}$ values.}
    \label{fig:phot_thetaE}
\end{figure}

\begin{figure*}
	\centering
    \includegraphics[width=1.9\columnwidth]{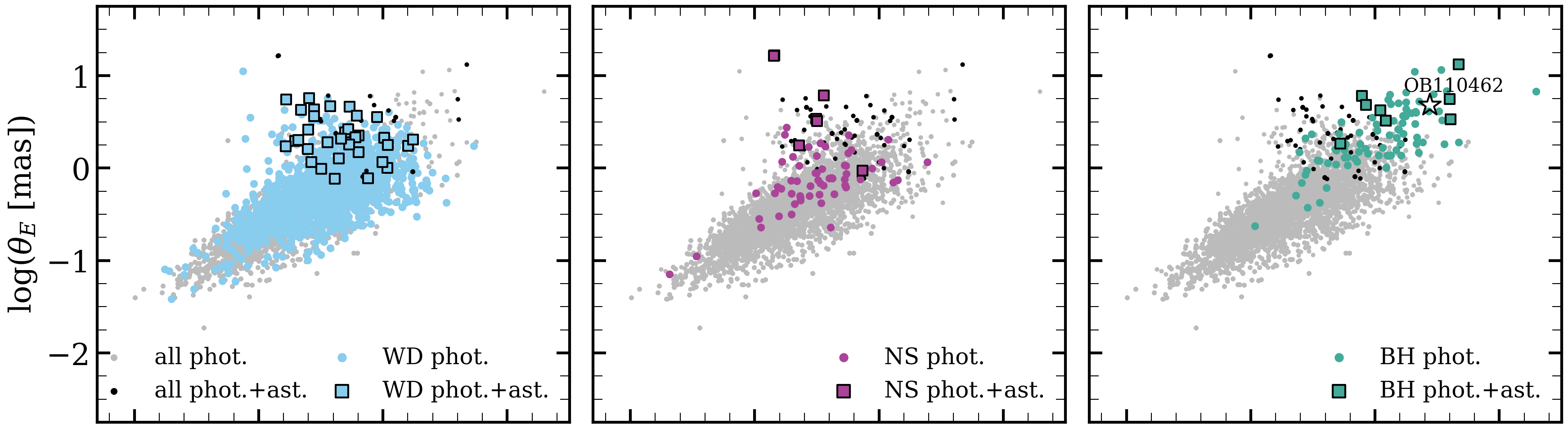}
    \includegraphics[width=1.9\columnwidth]{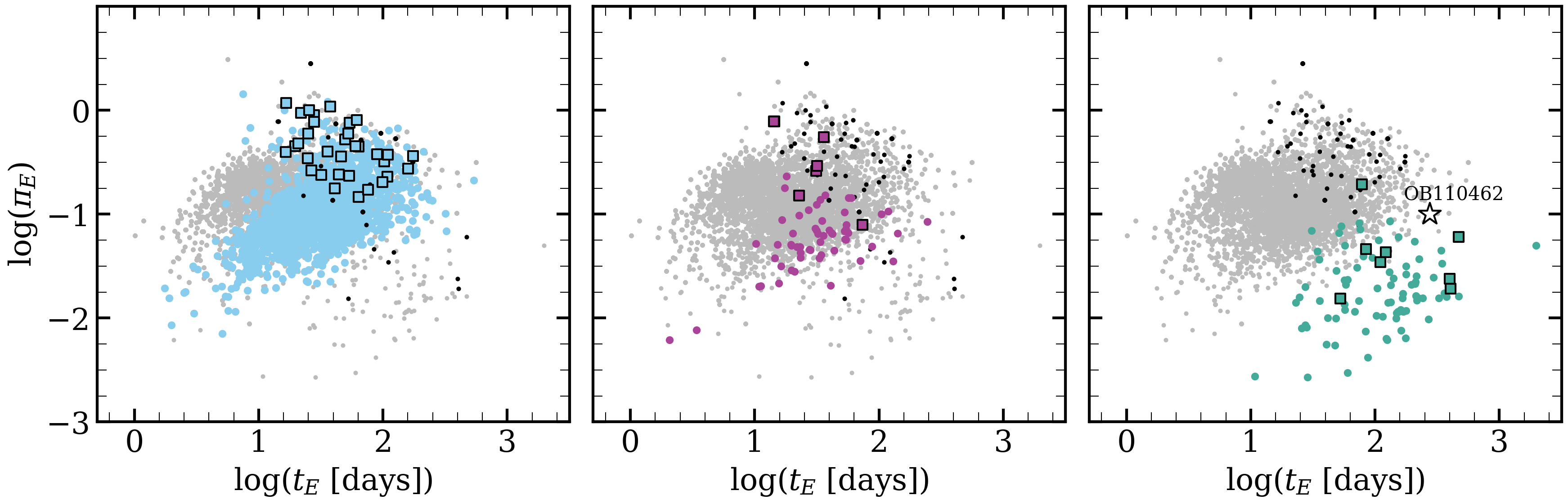}
    
    \caption{Parameter distributions of simulated photometric events, coloured by lens class and astrometric detectability. For each remnant lens class {(left: WD, center: NS, right: BH)}, events only detectable in photometry are marked with coloured circles, while events detectable in photometry and astrometry are marked with large coloured squares with black outlines. For comparison, events from all other lens classes are plotted with small grey (photometry-only) and black (astrometry) circles in the background. The top subplots show the $\log t_{\rm E}$ -- $\log \theta_{\rm E}$ distribution; the bottom subplots {show} the $\log t_{\rm E}$ -- $\log \pi_{\rm E}$ distribution. The only confirmed isolated stellar-mass BH, {OGLE-2011-BLG-0462 (abbreviated to OB110462 in the label)}, is {included in the right panel subplots} for comparison {(star, black edges)}.}
    \label{fig:populations_det}
\end{figure*}

In Fig.~\ref{fig:detectability_hist}, we show lens mass and baseline magnitude distributions of all {approaches where $\min(u) < 10$, photometric events, and }astrometric events. Both photometric and astrometric detectability show selection effects favouring bright sources, as \gaia 's precision is a function of $G$ magnitude{; these effects are particularly strong for astrometric detectability. Fractions of events detectable in photometry remain constant throughout the mass range, while astrometric detectability is strongly skewed towards higher lens masses. Remnant lenses contribute significant fractions of the astrometric event count, especially at the faint end.}

In Fig.~\ref{fig:thresholds_AL}, we compare \gaia 's precision to expected signals and estimate the $G$ levels at which photometric events from a given lens class have detectable astrometric counterparts. Taking the `typical' signal to be the median of the expected signal distribution, we {show} that a typical BH (NS, WD) lens in a photometric event produces {an} astrometric signal at the level \gaia~precision reaches at $G \leq$ 17 (15, 14){\,mag.} {A typical stellar lens causes an astrometric signal too small to be detected even at \gaia 's precision limit. (All stellar lenses that cause astrometric events in the simulated catalogue are outliers from the stellar lens $\theta_{\rm E}$ distribution, placing above the 93$^\text{th}$ percentile.)}

\subsection{Distributions of lensing events}
\label{sec:distrs}

In Fig.~\ref{fig:phot_thetaE}, we show the on-sky distribution of all photometric events. We argue that {while the {Galactic bulge} fields contain the bulk of microlensing events}, off-bulge events could be of particular interest {for remnant searches}, as they are strongly pre-selected for high $\theta_{\rm E}$ due to low background source density. Furthermore, these events could be uniquely accessible {to} \gaia~(as specialised microlensing surveys focus on the central Galactic regions) and have the best \gaia~data quality, as they are not impacted by the effects limiting astrometry in dense fields \citep[e.g.][]{CastroGinard2024}.

In Fig.~\ref{fig:populations_det}, we show distributions of photometric events in the space of observables{, focusing} on the detectability and signatures of stellar remnants. {The WD population has similar observables to those of stars; only the subset selected for the highest $\theta_E$, which also preferably has high $\pi_E$ (nearby lenses), can cause astrometric events. In the NS population, most NS lenses causing detectable astrometric signal belong to a feature outlying towards the upper-left of the main distribution in $\log t_{\rm E}$ -- $\log \theta_{\rm E}$, as reported by \citet{Kaczmarek2026}; however, in photometry, NSs are indistinguishable from stellar lenses.
The BH population is situated in the highest-signal zone (high $t_{\rm E}$, high $\theta_{\rm E}$), with many of the events causing detectable astrometric signal. Measured parameters of the only known isolated stellar-mass BH, {OGLE-2011-BLG-0462}, are consistent with our simulated BH distributions. We also find these parameter distributions consistent with other microlensing survey simulations in the literature \citep{Lam2020, Fardeen2024, Koshimoto2024, Kaczmarek2025, Sallaberry2025}, with the note that we see a relatively large spread in $\pi_{\rm E}$ values of our BH lenses.}

\subsection{Mock \gaia~observations and {astrometric} fits}
\label{sec:mock_obs}

\begin{figure}[h]
	\includegraphics[width=\columnwidth]{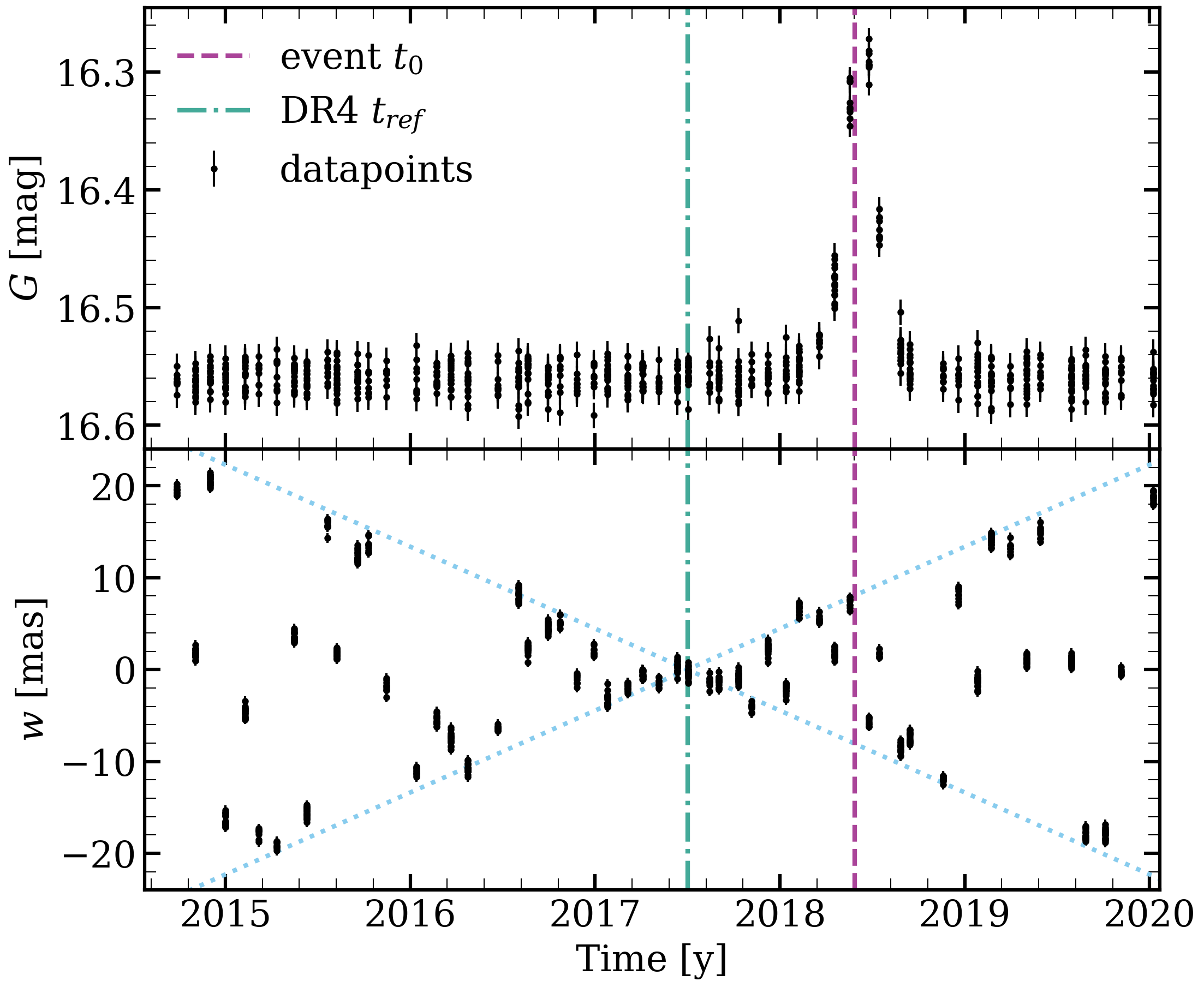}
    \caption{Mock \gaia~photometric and astrometric time series for an example event. The two subplots share the horizontal time axis, limited by the start and end dates of \gaia~DR4; vertical lines denote the closest lens-source linear motion approach $t_0$ (purple dashed) and the reference epoch of \gaia~DR4 (green dotted-dashed). \textit{Top:} Light curve of the simulated microlensing event. Black points with error bars represent mock photometric data for each CCD, grouped in series of {nine} near-simultaneous observations for \gaia 's {nine} consecutive CCDs. \textit{Bottom:} The astrometric time series of along-scan measurements for the simulated microlensing event. Black points with error bars represent observed along-scan positions with respect to the source's reference position, similarly in groups of {nine}. The light blue dotted lines represent $\pm \mu_{\rm S} \cdot (t - t_{\rm ref})$, {which is the angular} distance crossed by the source from its reference position {in linear motion}.}
    \label{fig:mockdata}
\end{figure}

\begin{figure}[h]
	\includegraphics[width=\columnwidth]{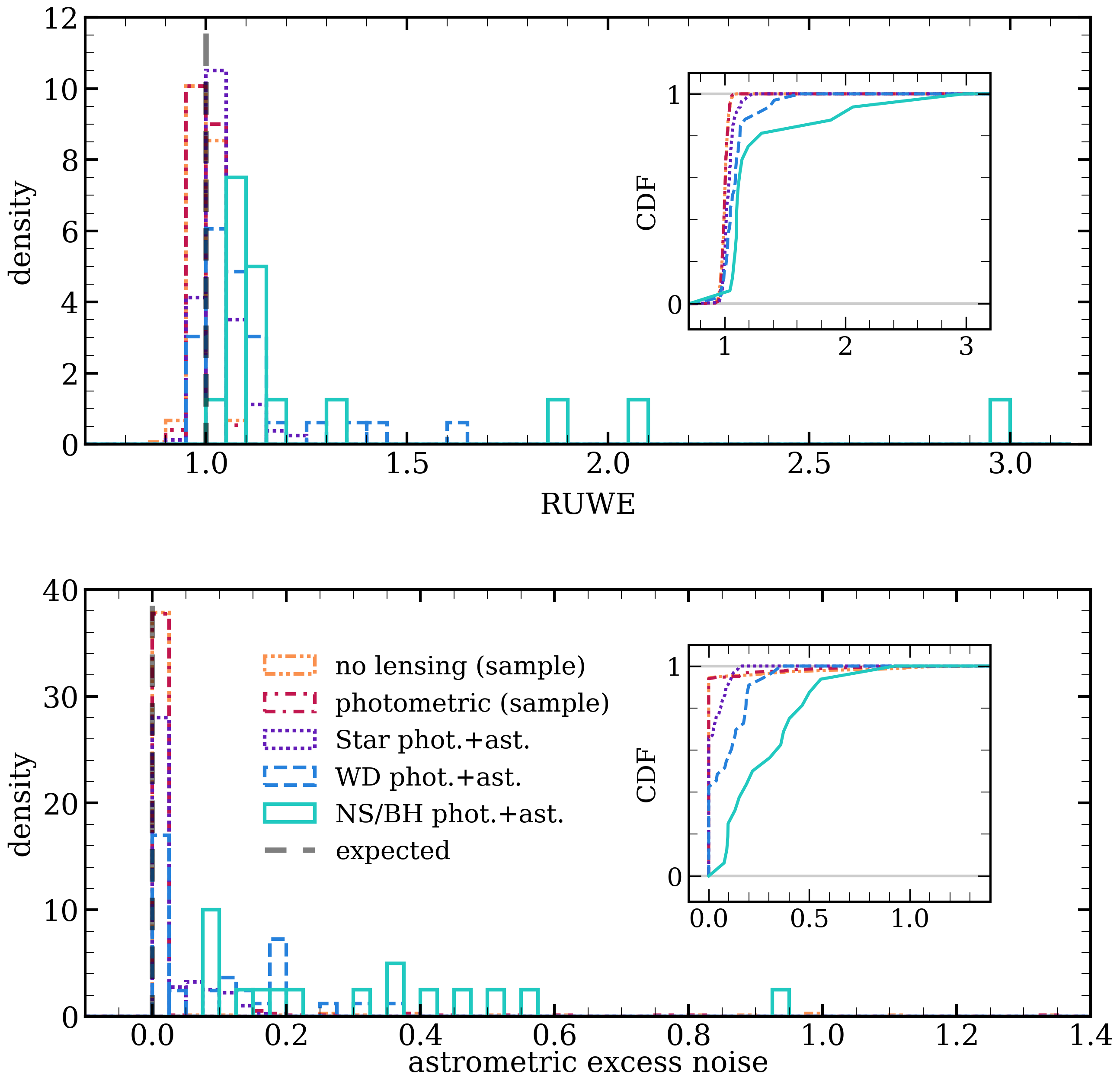}
    \caption{{Distributions of astrometric fit metrics (top: RUWE, bottom: astrometric excess noise). {Five} datasets of simulated tracks, which are described in detail in {Sect.}~\ref{sec:mock_obs}, are represented with histograms:} `no lensing' {(orange, dashed-dot-dotted line)}, {`photometric' (red, dashed-dotted), 'Star phot.+ast.' (indigo, dotted), 'WD phot.+ast.' (blue, dashed) and `NS/BH phot.+ast.'(teal, solid)}. {T}he expected value of {each metric} for a {normal} \gaia~source {is overplotted with a vertical line (grey, long dash).} All distributions are normalized. To help visualize whether distributions are consistent with each other, especially in the sparsely covered tails, the cumulative distribution function (CDF) over the same $x$-axis extent is also displayed in the inset of each subplot{, with 0 and 1 borders marked in light grey}.}
    \label{fig:ruwe}
\end{figure}

We {used} {\tt astromet} to generate mock \gaia~photometric and astrometric time series. For each event classified as both photometric and astrometric, we {generated} the full time series over 66 months, matching \gaia~DR4 coverage. As the {lens/source tracking interval $t$ ({Sect.} \ref{sec:matching}) greatly exceeded actual DR4 coverage}, we {replaced} the simulated $t_0$ with randomly drawn values within the bounds of \gaia~DR4; all other microlensing parameters {were} taken as derived from the simulated lens and source trajectories. {When generating the mock data, we fixed the actual (unlensed) position of the source at the reference time $t_{\text{ref}}$ to be the same as the reference position used for reporting $w$. This is not guaranteed for real data where small offsets may appear: for this reason, a constant position correction needs to be included in astrometric modelling ({as shown in} Table~\ref{tab:joint_priors} and {Sect.}~\ref{sec:models}).} {In the case of events with luminous lenses, we included} the effect of blending as implemented in {\tt astromet}, though we advise caution {as the actual DR4 data quality is likely to be lower for such events} ({we refer to} {Sect.}~\ref{sec:detectable} {for detailed discussion}). We {assumed} source visits {to} follow the nominal scanning law as implemented in {\tt gaiascanlaw}\footnote{\url{https://github.com/zpenoyre/gaiascanlaw}}, and each visit {to} result in simultaneous {astrometric and photometric} measurements over {nine} CCDs. We {added} Gaussian noise to the simulated $G$ and $w$ values, {scaled by the instrumental precision} defined as a function of $G$ in {Sect.} \ref{sec:detectable}. {We treated} $t, \psi, f_w, f_z$ {as} known exactly{, consistently with how \gaia~reports its observations and following the principle of a self-calibrating astrometric mission \citep{LindegrenBastian2010}}.

{We also simulated the \gaia~{five-parameter} fit results, using a close emulation \citep{Penoyre2022} of the Astrometric Global Iterative Solution pipeline \citep{Gaia_astrom_core}. The fit results include diagnostics like RUWE (Renormalized Unit Weight Error) and astrometric excess noise -- we refer to {Sect.} \href{https://gea.esac.esa.int/archive/documentation/GDR3/Gaia_archive/chap_datamodel/sec_dm_main_source_catalogue/ssec_dm_gaia_source.html}{20.1.1} of the \textit{Gaia} DR3 Documentation and \citet{Gaia_astrom_core} for their definitions and detailed discussion. Discrepancy of these diagnostics from values expected for normal sources can be used as an indicator of astrometric microlensing signal \citep[][]{Jablonska2022}.}

In Fig.~\ref{fig:mockdata}, we present the mock time series of an example event, chosen so that the maximum time $t_0$ is {not close to} the \gaia~DR4 reference time $t_{\rm ref}$. While the lensing effect around $t_0$ is {distinct} in the light curve, it is not at all discernible in the astrometric time series, which cannot be effectively interpreted without their respective scan directions and an assumed 2D model of motion. 

\begin{table*}[]
\begin{center}
\caption{Model parameters and their prior distributions used in the joint photometry + astrometry nested sampling modelling.}
\begin{tabular}{p{1.3 cm} p{4.9 cm} p{1.2cm} p{9cm}}
\hline \hline
Parameter & Prior distribution & Unit & Definition \\
\hline
$t_0$ & $\mathcal{U}(\max(t), \min(t))$ & yr & time of closest approach in straight-line motion\\
$|u_0|$ & $\log \mathcal{U}(0.005, 10)$ & -- & absolute value of the impact parameter; in units of $\theta_{\rm E}$\\
$t_{\rm E}$ & $\log \mathcal{U}(0.01, 3)$ & yr & event timescale\\
$\pi_{\rm E}$ & $\log \mathcal{N}(-0.5, 0.5; 0, 3)$ & -- & microlensing parallax; in units of $\theta_{\rm E}$ \\
$\varphi$ & $\mathcal{U}(0, 2\pi)$ & rad & direction of relative source-lens motion, measured North to East\\
$G_0$ & $\mathcal{N}(\text{med}(G), \text{std}(G);\newline \text{med}(G) - 0.5, \text{med}(G)+0.5)$ & mag & \gaia~magnitude at baseline\\ \hline
$\theta_{\rm E}$ & $\log \mathcal{U}(0.01, 30)$ & mas & Einstein radius -- angular event scale\\
$\Delta\alpha^*_{\rm 0, S}$ & $\mathcal{N}(0, \text{std}(|w|); -\max(|w|), \max(|w|))$ & mas & correction to source position at $t_{\rm ref}$ in right ascension\\
$\Delta\delta_{\rm 0, S}$ & $\mathcal{N}(0, \text{std}(|w|); -\max(|w|), \max(|w|))$ & mas & correction to source position at $t_{\rm ref}$ in declination\\
$\mu_{\alpha^*, \text{S}}$ & $\mathcal{N}(-3, 5; -18, 12)$ &mas\,yr$^{-1}$ & source proper motion in right ascension\\
$\mu_{\delta, \text{S}}$ & $\mathcal{N}(-5, 5; -20, 10)$ &mas\,yr$^{-1}$ & source proper motion in declination\\
$\varpi_{\rm S}$ & $\log \mathcal{U}(0.01,2)$ & mas & source parallax\\
\hline
\label{tab:joint_priors}
\end{tabular}
\end{center}
{\footnotesize \textbf{Notes:} $\mathcal{U}$ and $\log \mathcal{U}$ denote uniform and log-uniform priors, respectively, which are defined by their minimum and maximum bounds. $\mathcal{N}(\mu, \sigma; a,b)$ denotes a normal distribution $N(x, \mu, \sigma)$ with a mean $\mu$ and standard deviation $\sigma$, which is truncated to $x \in (a, b)$. Analogously, $\log \mathcal{N}(\mu, \sigma; a,b)$ denotes a normal distribution $N(\log x, \mu, \sigma)$, where $x \in (a, b)$. {$\alpha^*$ is a shorthand notation for $\alpha \cos \delta$.}}

\end{table*}

\begin{figure*}[h]
    \centering
	\includegraphics[width=2\columnwidth]{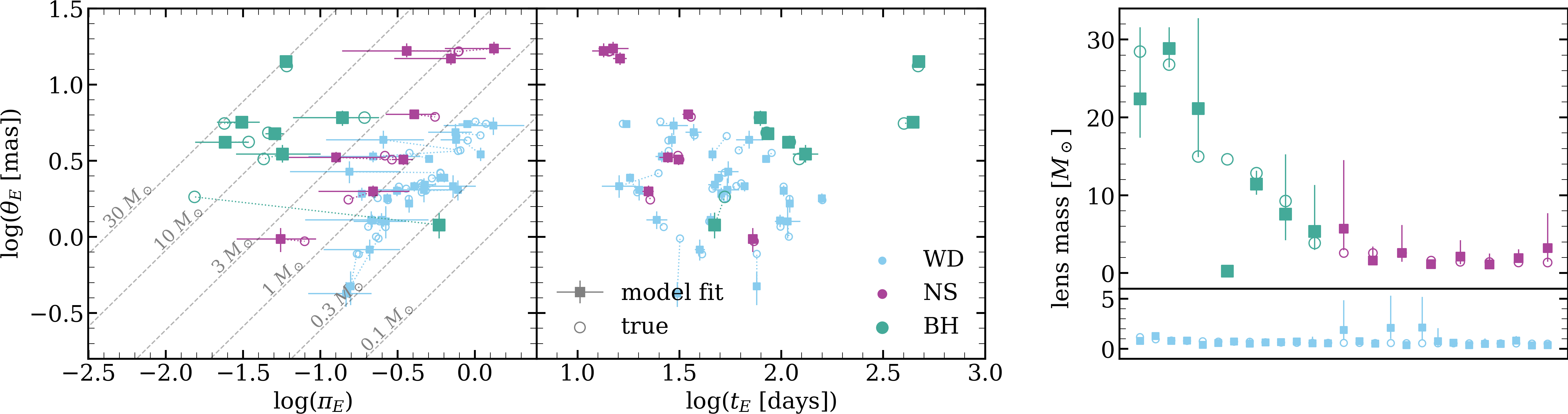}
    \caption{Results of modelling of all simulated events caused by remnants {where recovered astrometric signal is not consistent with 0 (i.e. excluding 1 BH and {6} WD events)}. \textit{Left:} True vs. estimated positions in the $\log \pi_{\rm E}$ -- $\log \theta_{\rm E}$ space, {which is} the space of parameters from which lens mass can be directly calculated. Dashed grey diagonal lines represent contours of equal lens mass. {True parameters (empty circles), and medians from model fit posterior samples (filled squares) with errorbars representing 16-84 percentile ranges, are presented for each event; dotted lines connect true parameters to model fit values.} \textit{Center:} As left (and sharing the same $y$ axis) but in the space of high-signal observables $\log t_{\rm E}$ -- $\log \theta_{\rm E}${; true event positions in this space are generally well-recovered, making this space potentially useful for lens classification}. \textit{Right:} As left, but in the 1D space of lens mass, calculated from samples using Eq.~{(}\ref{eq:lens_mass}{)}. The $x$-axis position has no physical meaning, and lenses are sorted by decreasing true mass for clarity of the figure. {WD lenses are displayed in an additional zoom-in inset in the bottom section of the plot to improve readability. Exceptions with high discrepancies between fitted masses and ground truth are} discussed in {Sect.}~\ref{sec:models}.}
    \label{fig:mass_space}
\end{figure*}

{We also analysed the five-parameter} fits {and found} that astrometric events caused by {remnants, especially NSs and BHs} show {anomalous values of} RUWE and astrometric excess noise. {Stellar lens events generally had non-anomalous five-parameter fits, though we note the pipeline performance is not fully realistic due to a simplified handling of blending effects in {\tt astromet} ({we refer to} {Sect.}~\ref{sec:detectable} {for a detailed discussion}).} {In Fig.~\ref{fig:ruwe}, we compare distributions of astrometric fit metrics for five sets of \gaia~sources. The `no lensing' dataset consists of tracks of a random sample of 300 source stars, where the source motion has been modelled without any effects of the lens. The source stars were chosen from the photometric dataset (without repetition and with probability of being chosen weighted by 1/{$f_{\rm sub}$}). The `photometric' dataset consists of tracks of 300 photometric events chosen with the same random procedure, where the mock data includes the lensing effect and blending with the lens where applicable. The photometric events mostly have non-anomalous {five-parameter} solutions and their metric distributions closely overlap with `no lensing'; this implies close pairs of a luminous lens and an amplified source creating a {variability-induced mover effect} \citep{Wielen1996} should be rare. Datasets of events that are photometrically and astrometrically detectable have been split into star, WD and BH+NS populations (as WDs cause systematically weaker astrometric anomalies than other remnants, while there is no statistically significant difference between fit metric distributions for NS and BH lens samples). These distributions are inherently noisy, as astrometric events are rare and their samples expected in DR4 are small. However, this analysis hints that selecting for events with obvious astrometric anomalies could yield the most promising candidates for remnant mass measurement. Fit metrics are reported as calculated by the emulated \gaia~pipeline implemented in {\tt astromet}.}

In addition to the mentioned quality metrics, parameter differences (e.g.~in proper motions) between \gaia~Data Releases can indicate astrometric signal beyond the {five-parameter} solution \citep[e.g.][]{Lehtinen2023}, similarly to \gaia--\textit{Hipparcos} proper motion anomalies \citep{Brandt2021, Kervella2022}, and have been leveraged to find and analyse astrometric disturbances caused by unresolved stellar companions or exoplanets \citep[e.g.][]{Feng2024, Ribas2025, Thompson2025, Thompson2026}. We tried this method by generating fits to our data over `DR3' and `DR4' baselines, but found no significant excess in fit differences for astrometrically lensed sources (as compared to regular {five-parameter} tracks of sources of similar $G$ magnitudes). The reason presumably lies in the fact that astrometrically lensed sources follow regular {five-parameter} tracks before and after the event, making proper motion fits over timescales of years largely unaffected by lensing effects.

\subsection{Modelling of the mock events}
\label{sec:models}

To test how well true lens parameters can be recovered for previously unknown stellar remnants, we {performed microlensing model fits to the} data generated in {Sect.}~\ref{sec:mock_obs} for {events with remnant lenses} (a sample of 17). {The method is as follows:} for a given set of parameters ({as listed in} Table~\ref{tab:joint_priors}), given the $(t, \psi, f_w, f_z)$ values reported by \gaia~for each observation, photometric ($G$) and astrometric ($w$) can be uniquely determined as defined in {Sect.} \ref{sec:methods}. We then evaluate{d} the Gaussian PDF log-likelihoods for the set of parameters, minimizing $-\log{\mathcal{L}_{\rm tot}} = - \ (\log\mathcal{L}_{G} + \log\mathcal{L}_{w})$.

We {implemented} this simultaneous photometry and astrometry fitting with a modified version of the {\tt nested\_ulens\_parallax}\footnote{\url{https://github.com/zofiakaczmarek/nested_ulens_parallax}} code \citep{Kaczmarek2022}, where we {adjusted} the astrometric data format to match \gaia's LPC observations. This code uses the dynamic nested sampling algorithm \citep{Higson2019} as implemented in {\tt dynesty} \citep{dynesty}. We {used} random walk sampling \citep{Skilling2006} with multiple bounding ellipsoids and 1000 initial live points. We {adopted} a stopping criterion in the remaining fractional evidence of 0.01, and {allocated} 100\% of weight to the posterior distributions. {We provide this code in the \href{https://github.com/zofiakaczmarek/gaia-gleams}{GitHub repository} accompanying this work} for application to future \gaia~data.

We {summarise} the fit parameters and their priors in Table \ref{tab:joint_priors}. {In early runs, we noticed the unstable behaviour of our fitting code with $u_0 \rightarrow 0$ (corresponding to $A \rightarrow \infty$), leading to physically implausible solutions and low reproducibility of results between runs. To remediate this issue, we modified the code to keep the $u_0$ sign fixed within a single run and model each source with $u_0 > 0$ and $u_0 < 0$ separately, eventually selecting the model with higher $\log{\mathcal{L}}$.}

{Overall, we fitted 12 free parameters (6 standard parameters defining the microlensing light curve, the source 5-parameter solution, and $\theta_{\rm E}$). We assumed no blending with neighbour stars (i.e. $G = G_0 - 2.5 \log  A$ and no interference from other objects in astrometry). For real DR4 data, we expect such a cut would be done at the stage of pre-selecting best candidates to model: upon adding a third object's trajectory, the dimensionality of the required fit parameter space would become very high given the limited amount of constraining information.}

\citet{Lam2022} {showed that} tension between astrometry and photometry can lead to different solutions depending on how datasets of vastly different sizes are weighted. In our mock dataset, photometric and astrometric data points are taken simultaneously, which should generally be the case for \gaia~data; this makes equal weighting of photometry and astrometry the natural solution. However, we note {the impact of weighting should be considered and tested {when} applying our code to diverse} datasets {(}e.g. \gaia~and dense ground-based follow-up{)}.

In Fig.~\ref{fig:mass_space}, we assess results of modelling for {remnant-lens events}. Specifically, we {present} lens parameter inference for the physically meaningful parameters ($\theta_{\rm E}$, $\pi_{\rm E}$, $t_{\rm E}$) in the context of lens classification and mass measurement. {To avoid overcrowding the figure with events for which no parameter constraints could be inferred, we only plot the events where astrometric signal beyond the 5-parameters solution is statistically significant in the fit results, {that is}, $\theta_{\rm E} - 3\sigma^-_{\theta_{\rm E}} > 0$. That eliminates 7 events out of 49 (1 BH and 6 WDs), which we discuss in detail in the following paragraph.} We {show} that $\theta_{\rm E}$ and $t_{\rm E}$ can be reliably determined from model fits; $\pi_{\rm E}$ measurements, while {generally} well consistent with the true values, have high uncertainties, which become the bottleneck for mass determination. The $\log t_{\rm E}$ -- $\log \theta_{\rm E}$ space may be used for event classification \citep[e.g.][]{Fardeen2024, Kaczmarek2026}, but event positions in this space do not directly correspond to lens mass, and some auxiliary information (from follow-up observations or Galactic models) must be employed to infer it.

Despite this limitation, for most events a mass estimate that is informative and consistent with the true mass can be obtained using the $\theta_{\rm E}$, $\pi_{\rm E}$ fit ({as presented in} Fig.~\ref{fig:mass_space}, right panel). The event with an estimate largely inconsistent with the true mass (BH33S{48}\footnote{{Event names are created by combining lens types and random indices with source IDs from GaiaEDR3mock, while their shorthand versions in this text use the first and last two digits{. For example}, this event is designated in the repository as \underline{BH33}6832\underline{S}40628527291548308\underline{48}}.}, {fourth} event from left) is entirely missing data points on the descent from the peak to the baseline, leading to very weak constraints on event parameters; despite a good $\theta_{\rm E}$ fit, $\pi_{\rm E}$ is drastically overestimated. {Similarly, all 7 events with $\theta_{\rm E}$ fits consistent with 0 can, via visual inspection, be associated with gaps in cadence -- either due to a complete mismatch of $t_0$ (WD20S28, WD35S84, WD66S04), or missing ascent {or} descent (BH23S20, WD25S08, WD40S52, WD76S04).} {These events typically have low $t_{\rm end} - t_0$ or $t_0 - t_{\rm start}$ in units of $t_{\rm E}$, meaning a fraction of the event happens outside survey bounds and a reliable astrometric baseline may not be established.} {All light curves and corner plots can be examined individually in the accompanying \href{https://github.com/zofiakaczmarek/gaia-gleams}{GitHub repository}.}

Due to its large size, the full visualization of a well-modelled example event, Fig.~\ref{fig:corner}, is {presented in} Appendix~\ref{sec:full_visualization}. This figure {showcases} results of modelling a simulated event caused by a 12.8\,$M_\odot$ BH, including fits to photometric and astrometric data and full posterior sample distributions. Analogous figures for all modelled events are available in the \href{https://github.com/zofiakaczmarek/gaia-gleams}{GitHub repository} for this work.

\section{Discussion and conclusions}
\label{sec:discussion} 

In this work we presented a detailed simulation of microlensing events, focusing especially on astrometric signals, as they would be seen in \gaia~DR4 data, namely:

\begin{itemize}

\item  We {have created} a new all-sky population model for microlensing studies, tailored for future \gaia~data and based on state-of-the-art Galactic models and software. In particular, we {have incorporated} a detailed treatment of remnants via tested initial-final mass relations and modelling the influence of natal kicks on present-day remnant distributions and dynamics. Our Galactic simulation predicts total populations of $9.1 \times 10^7$ stellar-origin BHs and $1.7 \times 10^8$ NSs gravitationally bound to the Milky Way. Predictions of microlensing event populations as seen by \gaia~will be needed to interpret the actually observed lensing events and adjust Galactic models. Therefore, our predictions will deliver their greatest impact once the DR4 data have been thoroughly scrutinized for microlensing events.\\

\item We {estimated} the event yields for \gaia~DR4. According to our simulation, \gaia~DR4 will contain of order $\sim${300} astrometric events with detectable signal, out of which $\sim${200} will have both photometric and astrometric signal. Among those samples, {$82$} and {$49$} events, respectively, are caused by stellar remnants. This set of {$49$} is composed of 8 events with BH lenses, 8 with NS lenses, and {33} with WD lenses. \\

\item We {demonstrated} that although off-bulge events are rare, they are strongly selected for higher $\theta_{\rm E}$, and therefore are promising candidates for mass determination and remnant searches. In the context of \gaia~as well as future astrometric missions, this is especially important, as lower source density corresponds to higher data quality.\\

\item We {assessed} selection effects on lens detectability. We {presented} how \gaia~precision corresponds to signal from various lens classes, estimating magnitude limits for their detection. {For example}: a typical BH (NS, WD) lens in a photometric microlensing event produces a signal that can be detected at $G \leq$ 17 (15, 14){\,mag}; typical signals of stellar lenses are not detectable by \gaia~at all.\\

\item We {generated} \gaia~astrometric fits for selected events and {compared} them to fits of normal stellar sources. We {found} that events caused by dark stellar remnants have significantly increased metrics indicating anomalous {five-parameter} fits (RUWE and excess noise), making these metrics potentially useful for candidate selection in real data. We {found} no meaningful lensing signatures in changes in fit values (e.g. proper motion anomaly) between Data Releases.\\

\item We {analysed} how well event properties, in particular the lens mass, can be recovered from \gaia -like data. We {used} the nested sampling method to fit models to simulated data. We {concluded that this modelling approach} provides precise and reliable estimates of $t_{\rm E}$ and $\theta_{\rm E}$, with {weak constraints on} $\pi_{\rm E}$ being the bottleneck for complete {event} resolution. We {found that the} event parameter estimates, including lens mass estimates, {were} consistent with ground truth for the bulk of events. {We found that incomplete temporal coverage was responsible for the exceptions to this consistency.}\\

\item 
We {concluded} that gaps in the \gaia~cadence are a serious limiting factor in event modelling. A promising finding is that, in most {of our simulated} cases, the astrometric signal {appears to be successfully recovered in the} \gaia~observations{, as $\theta_{\rm E}$ could} be measured ({as shown in} Fig.~\ref{fig:mass_space}). However, cadence gaps severely hinder parallax determination; as a result, mass constraints are limited even in case of well-measured $\theta_{\rm E}$. This impact of gaps should be mitigated where possible. While follow-up can no longer be triggered after \gaia 's end of operations, archival ground-based observations should be incorporated whenever available (e.g. follow-up observations conducted for \gaia~Science Alerts, or independent detections by other surveys) to enable high-cadence photometry modelling or space parallax determination. A $\pi_{\rm E}$ value obtained from such observations could then be combined with the $\theta_{\rm E}$ measured by \gaia~to obtain a robust mass measurement. Additionally, the $\log t_{\rm E}$ -- $\log \theta_{\rm E}$ space can be used for lens classification, and missing parameters may be inferred with the help of Galactic models \citep[e.g. similarly to][]{Howil2024, Kaczmarek2025}.\\

\item We {have developed} codes for modelling the events, as well as for visualizing \gaia~data and the corresponding model fits, in anticipation of DR4. We {have also created a} sample of mock \gaia~{astrometric and photometric} data and the true injected lens parameters. {We make those codes and mock datasets available in a \href{https://github.com/zofiakaczmarek/gaia-gleams}{GitHub repository}.} We encourage further experiments with this data. Some suggested avenues could include multi-stage modelling using priors from photometry as input for a later stage \citep[e.g.][]{Lam2022, Jablonska2022}, incorporating ground-based follow-up, or developing advanced classifiers (e.g. using machine learning) for \gaia 's microlensing events.

\end{itemize}

{T}his study presents a simplified model of the Galaxy, as well as of \gaia~observations, and as such it suffers from some limitations. Concerning the Galactic model, we {used} state-of-the-art tools including {\tt Galaxia} \citep{Sharma2011}, {\tt SPISEA} \citep{SPISEA} {and} {\tt StellarMortis} \citep{StellarMortis}. However, some parameters of the Milky Way (e.g. bar angle and pattern speed) remain highly uncertain \citep[e.g.][]{HuntVasiliev2025}. Stellar remnants, being very difficult to detect, are particularly impacted by model uncertainty regarding their numbers \citep[e.g.][]{Olejak2022, Underworld2022}, kick profiles \citep[e.g.][]{Igoshev2020, ODoherty2023}, or initial-final mass relations \citep[e.g.][]{Rose2022}. Regarding the \gaia~observations, the main limitation is that we {did} not model interference from the background, such as blending with neighbour stars, which impacts dense fields disproportionately \citep[e.g.][]{CastroGinard2024}. We also {did} not model complex instrumental effects or artifacts, assuming Gaussian noise for all observations. On the other hand, our assumed detection thresholds {were} chosen conservatively. Altogether, we expect that we {inferred} roughly realistic event rate{s}. We furthermore note that \gaia~DR5 will increase these rate{s} by at least a factor of {two}.

The field of astrometric microlensing is expected to grow significantly in the coming years. The \textit{Nancy Grace Roman Space Telescope} \citep{Spergel2015}, {successfully launched} {in August} {2026}, will be the first space mission optimized for conducting a microlensing survey. \textit{Roman}'s Galactic Bulge Time Domain Survey will cover a small (1.68 deg$^2$) footprint situated near the Galactic {c}enter with very high-cadence observations. \textit{Roman} will observe in the near-infrared, which is optimal for exploring dense, dust-obscured regions at low Galactic latitudes. For the next generation of space missions, a \gaia~successor in the near-infrared has been proposed, dubbed \textit{GaiaNIR} \citep{Hobbs2021}. \textit{GaiaNIR} would be well-suited for conducting a Galaxy-wide astrometric microlensing study, combining near-infrared reach and all-sky coverage. Between \gaia~DR4, the \gaia~legacy catalogue DR5, and these upcoming surveys, the future of astrometric microlensing studies is full of great promise.

\begin{acknowledgements}
The authors acknowledge support by the state of Baden-Württemberg through bwHPC. {We thank the anonymous referee for their comments which helped improve this paper.} We thank Joachim Wambsganss, Markus Hundertmark, Yiannis Tsapras, Christian Sorgenfrei, Markus Demleitner, Katarzyna Kruszyńska, Lucien Mauviard-Haag, Johannes Sahlmann, Alex Bombrun{,} and Elana Kane for helpful discussions. Z.K. is a Fellow of the International Max Planck Research School for Astronomy and Cosmic Physics at the University of Heidelberg (IMPRS-HD). 

{\L}.W. acknowledges support from the European Union’s Horizon Europe Research and Innovation Programme under Grant Agreement No. 101131928 (ACME) and the Polish National Science Centre DAINA grant No. 2024/52/L/ST9/00210.

We acknowledge the use of Paul Tol's colourblind-friendly \href{https://sronpersonalpages.nl/~pault/}{colour palettes}.
\end{acknowledgements}

\bibliographystyle{aa}
\bibliography{refs}

@ARTICLE{Klueter2020,
       author = {{Kl{\"u}ter}, J. and {Bastian}, U. and {Wambsganss}, J.},
        title = "{Expectations on mass determination using astrometric microlensing by Gaia}",
      journal = {\aap},
         year = 2020,
        month = aug,
       volume = {640},
          eid = {A83},
        pages = {A83},
          doi = {10.1051/0004-6361/201937061},
archivePrefix = {arXiv},
       eprint = {1911.02584},
 primaryClass = {astro-ph.IM},
       adsurl = {https://ui.adsabs.harvard.edu/abs/2020A&A...640A..83K}
}

@ARTICLE{Hobbs2021,
       author = {{Hobbs}, David and {Brown}, Anthony and {H{\o}g}, Erik and {Jordi}, Carme and {Kawata}, Daisuke and {Tanga}, Paolo and {Klioner}, Sergei and {Sozzetti}, Alessandro and {Wyrzykowski}, {\L}ukasz and {Walton}, Nicholas and {Vallenari}, Antonella and {Makarov}, Valeri and {Rybizki}, Jan and {Jim{\'e}nez-Esteban}, Fran and {Caballero}, Jos{\'e} A. and {McMillan}, Paul J. and {Secrest}, Nathan and {Mor}, Roger and {Andrews}, Jeff J. and {Zwitter}, Toma{\v{z}} and {Chiappini}, Cristina and {Fynbo}, Johan P.~U. and {Ting}, Yuan-Sen and {Hestroffer}, Daniel and {Lindegren}, Lennart and {McArthur}, Barbara and {Gouda}, Naoteru and {Moore}, Anna and {Gonzalez}, Oscar A. and {Vaccari}, Mattia},
        title = "{All-sky visible and near infrared space astrometry}",
      journal = {Experimental Astronomy},
         year = 2021,
        month = jun,
       volume = {51},
       number = {3},
        pages = {783-843},
          doi = {10.1007/s10686-021-09705-z},
archivePrefix = {arXiv},
       eprint = {1907.12535},
 primaryClass = {astro-ph.IM},
       adsurl = {https://ui.adsabs.harvard.edu/abs/2021ExA....51..783H}
}

@ARTICLE{HuntVasiliev2025,
       author = {{Hunt}, Jason A.~S. and {Vasiliev}, Eugene},
        title = "{Milky Way dynamics in light of Gaia}",
      journal = {\nar},
         year = 2025,
        month = jun,
       volume = {100},
          eid = {101721},
        pages = {101721},
          doi = {10.1016/j.newar.2024.101721},
archivePrefix = {arXiv},
       eprint = {2501.04075},
 primaryClass = {astro-ph.GA},
       adsurl = {https://ui.adsabs.harvard.edu/abs/2025NewAR.10001721H}
}

@ARTICLE{Rose2022,
       author = {{Rose}, Sam and {Lam}, Casey Y. and {Lu}, Jessica R. and {Medford}, Michael and {Hosek}, Matthew W. and {Abrams}, Natasha S. and {Ramey}, Emily and {Vasylyev}, Sergiy S.},
        title = "{The Impact of Initial-Final Mass Relations on Black Hole Microlensing}",
      journal = {\apj},
         year = 2022,
        month = dec,
       volume = {941},
       number = {2},
          eid = {116},
        pages = {116},
          doi = {10.3847/1538-4357/aca09d},
archivePrefix = {arXiv},
       eprint = {2211.04471},
 primaryClass = {astro-ph.HE},
       adsurl = {https://ui.adsabs.harvard.edu/abs/2022ApJ...941..116R}
}

@ARTICLE{Rybicki2018,
       author = {{Rybicki}, Krzysztof A. and {Wyrzykowski}, {\L}ukasz and {Klencki}, Jakub and {de Bruijne}, Jos and {Belczy{\'n}ski}, Krzysztof and {Chru{\'s}li{\'n}ska}, Martyna},
        title = "{On the accuracy of mass measurement for microlensing black holes as seen by Gaia and OGLE}",
      journal = {\mnras},
         year = 2018,
        month = may,
       volume = {476},
       number = {2},
        pages = {2013-2028},
          doi = {10.1093/mnras/sty356},
archivePrefix = {arXiv},
       eprint = {1802.03258},
 primaryClass = {astro-ph.SR},
       adsurl = {https://ui.adsabs.harvard.edu/abs/2018MNRAS.476.2013R}
}

@ARTICLE{CastroGinard2024,
       author = {{Castro-Ginard}, Alfred and {Penoyre}, Zephyr and {Casey}, Andrew R. and {Brown}, Anthony G.~A. and {Belokurov}, Vasily and {Cantat-Gaudin}, Tristan and {Drimmel}, Ronald and {Fouesneau}, Morgan and {Khanna}, Shourya and {Kurbatov}, Evgeny P. and {Price-Whelan}, Adrian M. and {Rix}, Hans-Walter and {Smart}, Richard L.},
        title = "{Gaia DR3 detectability of unresolved binary systems}",
      journal = {\aap},
         year = 2024,
        month = aug,
       volume = {688},
          eid = {A1},
        pages = {A1},
          doi = {10.1051/0004-6361/202450172},
archivePrefix = {arXiv},
       eprint = {2404.14127},
 primaryClass = {astro-ph.GA},
       adsurl = {https://ui.adsabs.harvard.edu/abs/2024A&A...688A...1C}
}

@ARTICLE{ODoherty2023,
       author = {{O'Doherty}, Tyrone N. and {Bahramian}, Arash and {Miller-Jones}, James C.~A. and {Goodwin}, Adelle J. and {Mandel}, Ilya and {Willcox}, Reinhold and {Atri}, Pikky and {Strader}, Jay},
        title = "{An observationally derived kick distribution for neutron stars in binary systems}",
      journal = {\mnras},
         year = 2023,
        month = may,
       volume = {521},
       number = {2},
        pages = {2504-2524},
          doi = {10.1093/mnras/stad680},
archivePrefix = {arXiv},
       eprint = {2303.01059},
 primaryClass = {astro-ph.HE},
       adsurl = {https://ui.adsabs.harvard.edu/abs/2023MNRAS.521.2504O}
}

@ARTICLE{GaiaNIR,
       author = {{Hobbs}, David and {H{\o}g}, Erik and {Mora}, Alcione and {Crowley}, Cian and {McMillan}, Paul and {Ranalli}, Piero and {Heiter}, Ulrike and {Jordi}, Carme and {Hambly}, Nigel and {Church}, Ross and {Anthony}, Brown and {Tanga}, Paolo and {Chemin}, Laurent and {Portell}, Jordi and {Jim{\'e}nez-Esteban}, Fran and {Klioner}, Sergei and {Mignard}, Francois and {Fynbo}, Johan and {Wyrzykowski}, {\L}ukasz and {Rybicki}, Krzysztof and {Anderson}, Richard I. and {Cellino}, Alberto and {Fabricius}, Claus and {Davidson}, Michael and {Lindegren}, Lennart},
        title = "{GaiaNIR: Combining optical and Near-Infra-Red (NIR) capabilities with Time-Delay-Integration (TDI) sensors for a future Gaia-like mission}",
      journal = {arXiv e-prints},
         year = 2016,
        month = sep,
          eid = {arXiv:1609.07325},
        pages = {arXiv:1609.07325},
          doi = {10.48550/arXiv.1609.07325},
archivePrefix = {arXiv},
       eprint = {1609.07325},
 primaryClass = {astro-ph.IM},
       adsurl = {https://ui.adsabs.harvard.edu/abs/2016arXiv160907325H}
}

@ARTICLE{Belokurov2002,
       author = {{Belokurov}, V.~A. and {Evans}, N.~W.},
        title = "{Astrometric microlensing with the GAIA satellite}",
      journal = {\mnras},
         year = 2002,
        month = apr,
       volume = {331},
       number = {3},
        pages = {649-665},
          doi = {10.1046/j.1365-8711.2002.05222.x},
archivePrefix = {arXiv},
       eprint = {astro-ph/0112243},
 primaryClass = {astro-ph},
       adsurl = {https://ui.adsabs.harvard.edu/abs/2002MNRAS.331..649B}
}

@ARTICLE{Lindegren2021,
       author = {{Lindegren}, L. and {Klioner}, S.~A. and {Hern{\'a}ndez}, J. and {Bombrun}, A. and {Ramos-Lerate}, M. and {Steidelm{\"u}ller}, H. and {Bastian}, U. and {Biermann}, M. and {de Torres}, A. and {Gerlach}, E. and {Geyer}, R. and {Hilger}, T. and {Hobbs}, D. and {Lammers}, U. and {McMillan}, P.~J. and {Stephenson}, C.~A. and {Casta{\~n}eda}, J. and {Davidson}, M. and {Fabricius}, C. and {Gracia-Abril}, G. and {Portell}, J. and {Rowell}, N. and {Teyssier}, D. and {Torra}, F. and {Bartolom{\'e}}, S. and {Clotet}, M. and {Garralda}, N. and {Gonz{\'a}lez-Vidal}, J.~J. and {Torra}, J. and {Abbas}, U. and {Altmann}, M. and {Anglada Varela}, E. and {Balaguer-N{\'u}{\~n}ez}, L. and {Balog}, Z. and {Barache}, C. and {Becciani}, U. and {Bernet}, M. and {Bertone}, S. and {Bianchi}, L. and {Bouquillon}, S. and {Brown}, A.~G.~A. and {Bucciarelli}, B. and {Busonero}, D. and {Butkevich}, A.~G. and {Buzzi}, R. and {Cancelliere}, R. and {Carlucci}, T. and {Charlot}, P. and {Cioni}, M. -R.~L. and {Crosta}, M. and {Crowley}, C. and {del Peloso}, E.~F. and {del Pozo}, E. and {Drimmel}, R. and {Esquej}, P. and {Fienga}, A. and {Fraile}, E. and {Gai}, M. and {Garcia-Reinaldos}, M. and {Guerra}, R. and {Hambly}, N.~C. and {Hauser}, M. and {Jan{\ss}en}, K. and {Jordan}, S. and {Kostrzewa-Rutkowska}, Z. and {Lattanzi}, M.~G. and {Liao}, S. and {Licata}, E. and {Lister}, T.~A. and {L{\"o}ffler}, W. and {Marchant}, J.~M. and {Masip}, A. and {Mignard}, F. and {Mints}, A. and {Molina}, D. and {Mora}, A. and {Morbidelli}, R. and {Murphy}, C.~P. and {Pagani}, C. and {Panuzzo}, P. and {Pe{\~n}alosa Esteller}, X. and {Poggio}, E. and {Re Fiorentin}, P. and {Riva}, A. and {Sagrist{\`a} Sell{\'e}s}, A. and {Sanchez Gimenez}, V. and {Sarasso}, M. and {Sciacca}, E. and {Siddiqui}, H.~I. and {Smart}, R.~L. and {Souami}, D. and {Spagna}, A. and {Steele}, I.~A. and {Taris}, F. and {Utrilla}, E. and {van Reeven}, W. and {Vecchiato}, A.},
        title = "{Gaia Early Data Release 3. The astrometric solution}",
      journal = {\aap},
         year = 2021,
        month = may,
       volume = {649},
          eid = {A2},
        pages = {A2},
          doi = {10.1051/0004-6361/202039709},
archivePrefix = {arXiv},
       eprint = {2012.03380},
 primaryClass = {astro-ph.IM},
       adsurl = {https://ui.adsabs.harvard.edu/abs/2021A&A...649A...2L}
}

@INPROCEEDINGS{LindegrenBastian2010,
       author = {{Lindegren}, L. and {Bastian}, U.},
        title = "{Basic principles of scanning space astrometry}",
    booktitle = {EAS Publications Series},
         year = 2010,
       editor = {{Turon}, C. and {Meynadier}, F. and {Arenou}, F.},
       series = {EAS Publications Series},
       volume = {45},
        month = jan,
    publisher = {EDP},
        pages = {109-114},
          doi = {10.1051/eas/1045018},
       adsurl = {https://ui.adsabs.harvard.edu/abs/2010EAS....45..109L}


}

@ARTICLE{Wielen1996,
       author = {{Wielen}, R.},
        title = "{Searching for VIMs: an astrometric method to detect the binary nature of double stars with a variable component}",
      journal = {\aap},
         year = 1996,
        month = oct,
       volume = {314},
        pages = {679},
       adsurl = {https://ui.adsabs.harvard.edu/abs/1996A&A...314..679W}
}

@INPROCEEDINGS{deBruijne2010,
       author = {{de Bruijne}, Jos and {Siddiqui}, Hassan and {Lammers}, Uwe and {Hoar}, John and {O'Mullane}, William and {Prusti}, Timo},
        title = "{Optimising the Gaia scanning law for relativity experiments}",
    booktitle = {Relativity in Fundamental Astronomy: Dynamics, Reference Frames, and Data Analysis},
         year = 2010,
       editor = {{Klioner}, Sergei A. and {Seidelmann}, P. Kenneth and {Soffel}, Michael H.},
       series = {IAU Symposium},
       volume = {261},
        month = jan,
        pages = {331-333},
          doi = {10.1017/S1743921309990597},
       adsurl = {https://ui.adsabs.harvard.edu/abs/2010IAUS..261..331D}
}

@ARTICLE{Cifuentes2025,
       author = {{Cifuentes}, C. and {Caballero}, J.~A. and {Gonz{\'a}lez-Payo}, J. and {Amado}, P.~J. and {B{\'e}jar}, V.~J.~S. and {Burgasser}, A.~J. and {Cort{\'e}s-Contreras}, M. and {Lodieu}, N. and {Montes}, D. and {Quirrenbach}, A. and {Reiners}, A. and {Ribas}, I. and {Sanz-Forcada}, J. and {Seifert}, W. and {Zapatero Osorio}, M.~R.},
        title = "{CARMENES input catalogue of M dwarfs: IX. Multiplicity from close spectroscopic binaries to ultra-wide systems}",
      journal = {\aap},
         year = 2025,
        month = jan,
       volume = {693},
          eid = {A228},
        pages = {A228},
          doi = {10.1051/0004-6361/202452527},
archivePrefix = {arXiv},
       eprint = {2412.12264},
 primaryClass = {astro-ph.SR},
       adsurl = {https://ui.adsabs.harvard.edu/abs/2025A&A...693A.228C}
}

@ARTICLE{GaiaEDR3,
       author = {{Gaia Collaboration} and {Brown}, A.~G.~A. and {Vallenari}, A. and {Prusti}, T. and {de Bruijne}, J.~H.~J. and {Babusiaux}, C. and {Biermann}, M. and {Creevey}, O.~L. and {Evans}, D.~W. and {Eyer}, L. and {Hutton}, A. and {Jansen}, F. and {Jordi}, C. and {Klioner}, S.~A. and {Lammers}, U. and {Lindegren}, L. and {Luri}, X. and {Mignard}, F. and {Panem}, C. and {Pourbaix}, D. and {Randich}, S. and {Sartoretti}, P. and {Soubiran}, C. and {Walton}, N.~A. and {Arenou}, F. and {Bailer-Jones}, C.~A.~L. and {Bastian}, U. and {Cropper}, M. and {Drimmel}, R. and {Katz}, D. and {Lattanzi}, M.~G. and {van Leeuwen}, F. and {Bakker}, J. and {Cacciari}, C. and {Casta{\~n}eda}, J. and {De Angeli}, F. and {Ducourant}, C. and {Fabricius}, C. and {Fouesneau}, M. and {Fr{\'e}mat}, Y. and {Guerra}, R. and {Guerrier}, A. and {Guiraud}, J. and {Jean-Antoine Piccolo}, A. and {Masana}, E. and {Messineo}, R. and {Mowlavi}, N. and {Nicolas}, C. and {Nienartowicz}, K. and {Pailler}, F. and {Panuzzo}, P. and {Riclet}, F. and {Roux}, W. and {Seabroke}, G.~M. and {Sordo}, R. and {Tanga}, P. and {Th{\'e}venin}, F. and {Gracia-Abril}, G. and {Portell}, J. and {Teyssier}, D. and {Altmann}, M. and {Andrae}, R. and {Bellas-Velidis}, I. and {Benson}, K. and {Berthier}, J. and {Blomme}, R. and {Brugaletta}, E. and {Burgess}, P.~W. and {Busso}, G. and {Carry}, B. and {Cellino}, A. and {Cheek}, N. and {Clementini}, G. and {Damerdji}, Y. and {Davidson}, M. and {Delchambre}, L. and {Dell'Oro}, A. and {Fern{\'a}ndez-Hern{\'a}ndez}, J. and {Galluccio}, L. and {Garc{\'\i}a-Lario}, P. and {Garcia-Reinaldos}, M. and {Gonz{\'a}lez-N{\'u}{\~n}ez}, J. and {Gosset}, E. and {Haigron}, R. and {Halbwachs}, J. -L. and {Hambly}, N.~C. and {Harrison}, D.~L. and {Hatzidimitriou}, D. and {Heiter}, U. and {Hern{\'a}ndez}, J. and {Hestroffer}, D. and {Hodgkin}, S.~T. and {Holl}, B. and {Jan{\ss}en}, K. and {Jevardat de Fombelle}, G. and {Jordan}, S. and {Krone-Martins}, A. and {Lanzafame}, A.~C. and {L{\"o}ffler}, W. and {Lorca}, A. and {Manteiga}, M. and {Marchal}, O. and {Marrese}, P.~M. and {Moitinho}, A. and {Mora}, A. and {Muinonen}, K. and {Osborne}, P. and {Pancino}, E. and {Pauwels}, T. and {Petit}, J. -M. and {Recio-Blanco}, A. and {Richards}, P.~J. and {Riello}, M. and {Rimoldini}, L. and {Robin}, A.~C. and {Roegiers}, T. and {Rybizki}, J. and {Sarro}, L.~M. and {Siopis}, C. and {Smith}, M. and {Sozzetti}, A. and {Ulla}, A. and {Utrilla}, E. and {van Leeuwen}, M. and {van Reeven}, W. and {Abbas}, U. and {Abreu Aramburu}, A. and {Accart}, S. and {Aerts}, C. and {Aguado}, J.~J. and {Ajaj}, M. and {Altavilla}, G. and {{\'A}lvarez}, M.~A. and {{\'A}lvarez Cid-Fuentes}, J. and {Alves}, J. and {Anderson}, R.~I. and {Anglada Varela}, E. and {Antoja}, T. and {Audard}, M. and {Baines}, D. and {Baker}, S.~G. and {Balaguer-N{\'u}{\~n}ez}, L. and {Balbinot}, E. and {Balog}, Z. and {Barache}, C. and {Barbato}, D. and {Barros}, M. and {Barstow}, M.~A. and {Bartolom{\'e}}, S. and {Bassilana}, J. -L. and {Bauchet}, N. and {Baudesson-Stella}, A. and {Becciani}, U. and {Bellazzini}, M. and {Bernet}, M. and {Bertone}, S. and {Bianchi}, L. and {Blanco-Cuaresma}, S. and {Boch}, T. and {Bombrun}, A. and {Bossini}, D. and {Bouquillon}, S. and {Bragaglia}, A. and {Bramante}, L. and {Breedt}, E. and {Bressan}, A. and {Brouillet}, N. and {Bucciarelli}, B. and {Burlacu}, A. and {Busonero}, D. and {Butkevich}, A.~G. and {Buzzi}, R. and {Caffau}, E. and {Cancelliere}, R. and {C{\'a}novas}, H. and {Cantat-Gaudin}, T. and {Carballo}, R. and {Carlucci}, T. and {Carnerero}, M.~I. and {Carrasco}, J.~M. and {Casamiquela}, L. and {Castellani}, M. and {Castro-Ginard}, A. and {Castro Sampol}, P. and {Chaoul}, L. and {Charlot}, P. and {Chemin}, L. and {Chiavassa}, A. and {Cioni}, M. -R.~L. and {Comoretto}, G. and {Cooper}, W.~J. and {Cornez}, T. and {Cowell}, S. and {Crifo}, F. and {Crosta}, M. and {Crowley}, C. and {Dafonte}, C. and {Dapergolas}, A. and {David}, M. and {David}, P. and {de Laverny}, P. and {De Luise}, F. and {De March}, R. and {De Ridder}, J. and {de Souza}, R. and {de Teodoro}, P. and {de Torres}, A. and {del Peloso}, E.~F. and {del Pozo}, E. and {Delbo}, M. and {Delgado}, A. and {Delgado}, H.~E. and {Delisle}, J. -B. and {Di Matteo}, P. and {Diakite}, S. and {Diener}, C. and {Distefano}, E. and {Dolding}, C. and {Eappachen}, D. and {Edvardsson}, B. and {Enke}, H. and {Esquej}, P. and {Fabre}, C. and {Fabrizio}, M. and {Faigler}, S. and {Fedorets}, G. and {Fernique}, P. and {Fienga}, A. and {Figueras}, F. and {Fouron}, C. and {Fragkoudi}, F. and {Fraile}, E. and {Franke}, F. and {Gai}, M. and {Garabato}, D. and {Garcia-Gutierrez}, A. and {Garc{\'\i}a-Torres}, M. and {Garofalo}, A. and {Gavras}, P. and {Gerlach}, E. and {Geyer}, R. and {Giacobbe}, P. and {Gilmore}, G. and {Girona}, S. and {Giuffrida}, G. and {Gomel}, R. and {Gomez}, A. and {Gonzalez-Santamaria}, I. and {Gonz{\'a}lez-Vidal}, J.~J. and {Granvik}, M. and {Guti{\'e}rrez-S{\'a}nchez}, R. and {Guy}, L.~P. and {Hauser}, M. and {Haywood}, M. and {Helmi}, A. and {Hidalgo}, S.~L. and {Hilger}, T. and {H{\l}adczuk}, N. and {Hobbs}, D. and {Holland}, G. and {Huckle}, H.~E. and {Jasniewicz}, G. and {Jonker}, P.~G. and {Juaristi Campillo}, J. and {Julbe}, F. and {Karbevska}, L. and {Kervella}, P. and {Khanna}, S. and {Kochoska}, A. and {Kontizas}, M. and {Kordopatis}, G. and {Korn}, A.~J. and {Kostrzewa-Rutkowska}, Z. and {Kruszy{\'n}ska}, K. and {Lambert}, S. and {Lanza}, A.~F. and {Lasne}, Y. and {Le Campion}, J. -F. and {Le Fustec}, Y. and {Lebreton}, Y. and {Lebzelter}, T. and {Leccia}, S. and {Leclerc}, N. and {Lecoeur-Taibi}, I. and {Liao}, S. and {Licata}, E. and {Lindstr{\o}m}, E.~P. and {Lister}, T.~A. and {Livanou}, E. and {Lobel}, A. and {Madrero Pardo}, P. and {Managau}, S. and {Mann}, R.~G. and {Marchant}, J.~M. and {Marconi}, M. and {Marcos Santos}, M.~M.~S. and {Marinoni}, S. and {Marocco}, F. and {Marshall}, D.~J. and {Martin Polo}, L. and {Mart{\'\i}n-Fleitas}, J.~M. and {Masip}, A. and {Massari}, D. and {Mastrobuono-Battisti}, A. and {Mazeh}, T. and {McMillan}, P.~J. and {Messina}, S. and {Michalik}, D. and {Millar}, N.~R. and {Mints}, A. and {Molina}, D. and {Molinaro}, R. and {Moln{\'a}r}, L. and {Montegriffo}, P. and {Mor}, R. and {Morbidelli}, R. and {Morel}, T. and {Morris}, D. and {Mulone}, A.~F. and {Munoz}, D. and {Muraveva}, T. and {Murphy}, C.~P. and {Musella}, I. and {Noval}, L. and {Ord{\'e}novic}, C. and {Orr{\`u}}, G. and {Osinde}, J. and {Pagani}, C. and {Pagano}, I. and {Palaversa}, L. and {Palicio}, P.~A. and {Panahi}, A. and {Pawlak}, M. and {Pe{\~n}alosa Esteller}, X. and {Penttil{\"a}}, A. and {Piersimoni}, A.~M. and {Pineau}, F. -X. and {Plachy}, E. and {Plum}, G. and {Poggio}, E. and {Poretti}, E. and {Poujoulet}, E. and {Pr{\v{s}}a}, A. and {Pulone}, L. and {Racero}, E. and {Ragaini}, S. and {Rainer}, M. and {Raiteri}, C.~M. and {Rambaux}, N. and {Ramos}, P. and {Ramos-Lerate}, M. and {Re Fiorentin}, P. and {Regibo}, S. and {Reyl{\'e}}, C. and {Ripepi}, V. and {Riva}, A. and {Rixon}, G. and {Robichon}, N. and {Robin}, C. and {Roelens}, M. and {Rohrbasser}, L. and {Romero-G{\'o}mez}, M. and {Rowell}, N. and {Royer}, F. and {Rybicki}, K.~A. and {Sadowski}, G. and {Sagrist{\`a} Sell{\'e}s}, A. and {Sahlmann}, J. and {Salgado}, J. and {Salguero}, E. and {Samaras}, N. and {Sanchez Gimenez}, V. and {Sanna}, N. and {Santove{\~n}a}, R. and {Sarasso}, M. and {Schultheis}, M. and {Sciacca}, E. and {Segol}, M. and {Segovia}, J.~C. and {S{\'e}gransan}, D. and {Semeux}, D. and {Shahaf}, S. and {Siddiqui}, H.~I. and {Siebert}, A. and {Siltala}, L. and {Slezak}, E. and {Smart}, R.~L. and {Solano}, E. and {Solitro}, F. and {Souami}, D. and {Souchay}, J. and {Spagna}, A. and {Spoto}, F. and {Steele}, I.~A. and {Steidelm{\"u}ller}, H. and {Stephenson}, C.~A. and {S{\"u}veges}, M. and {Szabados}, L. and {Szegedi-Elek}, E. and {Taris}, F. and {Tauran}, G. and {Taylor}, M.~B. and {Teixeira}, R. and {Thuillot}, W. and {Tonello}, N. and {Torra}, F. and {Torra}, J. and {Turon}, C. and {Unger}, N. and {Vaillant}, M. and {van Dillen}, E. and {Vanel}, O. and {Vecchiato}, A. and {Viala}, Y. and {Vicente}, D. and {Voutsinas}, S. and {Weiler}, M. and {Wevers}, T. and {Wyrzykowski}, {\L}. and {Yoldas}, A. and {Yvard}, P. and {Zhao}, H. and {Zorec}, J. and {Zucker}, S. and {Zurbach}, C. and {Zwitter}, T.},
        title = "{Gaia Early Data Release 3. Summary of the contents and survey properties}",
      journal = {\aap},
         year = 2021,
        month = may,
       volume = {649},
          eid = {A1},
        pages = {A1},
          doi = {10.1051/0004-6361/202039657},
archivePrefix = {arXiv},
       eprint = {2012.01533},
 primaryClass = {astro-ph.GA},
       adsurl = {https://ui.adsabs.harvard.edu/abs/2021A&A...649A...1G}
}

@ARTICLE{Rybizki2020,
       author = {{Rybizki}, Jan and {Demleitner}, Markus and {Bailer-Jones}, Coryn and {Tio}, Piero Dal and {Cantat-Gaudin}, Tristan and {Fouesneau}, Morgan and {Chen}, Yang and {Andrae}, Ren{\'e} and {Girardi}, L{\'e}o and {Sharma}, Sanjib},
        title = "{A Gaia Early DR3 Mock Stellar Catalog: Galactic Prior and Selection Function}",
      journal = {\pasp},
         year = 2020,
        month = jul,
       volume = {132},
       number = {1013},
          eid = {074501},
        pages = {074501},
          doi = {10.1088/1538-3873/ab8cb0},
archivePrefix = {arXiv},
       eprint = {2004.09991},
 primaryClass = {astro-ph.IM},
       adsurl = {https://ui.adsabs.harvard.edu/abs/2020PASP..132g4501R}
}

@ARTICLE{GaiaEDR3val,
       author = {{Fabricius}, C. and {Luri}, X. and {Arenou}, F. and {Babusiaux}, C. and {Helmi}, A. and {Muraveva}, T. and {Reyl{\'e}}, C. and {Spoto}, F. and {Vallenari}, A. and {Antoja}, T. and {Balbinot}, E. and {Barache}, C. and {Bauchet}, N. and {Bragaglia}, A. and {Busonero}, D. and {Cantat-Gaudin}, T. and {Carrasco}, J.~M. and {Diakit{\'e}}, S. and {Fabrizio}, M. and {Figueras}, F. and {Garcia-Gutierrez}, A. and {Garofalo}, A. and {Jordi}, C. and {Kervella}, P. and {Khanna}, S. and {Leclerc}, N. and {Licata}, E. and {Lambert}, S. and {Marrese}, P.~M. and {Masip}, A. and {Ramos}, P. and {Robichon}, N. and {Robin}, A.~C. and {Romero-G{\'o}mez}, M. and {Rubele}, S. and {Weiler}, M.},
        title = "{Gaia Early Data Release 3. Catalogue validation}",
      journal = {\aap},
         year = 2021,
        month = may,
       volume = {649},
          eid = {A5},
        pages = {A5},
          doi = {10.1051/0004-6361/202039834},
archivePrefix = {arXiv},
       eprint = {2012.06242},
 primaryClass = {astro-ph.GA},
       adsurl = {https://ui.adsabs.harvard.edu/abs/2021A&A...649A...5F}
}

@ARTICLE{gaia-alerts,
       author = {{Hodgkin}, S.~T. and {Harrison}, D.~L. and {Breedt}, E. and {Wevers}, T. and {Rixon}, G. and {Delgado}, A. and {Yoldas}, A. and {Kostrzewa-Rutkowska}, Z. and {Wyrzykowski}, {\L}. and {van Leeuwen}, M. and {Blagorodnova}, N. and {Campbell}, H. and {Eappachen}, D. and {Fraser}, M. and {Ihanec}, N. and {Koposov}, S.~E. and {Kruszy{\'n}ska}, K. and {Marton}, G. and {Rybicki}, K.~A. and {Brown}, A.~G.~A. and {Burgess}, P.~W. and {Busso}, G. and {Cowell}, S. and {De Angeli}, F. and {Diener}, C. and {Evans}, D.~W. and {Gilmore}, G. and {Holland}, G. and {Jonker}, P.~G. and {van Leeuwen}, F. and {Mignard}, F. and {Osborne}, P.~J. and {Portell}, J. and {Prusti}, T. and {Richards}, P.~J. and {Riello}, M. and {Seabroke}, G.~M. and {Walton}, N.~A. and {{\'A}brah{\'a}m}, P. and {Altavilla}, G. and {Baker}, S.~G. and {Bastian}, U. and {O'Brien}, P. and {de Bruijne}, J. and {Butterley}, T. and {Carrasco}, J.~M. and {Casta{\~n}eda}, J. and {Clark}, J.~S. and {Clementini}, G. and {Copperwheat}, C.~M. and {Cropper}, M. and {Damljanovic}, G. and {Davidson}, M. and {Davis}, C.~J. and {Dennefeld}, M. and {Dhillon}, V.~S. and {Dolding}, C. and {Dominik}, M. and {Esquej}, P. and {Eyer}, L. and {Fabricius}, C. and {Fridman}, M. and {Froebrich}, D. and {Garralda}, N. and {Gomboc}, A. and {Gonz{\'a}lez-Vidal}, J.~J. and {Guerra}, R. and {Hambly}, N.~C. and {Hardy}, L.~K. and {Holl}, B. and {Hourihane}, A. and {Japelj}, J. and {Kann}, D.~A. and {Kiss}, C. and {Knigge}, C. and {Kolb}, U. and {Komossa}, S. and {K{\'o}sp{\'a}l}, {\'A}. and {Kov{\'a}cs}, G. and {Kun}, M. and {Leto}, G. and {Lewis}, F. and {Littlefair}, S.~P. and {Mahabal}, A.~A. and {Mundell}, C.~G. and {Nagy}, Z. and {Padeletti}, D. and {Palaversa}, L. and {Pigulski}, A. and {Pretorius}, M.~L. and {van Reeven}, W. and {Ribeiro}, V.~A.~R.~M. and {Roelens}, M. and {Rowell}, N. and {Schartel}, N. and {Scholz}, A. and {Schwope}, A. and {Sip{\H{o}}cz}, B.~M. and {Smartt}, S.~J. and {Smith}, M.~D. and {Serraller}, I. and {Steeghs}, D. and {Sullivan}, M. and {Szabados}, L. and {Szegedi-Elek}, E. and {Tisserand}, P. and {Tomasella}, L. and {van Velzen}, S. and {Whitelock}, P.~A. and {Wilson}, R.~W. and {Young}, D.~R.},
        title = "{Gaia Early Data Release 3. Gaia photometric science alerts}",
      journal = {\aap},
         year = 2021,
        month = aug,
       volume = {652},
          eid = {A76},
        pages = {A76},
          doi = {10.1051/0004-6361/202140735},
archivePrefix = {arXiv},
       eprint = {2106.01394},
 primaryClass = {astro-ph.IM},
       adsurl = {https://ui.adsabs.harvard.edu/abs/2021A&A...652A..76H}
}

@ARTICLE{Fardeen2024,
       author = {{Fardeen}, James and {McGill}, Peter and {Perkins}, Scott E. and {Dawson}, William A. and {Abrams}, Natasha S. and {Lu}, Jessica R. and {Ho}, Ming-Feng and {Bird}, Simeon},
        title = "{Astrometric Microlensing by Primordial Black Holes with the Roman Space Telescope}",
      journal = {\apj},
         year = 2024,
        month = apr,
       volume = {965},
       number = {2},
          eid = {138},
        pages = {138},
          doi = {10.3847/1538-4357/ad3243},
archivePrefix = {arXiv},
       eprint = {2312.13249},
 primaryClass = {astro-ph.GA},
       adsurl = {https://ui.adsabs.harvard.edu/abs/2024ApJ...965..138F}
}

@ARTICLE{Luna2023,
       author = {{Luna}, Alonso and {Marchetti}, Tommaso and {Rejkuba}, Marina and {Minniti}, Dante},
        title = "{Astrometry in crowded fields towards the Galactic bulge}",
      journal = {\aap},
         year = 2023,
        month = sep,
       volume = {677},
          eid = {A185},
        pages = {A185},
          doi = {10.1051/0004-6361/202346257},
archivePrefix = {arXiv},
       eprint = {2307.13719},
 primaryClass = {astro-ph.GA},
       adsurl = {https://ui.adsabs.harvard.edu/abs/2023A&A...677A.185L}
}

@ARTICLE{GaiaverseV,
       author = {{Everall}, Andrew and {Boubert}, Douglas},
        title = "{Completeness of the Gaia verse - V. Astrometry and radial velocity sample selection functions in Gaia EDR3}",
      journal = {\mnras},
         year = 2022,
        month = feb,
       volume = {509},
       number = {4},
        pages = {6205-6224},
          doi = {10.1093/mnras/stab3262},
archivePrefix = {arXiv},
       eprint = {2111.04127},
 primaryClass = {astro-ph.GA},
       adsurl = {https://ui.adsabs.harvard.edu/abs/2022MNRAS.509.6205E}
}

@ARTICLE{Gaia19bld,
       author = {{Rybicki}, K.~A. and {Wyrzykowski}, {\L}. and {Bachelet}, E. and {Cassan}, A. and {Zieli{\'n}ski}, P. and {Gould}, A. and {Calchi Novati}, S. and {Yee}, J.~C. and {Ryu}, Y. -H. and {Gromadzki}, M. and {Miko{\l}ajczyk}, P. and {Ihanec}, N. and {Kruszy{\'n}ska}, K. and {Hambsch}, F. -J. and {Zo{\l}a}, S. and {Fossey}, S.~J. and {Awiphan}, S. and {Nakharutai}, N. and {Lewis}, F. and {Olivares E.}, F. and {Hodgkin}, S. and {Delgado}, A. and {Breedt}, E. and {Harrison}, D.~L. and {van Leeuwen}, M. and {Rixon}, G. and {Wevers}, T. and {Yoldas}, A. and {Udalski}, A. and {Szyma{\'n}ski}, M.~K. and {Soszy{\'n}ski}, I. and {Pietrukowicz}, P. and {Koz{\l}owski}, S. and {Skowron}, J. and {Poleski}, R. and {Ulaczyk}, K. and {Mr{\'o}z}, P. and {Iwanek}, P. and {Wrona}, M. and {Street}, R.~A. and {Tsapras}, Y. and {Hundertmark}, M. and {Dominik}, M. and {Beichman}, C. and {Bryden}, G. and {Carey}, S. and {Gaudi}, B.~S. and {Henderson}, C. and {Shvartzvald}, Y. and {Zang}, W. and {Zhu}, W. and {Christie}, G.~W. and {Green}, J. and {Hennerley}, S. and {McCormick}, J. and {Monard}, L.~A.~G. and {Natusch}, T. and {Pogge}, R.~W. and {Gezer}, I. and {Gurgul}, A. and {Kaczmarek}, Z. and {Konacki}, M. and {Lam}, M.~C. and {Maskoliunas}, M. and {Pakstiene}, E. and {Ratajczak}, M. and {Stankeviciute}, A. and {Zdanavicius}, J. and {Zi{\'o}{\l}kowska}, O.},
        title = "{Single-lens mass measurement in the high-magnification microlensing event Gaia19bld located in the Galactic disc}",
      journal = {\aap},
         year = 2022,
        month = jan,
       volume = {657},
          eid = {A18},
        pages = {A18},
          doi = {10.1051/0004-6361/202039542},
archivePrefix = {arXiv},
       eprint = {2112.01613},
 primaryClass = {astro-ph.GA},
       adsurl = {https://ui.adsabs.harvard.edu/abs/2022A&A...657A..18R}
}

@ARTICLE{Gaia16aye,
       author = {{Wyrzykowski}, {\L}. and {Mr{\'o}z}, P. and {Rybicki}, K.~A. and {Gromadzki}, M. and {Ko{\l}aczkowski}, Z. and {Zieli{\'n}ski}, M. and {Zieli{\'n}ski}, P. and {Britavskiy}, N. and {Gomboc}, A. and {Sokolovsky}, K. and {Hodgkin}, S.~T. and {Abe}, L. and {Aldi}, G.~F. and {AlMannaei}, A. and {Altavilla}, G. and {Al Qasim}, A. and {Anupama}, G.~C. and {Awiphan}, S. and {Bachelet}, E. and {Bak{\i}{\c{s}}}, V. and {Baker}, S. and {Bartlett}, S. and {Bendjoya}, P. and {Benson}, K. and {Bikmaev}, I.~F. and {Birenbaum}, G. and {Blagorodnova}, N. and {Blanco-Cuaresma}, S. and {Boeva}, S. and {Bonanos}, A.~Z. and {Bozza}, V. and {Bramich}, D.~M. and {Bruni}, I. and {Burenin}, R.~A. and {Burgaz}, U. and {Butterley}, T. and {Caines}, H.~E. and {Caton}, D.~B. and {Calchi Novati}, S. and {Carrasco}, J.~M. and {Cassan}, A. and {{\v{C}}epas}, V. and {Cropper}, M. and {Chru{\'s}li{\'n}ska}, M. and {Clementini}, G. and {Clerici}, A. and {Conti}, D. and {Conti}, M. and {Cross}, S. and {Cusano}, F. and {Damljanovic}, G. and {Dapergolas}, A. and {D'Ago}, G. and {de Bruijne}, J.~H.~J. and {Dennefeld}, M. and {Dhillon}, V.~S. and {Dominik}, M. and {Dziedzic}, J. and {Erece}, O. and {Eselevich}, M.~V. and {Esenoglu}, H. and {Eyer}, L. and {Figuera Jaimes}, R. and {Fossey}, S.~J. and {Galeev}, A.~I. and {Grebenev}, S.~A. and {Gupta}, A.~C. and {Gutaev}, A.~G. and {Hallakoun}, N. and {Hamanowicz}, A. and {Han}, C. and {Handzlik}, B. and {Haislip}, J.~B. and {Hanlon}, L. and {Hardy}, L.~K. and {Harrison}, D.~L. and {van Heerden}, H.~J. and {Hoette}, V.~L. and {Horne}, K. and {Hudec}, R. and {Hundertmark}, M. and {Ihanec}, N. and {Irtuganov}, E.~N. and {Itoh}, R. and {Iwanek}, P. and {Jovanovic}, M.~D. and {Janulis}, R. and {Jel{\'\i}nek}, M. and {Jensen}, E. and {Kaczmarek}, Z. and {Katz}, D. and {Khamitov}, I.~M. and {Kilic}, Y. and {Klencki}, J. and {Kolb}, U. and {Kopacki}, G. and {Kouprianov}, V.~V. and {Kruszy{\'n}ska}, K. and {Kurowski}, S. and {Latev}, G. and {Lee}, C. -H. and {Leonini}, S. and {Leto}, G. and {Lewis}, F. and {Li}, Z. and {Liakos}, A. and {Littlefair}, S.~P. and {Lu}, J. and {Manser}, C.~J. and {Mao}, S. and {Maoz}, D. and {Martin-Carrillo}, A. and {Marais}, J.~P. and {Maskoli{\={u}}nas}, M. and {Maund}, J.~R. and {Meintjes}, P.~J. and {Melnikov}, S.~S. and {Ment}, K. and {Miko{\l}ajczyk}, P. and {Morrell}, M. and {Mowlavi}, N. and {Mo{\'z}dzierski}, D. and {Murphy}, D. and {Nazarov}, S. and {Netzel}, H. and {Nesci}, R. and {Ngeow}, C. -C. and {Norton}, A.~J. and {Ofek}, E.~O. and {Pak{\v{s}}tien{\.{e}}}, E. and {Palaversa}, L. and {Pandey}, A. and {Paraskeva}, E. and {Pawlak}, M. and {Penny}, M.~T. and {Penprase}, B.~E. and {Piascik}, A. and {Prieto}, J.~L. and {Qvam}, J.~K.~T. and {Ranc}, C. and {Rebassa-Mansergas}, A. and {Reichart}, D.~E. and {Reig}, P. and {Rhodes}, L. and {Rivet}, J. -P. and {Rixon}, G. and {Roberts}, D. and {Rosi}, P. and {Russell}, D.~M. and {Zanmar Sanchez}, R. and {Scarpetta}, G. and {Seabroke}, G. and {Shappee}, B.~J. and {Schmidt}, R. and {Shvartzvald}, Y. and {Sitek}, M. and {Skowron}, J. and {{\'S}niegowska}, M. and {Snodgrass}, C. and {Soares}, P.~S. and {van Soelen}, B. and {Spetsieri}, Z.~T. and {Stankevi{\v{c}}i{\={u}}t{\.{e}}}, A. and {Steele}, I.~A. and {Street}, R.~A. and {Strobl}, J. and {Strubble}, E. and {Szegedi}, H. and {Tinjaca Ramirez}, L.~M. and {Tomasella}, L. and {Tsapras}, Y. and {Vernet}, D. and {Villanueva}, S. and {Vince}, O. and {Wambsganss}, J. and {van der Westhuizen}, I.~P. and {Wiersema}, K. and {Wium}, D. and {Wilson}, R.~W. and {Yoldas}, A. and {Zhuchkov}, R. Ya. and {Zhukov}, D.~G. and {Zdanavi{\v{c}}ius}, J. and {Zo{\l}a}, S. and {Zubareva}, A.},
        title = "{Full orbital solution for the binary system in the northern Galactic disc microlensing event Gaia16aye}",
      journal = {\aap},
         year = 2020,
        month = jan,
       volume = {633},
          eid = {A98},
        pages = {A98},
          doi = {10.1051/0004-6361/201935097},
archivePrefix = {arXiv},
       eprint = {1901.07281},
 primaryClass = {astro-ph.SR},
       adsurl = {https://ui.adsabs.harvard.edu/abs/2020A&A...633A..98W}
}

@ARTICLE{Gaia22dkv,
       author = {{Wu}, Zexuan and {Dong}, Subo and {Yi}, Tuan and {Liu}, Zhuokai and {El-Badry}, Kareem and {Gould}, Andrew and {Wyrzykowski}, L. and {Rybicki}, K.~A. and {Bachelet}, Etienne and {Christie}, Grant W. and {de Almeida}, L. and {Monard}, L.~A.~G. and {McCormick}, J. and {Natusch}, Tim and {Zieli{\'n}ski}, P. and {Chen}, Huiling and {Huang}, Yang and {Liu}, Chang and {M{\'e}rand}, A. and {Mr{\'o}z}, Przemek and {Shangguan}, Jinyi and {Udalski}, Andrzej and {Woillez}, J. and {Zhang}, Huawei and {Hambsch}, Franz-Josef and {Miko{\l}ajczyk}, P.~J. and {Gromadzki}, M. and {Ratajczak}, M. and {Kruszy{\'n}ska}, Katarzyna and {Ihanec}, N. and {Pylypenko}, Uliana and {Sitek}, M. and {Howil}, K. and {Zola}, Staszek and {Michniewicz}, Olga and {Zejmo}, Michal and {Lewis}, Fraser and {Bronikowski}, Mateusz and {Potter}, Stephen and {Andrzejewski}, Jan and {Merc}, Jaroslav and {Street}, Rachel and {Fukui}, Akihiko and {Figuera Jaimes}, R. and {Bozza}, V. and {Rota}, P. and {Cassan}, A. and {Dominik}, M. and {Tsapras}, Y. and {Hundertmark}, M. and {Wambsganss}, J. and {B{\k{a}}kowska}, K. and {S{\l}owikowska}, A.},
        title = "{Gaia22dkvLb: A Microlensing Planet Potentially Accessible to Radial-velocity Characterization}",
      journal = {\aj},
         year = 2024,
        month = aug,
       volume = {168},
       number = {2},
          eid = {62},
        pages = {62},
          doi = {10.3847/1538-3881/ad5203},
archivePrefix = {arXiv},
       eprint = {2309.03944},
 primaryClass = {astro-ph.EP},
       adsurl = {https://ui.adsabs.harvard.edu/abs/2024AJ....168...62W}
}

@ARTICLE{Gaia21blx,
       author = {{Rota}, P. and {Bozza}, V. and {Hundertmark}, M. and {Bachelet}, E. and {Street}, R. and {Tsapras}, Y. and {Cassan}, A. and {Dominik}, M. and {Figuera Jaimes}, R. and {Rybicki}, K.~A. and {Wambsganss}, J. and {Wyrzykowski}, {\L}. and {Zieli{\'n}ski}, P. and {OMEGA Key Project} and {Bonavita}, M. and {Hinse}, T.~C. and {J{\o}rgensen}, U.~G. and {Khalouei}, E. and {Korhonen}, H. and {Longa-Pe{\~n}a}, P. and {Peixinho}, N. and {Rahvar}, S. and {Sajadian}, S. and {Skottfelt}, J. and {Snodgrass}, C. and {Tregolan-Reed}, J. and {MiNDSTEp Consortium}},
        title = "{Gaia21blx: Complete resolution of a binary microlensing event in the Galactic disk}",
      journal = {\aap},
         year = 2024,
        month = jun,
       volume = {686},
          eid = {A173},
        pages = {A173},
          doi = {10.1051/0004-6361/202347807},
archivePrefix = {arXiv},
       eprint = {2404.05078},
 primaryClass = {astro-ph.SR},
       adsurl = {https://ui.adsabs.harvard.edu/abs/2024A&A...686A.173R}
}

@ARTICLE{Koshimoto2024,
       author = {{Koshimoto}, Naoki and {Kawanaka}, Norita and {Tsuna}, Daichi},
        title = "{Influence of Black Hole Kick Velocity on Microlensing Distributions}",
      journal = {\apj},
         year = 2024,
        month = sep,
       volume = {973},
       number = {1},
          eid = {5},
        pages = {5},
          doi = {10.3847/1538-4357/ad5feb},
archivePrefix = {arXiv},
       eprint = {2405.07502},
 primaryClass = {astro-ph.GA},
       adsurl = {https://ui.adsabs.harvard.edu/abs/2024ApJ...973....5K}
}

@ARTICLE{DR3lenses,
       author = {{Wyrzykowski}, {\L}. and {Kruszy{\'n}ska}, K. and {Rybicki}, K.~A. and {Holl}, B. and {Lec{\oe}ur-Ta{\"\i}bi}, I. and {Mowlavi}, N. and {Nienartowicz}, K. and {Jevardat de Fombelle}, G. and {Rimoldini}, L. and {Audard}, M. and {Garcia-Lario}, P. and {Gavras}, P. and {Evans}, D.~W. and {Hodgkin}, S.~T. and {Eyer}, L.},
        title = "{Gaia Data Release 3. Microlensing events from all over the sky}",
      journal = {\aap},
         year = 2023,
        month = jun,
       volume = {674},
          eid = {A23},
        pages = {A23},
          doi = {10.1051/0004-6361/202243756},
archivePrefix = {arXiv},
       eprint = {2206.06121},
 primaryClass = {astro-ph.SR},
       adsurl = {https://ui.adsabs.harvard.edu/abs/2023A&A...674A..23W}
}

@ARTICLE{Spergel2015,
       author = {{Spergel}, D. and {Gehrels}, N. and {Baltay}, C. and {Bennett}, D. and {Breckinridge}, J. and {Donahue}, M. and {Dressler}, A. and {Gaudi}, B.~S. and {Greene}, T. and {Guyon}, O. and {Hirata}, C. and {Kalirai}, J. and {Kasdin}, N.~J. and {Macintosh}, B. and {Moos}, W. and {Perlmutter}, S. and {Postman}, M. and {Rauscher}, B. and {Rhodes}, J. and {Wang}, Y. and {Weinberg}, D. and {Benford}, D. and {Hudson}, M. and {Jeong}, W. -S. and {Mellier}, Y. and {Traub}, W. and {Yamada}, T. and {Capak}, P. and {Colbert}, J. and {Masters}, D. and {Penny}, M. and {Savransky}, D. and {Stern}, D. and {Zimmerman}, N. and {Barry}, R. and {Bartusek}, L. and {Carpenter}, K. and {Cheng}, E. and {Content}, D. and {Dekens}, F. and {Demers}, R. and {Grady}, K. and {Jackson}, C. and {Kuan}, G. and {Kruk}, J. and {Melton}, M. and {Nemati}, B. and {Parvin}, B. and {Poberezhskiy}, I. and {Peddie}, C. and {Ruffa}, J. and {Wallace}, J.~K. and {Whipple}, A. and {Wollack}, E. and {Zhao}, F.},
        title = "{Wide-Field InfrarRed Survey Telescope-Astrophysics Focused Telescope Assets WFIRST-AFTA 2015 Report}",
      journal = {arXiv e-prints},
         year = 2015,
        month = mar,
          eid = {arXiv:1503.03757},
        pages = {arXiv:1503.03757},
          doi = {10.48550/arXiv.1503.03757},
archivePrefix = {arXiv},
       eprint = {1503.03757},
 primaryClass = {astro-ph.IM},
       adsurl = {https://ui.adsabs.harvard.edu/abs/2015arXiv150303757S}
}

@ARTICLE{OGLEIV,
       author = {{Udalski}, A. and {Szyma{\'n}ski}, M.~K. and {Szyma{\'n}ski}, G.},
        title = "{OGLE-IV: Fourth Phase of the Optical Gravitational Lensing Experiment}",
      journal = {\actaa},
         year = 2015,
        month = mar,
       volume = {65},
       number = {1},
        pages = {1-38},
          doi = {10.48550/arXiv.1504.05966},
archivePrefix = {arXiv},
       eprint = {1504.05966},
 primaryClass = {astro-ph.SR},
       adsurl = {https://ui.adsabs.harvard.edu/abs/2015AcA....65....1U}
}

@ARTICLE{KMTNet,
       author = {{Kim}, Seung-Lee and {Lee}, Chung-Uk and {Park}, Byeong-Gon and {Kim}, Dong-Jin and {Cha}, Sang-Mok and {Lee}, Yongseok and {Han}, Cheongho and {Chun}, Moo-Young and {Yuk}, Insoo},
        title = "{KMTNET: A Network of 1.6 m Wide-Field Optical Telescopes Installed at Three Southern Observatories}",
      journal = {Journal of Korean Astronomical Society},
         year = 2016,
        month = feb,
       volume = {49},
       number = {1},
        pages = {37-44},
          doi = {10.5303/JKAS.2016.49.1.37},
       adsurl = {https://ui.adsabs.harvard.edu/abs/2016JKAS...49...37K}
}

@ARTICLE{Wyrzykowski2016,
       author = {{Wyrzykowski}, {\L}. and {Kostrzewa-Rutkowska}, Z. and {Skowron}, J. and {Rybicki}, K.~A. and {Mr{\'o}z}, P. and {Koz{\l}owski}, S. and {Udalski}, A. and {Szyma{\'n}ski}, M.~K. and {Pietrzy{\'n}ski}, G. and {Soszy{\'n}ski}, I. and {Ulaczyk}, K. and {Pietrukowicz}, P. and {Poleski}, R. and {Pawlak}, M. and {I{\l}kiewicz}, K. and {Rattenbury}, N.~J.},
        title = "{Black hole, neutron star and white dwarf candidates from microlensing with OGLE-III}",
      journal = {\mnras},
         year = 2016,
        month = may,
       volume = {458},
       number = {3},
        pages = {3012-3026},
          doi = {10.1093/mnras/stw426},
archivePrefix = {arXiv},
       eprint = {1509.04899},
 primaryClass = {astro-ph.SR},
       adsurl = {https://ui.adsabs.harvard.edu/abs/2016MNRAS.458.3012W}
}

@ARTICLE{Sallaberry2025,
       author = {{Sallaberry}, Greg and {Kaczmarek}, Zofia and {McGill}, Peter and {Perkins}, Scott and {Dawson}, William and {Begbie}, Caitlin},
        title = "{popclass: A Python Package for Classifying Microlensing Events}",
      journal = {The Journal of Open Source Software},
         year = 2025,
        month = may,
       volume = {10},
       number = {109},
          eid = {7769},
        pages = {7769},
          doi = {10.21105/joss.07769},
archivePrefix = {arXiv},
       eprint = {2410.14076},
 primaryClass = {astro-ph.IM},
       adsurl = {https://ui.adsabs.harvard.edu/abs/2025JOSS...10.7769S}
}

@ARTICLE{DR1_preprocessing,
       author = {{Fabricius}, C. and {Bastian}, U. and {Portell}, J. and {Casta{\~n}eda}, J. and {Davidson}, M. and {Hambly}, N.~C. and {Clotet}, M. and {Biermann}, M. and {Mora}, A. and {Busonero}, D. and {Riva}, A. and {Brown}, A.~G.~A. and {Smart}, R. and {Lammers}, U. and {Torra}, J. and {Drimmel}, R. and {Gracia}, G. and {L{\"o}ffler}, W. and {Spagna}, A. and {Lindegren}, L. and {Klioner}, S. and {Andrei}, A. and {Bach}, N. and {Bramante}, L. and {Br{\"u}semeister}, T. and {Busso}, G. and {Carrasco}, J.~M. and {Gai}, M. and {Garralda}, N. and {Gonz{\'a}lez-Vidal}, J.~J. and {Guerra}, R. and {Hauser}, M. and {Jordan}, S. and {Jordi}, C. and {Lenhardt}, H. and {Mignard}, F. and {Messineo}, R. and {Mulone}, A. and {Serraller}, I. and {Stampa}, U. and {Tanga}, P. and {van Elteren}, A. and {van Reeven}, W. and {Voss}, H. and {Abbas}, U. and {Allasia}, W. and {Altmann}, M. and {Anton}, S. and {Barache}, C. and {Becciani}, U. and {Berthier}, J. and {Bianchi}, L. and {Bombrun}, A. and {Bouquillon}, S. and {Bourda}, G. and {Bucciarelli}, B. and {Butkevich}, A. and {Buzzi}, R. and {Cancelliere}, R. and {Carlucci}, T. and {Charlot}, P. and {Collins}, R. and {Comoretto}, G. and {Cross}, N. and {Crosta}, M. and {de Felice}, F. and {Fienga}, A. and {Figueras}, F. and {Fraile}, E. and {Geyer}, R. and {Hernandez}, J. and {Hobbs}, D. and {Hofmann}, W. and {Liao}, S. and {Licata}, E. and {Martino}, M. and {McMillan}, P.~J. and {Michalik}, D. and {Morbidelli}, R. and {Parsons}, P. and {Pecoraro}, M. and {Ramos-Lerate}, M. and {Sarasso}, M. and {Siddiqui}, H. and {Steele}, I. and {Steidelm{\"u}ller}, H. and {Taris}, F. and {Vecchiato}, A. and {Abreu}, A. and {Anglada}, E. and {Boudreault}, S. and {Cropper}, M. and {Holl}, B. and {Cheek}, N. and {Crowley}, C. and {Fleitas}, J.~M. and {Hutton}, A. and {Osinde}, J. and {Rowell}, N. and {Salguero}, E. and {Utrilla}, E. and {Blagorodnova}, N. and {Soffel}, M. and {Osorio}, J. and {Vicente}, D. and {Cambras}, J. and {Bernstein}, H. -H.},
        title = "{Gaia Data Release 1. Pre-processing and source list creation}",
      journal = {\aap},
         year = 2016,
        month = nov,
       volume = {595},
          eid = {A3},
        pages = {A3},
          doi = {10.1051/0004-6361/201628643},
archivePrefix = {arXiv},
       eprint = {1609.04273},
 primaryClass = {astro-ph.IM},
       adsurl = {https://ui.adsabs.harvard.edu/abs/2016A&A...595A...3F}
}

@ARTICLE{Husseiniova2021,
       author = {{Husseiniova}, Andrea and {McGill}, Peter and {Smith}, Leigh C. and {Evans}, N. Wyn},
        title = "{A microlensing search of 700 million VVV light curves}",
      journal = {\mnras},
         year = 2021,
        month = sep,
       volume = {506},
       number = {2},
        pages = {2482-2502},
          doi = {10.1093/mnras/stab1882},
archivePrefix = {arXiv},
       eprint = {2106.15617},
 primaryClass = {astro-ph.GA},
       adsurl = {https://ui.adsabs.harvard.edu/abs/2021MNRAS.506.2482H}
}

@ARTICLE{Lam2023,
       author = {{Lam}, Casey Y. and {Lu}, Jessica R.},
        title = "{A Reanalysis of the Isolated Black Hole Candidate OGLE-2011-BLG-0462/MOA-2011-BLG-191}",
      journal = {\apj},
         year = 2023,
        month = oct,
       volume = {955},
       number = {2},
          eid = {116},
        pages = {116},
          doi = {10.3847/1538-4357/aced4a},
archivePrefix = {arXiv},
       eprint = {2308.03302},
 primaryClass = {astro-ph.SR},
       adsurl = {https://ui.adsabs.harvard.edu/abs/2023ApJ...955..116L}
}

@ARTICLE{Lam2022,
       author = {{Lam}, Casey Y. and {Lu}, Jessica R. and {Udalski}, Andrzej and {Bond}, Ian and {Bennett}, David P. and {Skowron}, Jan and {Mr{\'o}z}, Przemek and {Poleski}, Radek and {Sumi}, Takahiro and {Szyma{\'n}ski}, Micha{\l} K. and {Koz{\l}owski}, Szymon and {Pietrukowicz}, Pawe{\l} and {Soszy{\'n}ski}, Igor and {Ulaczyk}, Krzysztof and {Wyrzykowski}, {\L}ukasz and {Miyazaki}, Shota and {Suzuki}, Daisuke and {Koshimoto}, Naoki and {Rattenbury}, Nicholas J. and {Hosek}, Matthew W. and {Abe}, Fumio and {Barry}, Richard and {Bhattacharya}, Aparna and {Fukui}, Akihiko and {Fujii}, Hirosane and {Hirao}, Yuki and {Itow}, Yoshitaka and {Kirikawa}, Rintaro and {Kondo}, Iona and {Matsubara}, Yutaka and {Matsumoto}, Sho and {Muraki}, Yasushi and {Olmschenk}, Greg and {Ranc}, Cl{\'e}ment and {Okamura}, Arisa and {Satoh}, Yuki and {Silva}, Stela Ishitani and {Toda}, Taiga and {Tristram}, Paul J. and {Vandorou}, Aikaterini and {Yama}, Hibiki and {Abrams}, Natasha S. and {Agarwal}, Shrihan and {Rose}, Sam and {Terry}, Sean K.},
        title = "{An Isolated Mass-gap Black Hole or Neutron Star Detected with Astrometric Microlensing}",
      journal = {\apjl},
         year = 2022,
        month = jul,
       volume = {933},
       number = {1},
          eid = {L23},
        pages = {L23},
          doi = {10.3847/2041-8213/ac7442},
archivePrefix = {arXiv},
       eprint = {2202.01903},
 primaryClass = {astro-ph.GA},
       adsurl = {https://ui.adsabs.harvard.edu/abs/2022ApJ...933L..23L}
}

@ARTICLE{Sahu2022,
       author = {{Sahu}, Kailash C. and {Anderson}, Jay and {Casertano}, Stefano and {Bond}, Howard E. and {Udalski}, Andrzej and {Dominik}, Martin and {Calamida}, Annalisa and {Bellini}, Andrea and {Brown}, Thomas M. and {Rejkuba}, Marina and {Bajaj}, Varun and {Kains}, No{\'e} and {Ferguson}, Henry C. and {Fryer}, Chris L. and {Yock}, Philip and {Mr{\'o}z}, Przemek and {Koz{\l}owski}, Szymon and {Pietrukowicz}, Pawe{\l} and {Poleski}, Radek and {Skowron}, Jan and {Soszy{\'n}ski}, Igor and {Szyma{\'n}ski}, Micha{\l} K. and {Ulaczyk}, Krzysztof and {Wyrzykowski}, {\L}ukasz and {Barry}, Richard K. and {Bennett}, David P. and {Bond}, Ian A. and {Hirao}, Yuki and {Silva}, Stela Ishitani and {Kondo}, Iona and {Koshimoto}, Naoki and {Ranc}, Cl{\'e}ment and {Rattenbury}, Nicholas J. and {Sumi}, Takahiro and {Suzuki}, Daisuke and {Tristram}, Paul J. and {Vandorou}, Aikaterini and {Beaulieu}, Jean-Philippe and {Marquette}, Jean-Baptiste and {Cole}, Andrew and {Fouqu{\'e}}, Pascal and {Hill}, Kym and {Dieters}, Stefan and {Coutures}, Christian and {Dominis-Prester}, Dijana and {Bennett}, Clara and {Bachelet}, Etienne and {Menzies}, John and {Albrow}, Michael and {Pollard}, Karen and {Gould}, Andrew and {Yee}, Jennifer C. and {Allen}, William and {Almeida}, Leonardo A. and {Christie}, Grant and {Drummond}, John and {Gal-Yam}, Avishay and {Gorbikov}, Evgeny and {Jablonski}, Francisco and {Lee}, Chung-Uk and {Maoz}, Dan and {Manulis}, Ilan and {McCormick}, Jennie and {Natusch}, Tim and {Pogge}, Richard W. and {Shvartzvald}, Yossi and {J{\o}rgensen}, Uffe G. and {Alsubai}, Khalid A. and {Andersen}, Michael I. and {Bozza}, Valerio and {Novati}, Sebastiano Calchi and {Burgdorf}, Martin and {Hinse}, Tobias C. and {Hundertmark}, Markus and {Husser}, Tim-Oliver and {Kerins}, Eamonn and {Longa-Pe{\~n}a}, Penelope and {Mancini}, Luigi and {Penny}, Matthew and {Rahvar}, Sohrab and {Ricci}, Davide and {Sajadian}, Sedighe and {Skottfelt}, Jesper and {Snodgrass}, Colin and {Southworth}, John and {Tregloan-Reed}, Jeremy and {Wambsganss}, Joachim and {Wertz}, Olivier and {Tsapras}, Yiannis and {Street}, Rachel A. and {Bramich}, D.~M. and {Horne}, Keith and {Steele}, Iain A. and {RoboNet Collaboration}},
        title = "{An Isolated Stellar-mass Black Hole Detected through Astrometric Microlensing}",
      journal = {\apj},
         year = 2022,
        month = jul,
       volume = {933},
       number = {1},
          eid = {83},
        pages = {83},
          doi = {10.3847/1538-4357/ac739e},
archivePrefix = {arXiv},
       eprint = {2201.13296},
 primaryClass = {astro-ph.SR},
       adsurl = {https://ui.adsabs.harvard.edu/abs/2022ApJ...933...83S}
}

@ARTICLE{McGill2023,
       author = {{McGill}, Peter and {Anderson}, Jay and {Casertano}, Stefano and {Sahu}, Kailash C. and {Bergeron}, Pierre and {Blouin}, Simon and {Dufour}, Patrick and {Smith}, Leigh C. and {Evans}, N. Wyn and {Belokurov}, Vasily and {Smart}, Richard L. and {Bellini}, Andrea and {Calamida}, Annalisa and {Dominik}, Martin and {Kains}, No{\'e} and {Kl{\"u}ter}, Jonas and {Nielsen}, Martin Bo and {Wambsganss}, Joachim},
        title = "{First semi-empirical test of the white dwarf mass-radius relationship using a single white dwarf via astrometric microlensing}",
      journal = {\mnras},
         year = 2023,
        month = mar,
       volume = {520},
       number = {1},
        pages = {259-280},
          doi = {10.1093/mnras/stac3532},
archivePrefix = {arXiv},
       eprint = {2206.01814},
 primaryClass = {astro-ph.SR},
       adsurl = {https://ui.adsabs.harvard.edu/abs/2023MNRAS.520..259M}
}

@ARTICLE{Sahu2017,
       author = {{Sahu}, Kailash C. and {Anderson}, Jay and {Casertano}, Stefano and {Bond}, Howard E. and {Bergeron}, Pierre and {Nelan}, Edmund P. and {Pueyo}, Laurent and {Brown}, Thomas M. and {Bellini}, Andrea and {Levay}, Zoltan G. and {Sokol}, Joshua and {Dominik}, Martin and {Calamida}, Annalisa and {Kains}, No{\'e} and {Livio}, Mario},
        title = "{Relativistic deflection of background starlight measures the mass of a nearby white dwarf star}",
      journal = {Science},
         year = 2017,
        month = jun,
       volume = {356},
       number = {6342},
        pages = {1046-1050},
          doi = {10.1126/science.aal2879},
archivePrefix = {arXiv},
       eprint = {1706.02037},
 primaryClass = {astro-ph.SR},
       adsurl = {https://ui.adsabs.harvard.edu/abs/2017Sci...356.1046S}
}

@ARTICLE{Zurlo2018,
       author = {{Zurlo}, A. and {Gratton}, R. and {Mesa}, D. and {Desidera}, S. and {Enia}, A. and {Sahu}, K. and {Almenara}, J. -M. and {Kervella}, P. and {Avenhaus}, H. and {Girard}, J. and {Janson}, M. and {Lagadec}, E. and {Langlois}, M. and {Milli}, J. and {Perrot}, C. and {Schlieder}, J. -E. and {Thalmann}, C. and {Vigan}, A. and {Giro}, E. and {Gluck}, L. and {Ramos}, J. and {Roux}, A.},
        title = "{The gravitational mass of Proxima Centauri measured with SPHERE from a microlensing event}",
      journal = {\mnras},
         year = 2018,
        month = oct,
       volume = {480},
       number = {1},
        pages = {236-244},
          doi = {10.1093/mnras/sty1805},
archivePrefix = {arXiv},
       eprint = {1807.01318},
 primaryClass = {astro-ph.SR},
       adsurl = {https://ui.adsabs.harvard.edu/abs/2018MNRAS.480..236Z}
}

@ARTICLE{Bellini2011,
       author = {{Bellini}, A. and {Anderson}, J. and {Bedin}, L.~R.},
        title = "{Astrometry and Photometry with HST WFC3. II. Improved Geometric-Distortion Corrections for 10 Filters of the UVIS Channel}",
      journal = {\pasp},
         year = 2011,
        month = may,
       volume = {123},
       number = {903},
        pages = {622},
          doi = {10.1086/659878},
archivePrefix = {arXiv},
       eprint = {1102.5218},
 primaryClass = {astro-ph.IM},
       adsurl = {https://ui.adsabs.harvard.edu/abs/2011PASP..123..622B}
}

@article{LPC,
        author={{Lindegren}, L.},
        journal={GAIA-C3-TN-LU-LL-061},
        title={{L}ocal plane coordinates for the detailed analysis of complex {G}aia sources},
        year = 2022,
        url = {https://dms.cosmos.esa.int/COSMOS/doc_fetch.php?id=504573},
        type={Technical Note}
    }

@article{gaiawhitebook,
        author={{European Space Agency}},
        journal={ESA-SCI(2000)4},
        title={GAIA Concept and Technology Study Report: Composition, Formation and Evolution of the Galaxy},
        year = 2000,
        url = {https://www.cosmos.esa.int/documents/29201/297049/report-science.pdf},
    }

@ARTICLE{Gaia_astrom_core,
       author = {{Lindegren}, L. and {Lammers}, U. and {Hobbs}, D. and {O'Mullane}, W. and {Bastian}, U. and {Hern{\'a}ndez}, J.},
        title = "{The astrometric core solution for the Gaia mission. Overview of models, algorithms, and software implementation}",
      journal = {\aap},
         year = 2012,
        month = feb,
       volume = {538},
          eid = {A78},
        pages = {A78},
          doi = {10.1051/0004-6361/201117905},
archivePrefix = {arXiv},
       eprint = {1112.4139},
 primaryClass = {astro-ph.IM},
       adsurl = {https://ui.adsabs.harvard.edu/abs/2012A&A...538A..78L}
}

@ARTICLE{Hog1995,
       author = {{H{\o}g}, E. and {Novikov}, I.~D. and {Polnarev}, A.~G.},
        title = "{MACHO photometry and astrometry.}",
      journal = {\aap},
         year = 1995,
        month = feb,
       volume = {294},
        pages = {287-294},
       adsurl = {https://ui.adsabs.harvard.edu/abs/1995A&A...294..287H}
}

@ARTICLE{Walker1995,
       author = {{Walker}, Mark A.},
        title = "{Microlensed Image Motions}",
      journal = {\apj},
         year = 1995,
        month = nov,
       volume = {453},
        pages = {37},
          doi = {10.1086/176367},
       adsurl = {https://ui.adsabs.harvard.edu/abs/1995ApJ...453...37W}
}

@ARTICLE{Miyamoto1995,
       author = {{Miyamoto}, M. and {Yoshii}, Y.},
        title = "{Astrometry for Determining the MACHO Mass and Trajectory}",
      journal = {\aj},
         year = 1995,
        month = sep,
       volume = {110},
        pages = {1427},
          doi = {10.1086/117616},
       adsurl = {https://ui.adsabs.harvard.edu/abs/1995AJ....110.1427M}
}

@ARTICLE{Underworld2024,
       author = {{Sweeney}, David and {Tuthill}, Peter and {Krone-Martins}, Alberto and {M{\'e}rand}, Antoine and {Scalzo}, Richard and {Martinod}, Marc-Antoine},
        title = "{Observing the galactic underworld: predicting photometry and astrometry from compact remnant microlensing events}",
      journal = {\mnras},
         year = 2024,
        month = jun,
       volume = {531},
       number = {2},
        pages = {2433-2447},
          doi = {10.1093/mnras/stae1302},
archivePrefix = {arXiv},
       eprint = {2403.14612},
 primaryClass = {astro-ph.GA},
       adsurl = {https://ui.adsabs.harvard.edu/abs/2024MNRAS.531.2433S}
}

@ARTICLE{Underworld2022,
       author = {{Sweeney}, David and {Tuthill}, Peter and {Sharma}, Sanjib and {Hirai}, Ryosuke},
        title = "{The Galactic underworld: the spatial distribution of compact remnants}",
      journal = {\mnras},
         year = 2022,
        month = nov,
       volume = {516},
       number = {4},
        pages = {4971-4979},
          doi = {10.1093/mnras/stac2092},
archivePrefix = {arXiv},
       eprint = {2210.04241},
 primaryClass = {astro-ph.GA},
       adsurl = {https://ui.adsabs.harvard.edu/abs/2022MNRAS.516.4971S}
}

@ARTICLE{Kaczmarek2022,
       author = {{Kaczmarek}, Zofia and {McGill}, Peter and {Evans}, N. Wyn and {Smith}, Leigh C. and {Wyrzykowski}, {\L}ukasz and {Howil}, Kornel and {Jab{\l}o{\'n}ska}, Maja},
        title = "{Dark lenses through the dust: parallax microlensing events in the VVV}",
      journal = {\mnras},
         year = 2022,
        month = aug,
       volume = {514},
       number = {4},
        pages = {4845-4860},
          doi = {10.1093/mnras/stac1507},
archivePrefix = {arXiv},
       eprint = {2205.07922},
 primaryClass = {astro-ph.GA},
       adsurl = {https://ui.adsabs.harvard.edu/abs/2022MNRAS.514.4845K}
}

@ARTICLE{BailerJones2021,
       author = {{Bailer-Jones}, C.~A.~L. and {Rybizki}, J. and {Fouesneau}, M. and {Demleitner}, M. and {Andrae}, R.},
        title = "{Estimating Distances from Parallaxes. V. Geometric and Photogeometric Distances to 1.47 Billion Stars in Gaia Early Data Release 3}",
      journal = {\aj},
         year = 2021,
        month = mar,
       volume = {161},
       number = {3},
          eid = {147},
        pages = {147},
          doi = {10.3847/1538-3881/abd806},
archivePrefix = {arXiv},
       eprint = {2012.05220},
 primaryClass = {astro-ph.SR},
       adsurl = {https://ui.adsabs.harvard.edu/abs/2021AJ....161..147B}
}

@ARTICLE{Jablonska2022,
       author = {{Jab{\l}o{\'n}ska}, Maja and {Wyrzykowski}, {\L}ukasz and {Rybicki}, Krzysztof A. and {Kruszy{\'n}ska}, Katarzyna and {Kaczmarek}, Zofia and {Penoyre}, Zephyr},
        title = "{A possible nearby microlensing stellar remnant hiding in Gaia DR3 astrometry}",
      journal = {\aap},
         year = 2022,
        month = oct,
       volume = {666},
          eid = {L16},
        pages = {L16},
          doi = {10.1051/0004-6361/202244656},
archivePrefix = {arXiv},
       eprint = {2206.11342},
 primaryClass = {astro-ph.SR},
       adsurl = {https://ui.adsabs.harvard.edu/abs/2022A&A...666L..16J}
}

@ARTICLE{dynesty,
       author = {{Speagle}, Joshua S.},
        title = "{DYNESTY: a dynamic nested sampling package for estimating Bayesian posteriors and evidences}",
      journal = {\mnras},
         year = 2020,
        month = apr,
       volume = {493},
       number = {3},
        pages = {3132-3158},
          doi = {10.1093/mnras/staa278},
archivePrefix = {arXiv},
       eprint = {1904.02180},
 primaryClass = {astro-ph.IM},
       adsurl = {https://ui.adsabs.harvard.edu/abs/2020MNRAS.493.3132S}
}

@article{Higson2019,
  title={Dynamic nested sampling: an improved algorithm for parameter estimation and evidence calculation},
  author={Higson, Edward and Handley, Will and Hobson, Michael and Lasenby, Anthony},
  journal={Statistics and Computing},
  volume={29},
  number={5},
  pages={891--913},
  year={2019},
  publisher={Springer}
}

@article{Skilling2006,
  title={Nested sampling for general Bayesian computation},
  author={Skilling, John and others},
  journal={Bayesian analysis},
  volume={1},
  number={4},
  pages={833--859},
  year={2006},
  publisher={International Society for Bayesian Analysis}
}

@ARTICLE{delPino2022,
       author = {{del Pino}, Andr{\'e}s and {Libralato}, Mattia and {van der Marel}, Roeland P. and {Bennet}, Paul and {Fardal}, Mark A. and {Anderson}, Jay and {Bellini}, Andrea and {Tony Sohn}, Sangmo and {Watkins}, Laura L.},
        title = "{GaiaHub: A Method for Combining Data from the Gaia and Hubble Space Telescopes to Derive Improved Proper Motions for Faint Stars}",
      journal = {\apj},
         year = 2022,
        month = jul,
       volume = {933},
       number = {1},
          eid = {76},
        pages = {76},
          doi = {10.3847/1538-4357/ac70cf},
archivePrefix = {arXiv},
       eprint = {2205.08009},
 primaryClass = {astro-ph.GA},
       adsurl = {https://ui.adsabs.harvard.edu/abs/2022ApJ...933...76D}
}

@ARTICLE{Penoyre2022,
       author = {{Penoyre}, Zephyr and {Belokurov}, Vasily and {Evans}, N. Wyn},
        title = "{Astrometric identification of nearby binary stars - I. Predicted astrometric signals}",
      journal = {\mnras},
         year = 2022,
        month = jun,
       volume = {513},
       number = {2},
        pages = {2437-2456},
          doi = {10.1093/mnras/stac959},
archivePrefix = {arXiv},
       eprint = {2111.10380},
 primaryClass = {astro-ph.SR},
       adsurl = {https://ui.adsabs.harvard.edu/abs/2022MNRAS.513.2437P}
}

@ARTICLE{Penoyre2022b,
       author = {{Penoyre}, Zephyr and {Belokurov}, Vasily and {Evans}, N. Wyn},
        title = "{Astrometric identification of nearby binary stars - II. Astrometric binaries in the Gaia Catalogue of Nearby Stars}",
      journal = {\mnras},
         year = 2022,
        month = jul,
       volume = {513},
       number = {4},
        pages = {5270-5289},
          doi = {10.1093/mnras/stac1147},
archivePrefix = {arXiv},
       eprint = {2202.06963},
 primaryClass = {astro-ph.SR},
       adsurl = {https://ui.adsabs.harvard.edu/abs/2022MNRAS.513.5270P}
}

@ARTICLE{Nunota2025,
       author = {{Nunota}, Kansuke and {Sumi}, Takahiro and {Koshimoto}, Naoki and {Rattenbury}, Nicholas J. and {Abe}, Fumio and {Barry}, Richard and {Bennett}, David P. and {Bhattacharya}, Aparna and {Fukui}, Akihiko and {Hamada}, Ryusei and {Hamada}, Shunya and {Hamasaki}, Naoto and {Hirao}, Yuki and {Ishitani Silva}, Stela and {Itow}, Yoshitaka and {Matsubara}, Yutaka and {Miyazaki}, Shota and {Muraki}, Yasushi and {Nagai}, Tsutsumi and {Olmschenk}, Greg and {Ranc}, Clement and {Satoh}, Yuki K. and {Suzuki}, Daisuke and {Tristram}, Paul. J. and {Vandorou}, Aikaterini and {Yama}, Hibiki and {MOA Collaboration}},
        title = "{The Microlensing Event Rate and Optical Depth from MOA-II 9 Yr Survey Toward the Galactic Bulge}",
      journal = {\apj},
         year = 2025,
        month = feb,
       volume = {979},
       number = {2},
          eid = {123},
        pages = {123},
          doi = {10.3847/1538-4357/ada352},
archivePrefix = {arXiv},
       eprint = {2410.23553},
 primaryClass = {astro-ph.GA},
       adsurl = {https://ui.adsabs.harvard.edu/abs/2025ApJ...979..123N}
}

@ARTICLE{Robin2003,
       author = {{Robin}, A.~C. and {Reyl{\'e}}, C. and {Derri{\`e}re}, S. and {Picaud}, S.},
        title = "{A synthetic view on structure and evolution of the Milky Way}",
      journal = {\aap},
         year = 2003,
        month = oct,
       volume = {409},
        pages = {523-540},
          doi = {10.1051/0004-6361:20031117},
       adsurl = {https://ui.adsabs.harvard.edu/abs/2003A&A...409..523R}
}

@ARTICLE{Robin2004,
       author = {{Robin}, A.~C. and {Reyl{\'e}}, C. and {Derri{\`e}re}, S. and {Picaud}, S.},
        title = "{Erratum: A synthetic view on structure and evolution of the Milky Way}",
      journal = {\aap},
         year = 2004,
        month = mar,
       volume = {416},
        pages = {157},
          doi = {10.1051/0004-6361:20040968},
archivePrefix = {arXiv},
       eprint = {astro-ph/0401052},
 primaryClass = {astro-ph},
       adsurl = {https://ui.adsabs.harvard.edu/abs/2004A&A...416..157R}
}

@ARTICLE{Koshimoto2021,
       author = {{Koshimoto}, Naoki and {Baba}, Junichi and {Bennett}, David P.},
        title = "{A Parametric Galactic Model toward the Galactic Bulge Based on Gaia and Microlensing Data}",
      journal = {\apj},
         year = 2021,
        month = aug,
       volume = {917},
       number = {2},
          eid = {78},
        pages = {78},
          doi = {10.3847/1538-4357/ac07a8},
archivePrefix = {arXiv},
       eprint = {2104.03306},
 primaryClass = {astro-ph.GA},
       adsurl = {https://ui.adsabs.harvard.edu/abs/2021ApJ...917...78K}
}

@ARTICLE{Thompson2025,
       author = {{Thompson}, William and {Nielsen}, Eric L. and {Ruffio}, Jean-Baptiste and {Blunt}, Sarah and {Marois}, Christian},
        title = "{Revised Mass and Orbit of {\ensuremath{\in}} Eridani b: A 1 M$_{Jup}$ Planet on a Near-circular Orbit}",
      journal = {\aj},
         year = 2025,
        month = dec,
       volume = {170},
       number = {6},
          eid = {301},
        pages = {301},
          doi = {10.3847/1538-3881/ae0cbd},
archivePrefix = {arXiv},
       eprint = {2502.20561},
 primaryClass = {astro-ph.EP},
       adsurl = {https://ui.adsabs.harvard.edu/abs/2025AJ....170..301T}
}

@ARTICLE{Thompson2026,
       author = {{Thompson}, William and {Blakely}, Dori and {Xuan}, Jerry W. and {Blouin}, Simon and {Zhang}, Jingwen and {Johnstone}, Doug and {Ruffio}, Jean-Baptiste and {Nielsen}, Eric and {Speedie}, Jessica and {Bowler}, Brendan P. and {Bouchard-C{\^o}t{\'e}}, Alexandre and {Franson}, Kyle and {Blunt}, Sarah and {Roberson}, William and {Cloutier}, Ryan and {Fogal}, Andre and {Hessel}, Kaitlyn and {Marois}, Christian and {Rochon}, Alexandra},
        title = "{Detecting and Characterizing Companions with a Calibrated Gaia DR2, DR3, and Hipparcos Catalog (G23H)}",
      journal = {\aj},
         year = 2026,
        month = jul,
       volume = {172},
       number = {1},
          eid = {53},
        pages = {53},
          doi = {10.3847/1538-3881/ae64ed},
       adsurl = {https://ui.adsabs.harvard.edu/abs/2026AJ....172...53T}
}

@ARTICLE{Feng2024,
       author = {{Feng}, Fabo and {Rui}, Yicheng and {Xuan}, Yifan and {Jones}, Hugh},
        title = "{Modeling and Calibration of Gaia, Hipparcos, and Tycho-2 Astrometric Data for the Detection of Dark Companions}",
      journal = {\apjs},
         year = 2024,
        month = apr,
       volume = {271},
       number = {2},
          eid = {50},
        pages = {50},
          doi = {10.3847/1538-4365/ad27d2},
archivePrefix = {arXiv},
       eprint = {2402.04919},
 primaryClass = {astro-ph.SR},
       adsurl = {https://ui.adsabs.harvard.edu/abs/2024ApJS..271...50F}
}

@ARTICLE{Ribas2025,
       author = {{Ribas}, {\'A}lvaro and {Vioque}, Miguel and {Zagaria}, Francesco and {Longarini}, Cristiano and {Mac{\'\i}as}, Enrique and {Clarke}, Cathie J. and {P{\'e}rez}, Sebasti{\'a}n and {Carpenter}, John and {Cuello}, Nicol{\'a}s and {de Gregorio-Monsalvo}, Itziar},
        title = "{A young gas giant and hidden substructures in a protoplanetary disk}",
      journal = {Nature Astronomy},
         year = 2025,
        month = aug,
       volume = {9},
        pages = {1176-1183},
          doi = {10.1038/s41550-025-02576-w},
archivePrefix = {arXiv},
       eprint = {2507.11612},
 primaryClass = {astro-ph.EP},
       adsurl = {https://ui.adsabs.harvard.edu/abs/2025NatAs...9.1176R}
}

@ARTICLE{Brandt2021,
       author = {{Brandt}, Timothy D.},
        title = "{The Hipparcos-Gaia Catalog of Accelerations: Gaia EDR3 Edition}",
      journal = {\apjs},
         year = 2021,
        month = jun,
       volume = {254},
       number = {2},
          eid = {42},
        pages = {42},
          doi = {10.3847/1538-4365/abf93c},
archivePrefix = {arXiv},
       eprint = {2105.11662},
 primaryClass = {astro-ph.GA},
       adsurl = {https://ui.adsabs.harvard.edu/abs/2021ApJS..254...42B}
}

@ARTICLE{Kervella2022,
       author = {{Kervella}, Pierre and {Arenou}, Fr{\'e}d{\'e}ric and {Th{\'e}venin}, Fr{\'e}d{\'e}ric},
        title = "{Stellar and substellar companions from Gaia EDR3. Proper-motion anomaly and resolved common proper-motion pairs}",
      journal = {\aap},
         year = 2022,
        month = jan,
       volume = {657},
          eid = {A7},
        pages = {A7},
          doi = {10.1051/0004-6361/202142146},
archivePrefix = {arXiv},
       eprint = {2109.10912},
 primaryClass = {astro-ph.SR},
       adsurl = {https://ui.adsabs.harvard.edu/abs/2022A&A...657A...7K}
}

@ARTICLE{Sartore2010,
       author = {{Sartore}, N. and {Ripamonti}, E. and {Treves}, A. and {Turolla}, R.},
        title = "{Galactic neutron stars. I. Space and velocity distributions in the disk and in the halo}",
      journal = {\aap},
         year = 2010,
        month = feb,
       volume = {510},
          eid = {A23},
        pages = {A23},
          doi = {10.1051/0004-6361/200912222},
archivePrefix = {arXiv},
       eprint = {0908.3182},
 primaryClass = {astro-ph.GA},
       adsurl = {https://ui.adsabs.harvard.edu/abs/2010A&A...510A..23S}
}

@ARTICLE{Lehtinen2023,
       author = {{Lehtinen}, K. and {Prusti}, T. and {de Bruijne}, J. and {Lammers}, U. and {Manara}, C.~F. and {Ness}, J.-U. and {Siddiqui}, H. and {Poutanen}, M. and {Muinonen}, K. and {Morrison}, O.},
        title = "{Carte du Ciel and Gaia. I. Astrometry}",
      journal = {\aap},
         year = 2023,
        month = mar,
       volume = {671},
          eid = {A16},
        pages = {A16},
          doi = {10.1051/0004-6361/202142929},
       adsurl = {https://ui.adsabs.harvard.edu/abs/2023A&A...671A..16L}
}

@ARTICLE{Mroz2024,
       author = {{Mr{\'o}z}, Przemek and {Udalski}, Andrzej and {Szyma{\'n}ski}, Micha{\l} K. and {Kapusta}, Mateusz and {Soszy{\'n}ski}, Igor and {Wyrzykowski}, {\L}ukasz and {Pietrukowicz}, Pawe{\l} and {Koz{\l}owski}, Szymon and {Poleski}, Rados{\l}aw and {Skowron}, Jan and {Skowron}, Dorota and {Ulaczyk}, Krzysztof and {Gromadzki}, Mariusz and {Rybicki}, Krzysztof and {Iwanek}, Patryk and {Wrona}, Marcin and {Ratajczak}, Milena},
        title = "{Microlensing Optical Depth and Event Rate toward the Large Magellanic Cloud Based on 20 yr of OGLE Observations}",
      journal = {\apjs},
         year = 2024,
        month = jul,
       volume = {273},
       number = {1},
          eid = {4},
        pages = {4},
          doi = {10.3847/1538-4365/ad452e},
archivePrefix = {arXiv},
       eprint = {2403.02398},
 primaryClass = {astro-ph.GA},
       adsurl = {https://ui.adsabs.harvard.edu/abs/2024ApJS..273....4M}
}

@ARTICLE{DominikSahu2000,
       author = {{Dominik}, Martin and {Sahu}, Kailash C.},
        title = "{Astrometric Microlensing of Stars}",
      journal = {\apj},
         year = 2000,
        month = may,
       volume = {534},
       number = {1},
        pages = {213-226},
          doi = {10.1086/308716},
       adsurl = {https://ui.adsabs.harvard.edu/abs/2000ApJ...534..213D}
}

@ARTICLE{Kruszynska2024,
       author = {{Kruszy{\'n}ska}, K. and {Wyrzykowski}, {\L}. and {Rybicki}, K.~A. and {Howil}, K. and {Jab{\l}o{\'n}ska}, M. and {Kaczmarek}, Z. and {Ihanec}, N. and {Maskoli{\={u}}nas}, M. and {Bronikowski}, M. and {Pylypenko}, U. and {Udalski}, A. and {Mr{\'o}z}, P. and {Poleski}, R. and {Skowron}, J. and {Szyma{\'n}ski}, M.~K. and {Soszy{\'n}ski}, I. and {Pietrukowicz}, P. and {Koz{\l}owski}, S. and {Ulaczyk}, K. and {Iwanek}, P. and {Wrona}, M. and {Gromadzki}, M. and {Mr{\'o}z}, M.~J. and {Abe}, F. and {Bando}, K. and {Barry}, R. and {Bennett}, D.~P. and {Bhattacharya}, A. and {Bond}, I.~A. and {Fukui}, A. and {Hamada}, R. and {Hamada}, S. and {Hamasaki}, N. and {Hirao}, Y. and {Ishitani Silva}, S. and {Itow}, Y. and {Koshimoto}, N. and {Matsubara}, Y. and {Miyazaki}, S. and {Muraki}, Y. and {Nagai}, T. and {Nunota}, K. and {Olmschenk}, G. and {Ranc}, C. and {Rattenbury}, N.~J. and {Satoh}, Y. and {Sumi}, T. and {Suzuki}, D. and {Tristram}, P.~J. and {Vandorou}, A. and {Yama}, H.},
        title = "{Dark lens candidates from Gaia Data Release 3}",
      journal = {\aap},
         year = 2024,
        month = dec,
       volume = {692},
          eid = {A28},
        pages = {A28},
          doi = {10.1051/0004-6361/202449322},
archivePrefix = {arXiv},
       eprint = {2401.13759},
 primaryClass = {astro-ph.GA},
       adsurl = {https://ui.adsabs.harvard.edu/abs/2024A&A...692A..28K}
}

@ARTICLE{Howil2024,
       author = {{Howil}, K. and {Wyrzykowski}, {\L}. and {Kruszy{\'n}ska}, K. and {Zieli{\'n}ski}, P. and {Bachelet}, E. and {Gromadzki}, M. and {Miko{\l}ajczyk}, P.~J. and {Kotysz}, K. and {Jab{\l}o{\'n}ska}, M. and {Kaczmarek}, Z. and {Mr{\'o}z}, P. and {Ihanec}, N. and {Ratajczak}, M. and {Pylypenko}, U. and {Rybicki}, K. and {Sweeney}, D. and {Hodgkin}, S.~T. and {Larma}, M. and {Carrasco}, J.~M. and {Burgaz}, U. and {Godunova}, V. and {Simon}, A. and {Cusano}, F. and {Jelinek}, M. and {{\v{S}}trobl}, J. and {Hudec}, R. and {Merc}, J. and {Ku{\v{c}}{\'a}kov{\'a}}, H. and {Erece}, O. and {Kilic}, Y. and {Olivares}, F. and {Morrell}, M. and {Wicker}, M.},
        title = "{Uncovering the invisible: A study of Gaia18ajz, a candidate black hole revealed by microlensing}",
      journal = {\aap},
         year = 2025,
        month = feb,
       volume = {694},
          eid = {A94},
        pages = {A94},
          doi = {10.1051/0004-6361/202451046},
archivePrefix = {arXiv},
       eprint = {2403.09006},
 primaryClass = {astro-ph.GA},
       adsurl = {https://ui.adsabs.harvard.edu/abs/2025A&A...694A..94H}
}

@ARTICLE{Bachelet2024,
       author = {{Bachelet}, E. and {Hundertmark}, M. and {Calchi Novati}, S.},
        title = "{Estimating Microlensing Parameters from Observables and Stellar Isochrones with pyLIMASS}",
      journal = {\aj},
         year = 2024,
        month = jul,
       volume = {168},
       number = {1},
          eid = {24},
        pages = {24},
          doi = {10.3847/1538-3881/ad4862},
archivePrefix = {arXiv},
       eprint = {2405.02230},
 primaryClass = {astro-ph.IM},
       adsurl = {https://ui.adsabs.harvard.edu/abs/2024AJ....168...24B}
}

@ARTICLE{Sm03,
       author = {{Smith}, Martin C. and {Mao}, Shude and {Paczy{\'n}ski}, Bohdan},
        title = "{Acceleration and parallax effects in gravitational microlensing}",
      journal = {\mnras},
         year = 2003,
        month = mar,
       volume = {339},
       number = {4},
        pages = {925-936},
          doi = {10.1046/j.1365-8711.2003.06183.x},
archivePrefix = {arXiv},
       eprint = {astro-ph/0210370},
 primaryClass = {astro-ph},
       adsurl = {https://ui.adsabs.harvard.edu/abs/2003MNRAS.339..925S}
}

@ARTICLE{Gould2004,
       author = {{Gould}, Andrew},
        title = "{Resolution of the MACHO-LMC-5 Puzzle: The Jerk-Parallax Microlens Degeneracy}",
      journal = {\apj},
         year = 2004,
        month = may,
       volume = {606},
       number = {1},
        pages = {319-325},
          doi = {10.1086/382782},
archivePrefix = {arXiv},
       eprint = {astro-ph/0311548},
 primaryClass = {astro-ph},
       adsurl = {https://ui.adsabs.harvard.edu/abs/2004ApJ...606..319G}
}

@ARTICLE{WyrzMandel2020,
       author = {{Wyrzykowski}, {\L}ukasz and {Mandel}, Ilya},
        title = "{Constraining the masses of microlensing black holes and the mass gap with Gaia DR2}",
      journal = {\aap},
         year = 2020,
        month = apr,
       volume = {636},
          eid = {A20},
        pages = {A20},
          doi = {10.1051/0004-6361/201935842},
archivePrefix = {arXiv},
       eprint = {1904.07789},
 primaryClass = {astro-ph.SR},
       adsurl = {https://ui.adsabs.harvard.edu/abs/2020A&A...636A..20W}
}

@ARTICLE{IKChen2023,
       author = {{Chen}, I. -Kai and {Kongsore}, Marius and {Tilburg}, Ken Van},
        title = "{Detecting dark compact objects in Gaia DR4: A data analysis pipeline for transient astrometric lensing searches}",
      journal = {\jcap},
         year = 2023,
        month = jul,
       volume = {2023},
       number = {7},
          eid = {037},
        pages = {037},
          doi = {10.1088/1475-7516/2023/07/037},
archivePrefix = {arXiv},
       eprint = {2301.00822},
 primaryClass = {astro-ph.GA},
       adsurl = {https://ui.adsabs.harvard.edu/abs/2023JCAP...07..037C}
}

@ARTICLE{Han2008,
       author = {{Han}, Cheongho},
        title = "{Near-Field Microlensing from Wide-Field Surveys}",
      journal = {\apj},
         year = 2008,
        month = jul,
       volume = {681},
       number = {2},
        pages = {806-813},
          doi = {10.1086/588083},
archivePrefix = {arXiv},
       eprint = {0708.1215},
 primaryClass = {astro-ph},
       adsurl = {https://ui.adsabs.harvard.edu/abs/2008ApJ...681..806H}
}

@ARTICLE{Kaczmarek2024,
       author = {{Kaczmarek}, Zofia and {McGill}, Peter and {Evans}, N. Wyn and {Smith}, Leigh C. and {Golovich}, Nathan and {Kerins}, Eamonn and {Specht}, David and {Dawson}, William A.},
        title = "{Spatially resolved microlensing time-scale distributions across the Galactic bulge with the VVV survey}",
      journal = {\mnras},
         year = 2024,
        month = apr,
       volume = {529},
       number = {2},
        pages = {1308-1320},
          doi = {10.1093/mnras/stae445},
archivePrefix = {arXiv},
       eprint = {2312.11667},
 primaryClass = {astro-ph.GA},
       adsurl = {https://ui.adsabs.harvard.edu/abs/2024MNRAS.529.1308K}
}

@ARTICLE{Paczynski1991,
       author = {{Paczynski}, B.},
        title = "{Gravitational Microlensing of the Galactic Bulge Stars}",
      journal = {\apjl},
         year = 1991,
        month = apr,
       volume = {371},
        pages = {L63},
          doi = {10.1086/186003},
       adsurl = {https://ui.adsabs.harvard.edu/abs/1991ApJ...371L..63P}
}

@ARTICLE{Griest1991,
       author = {{Griest}, Kim and {Alcock}, Charles and {Axelrod}, Timothy S. and {Bennett}, David P. and {Cook}, Kem H. and {Freeman}, Kenneth C. and {Park}, Hye-Sook and {Perlmutter}, Saul and {Peterson}, Bruce A. and {Quinn}, Peter J. and {Rodgers}, Alexander W. and {Stubbs}, Christopher W. and {MACHO Collaboration}},
        title = "{Gravitational Microlensing as a Method of Detecting Disk Dark Matter and Faint Disk Stars}",
      journal = {\apjl},
         year = 1991,
        month = may,
       volume = {372},
        pages = {L79},
          doi = {10.1086/186028},
       adsurl = {https://ui.adsabs.harvard.edu/abs/1991ApJ...372L..79G}
}

@ARTICLE{Sumi2013,
       author = {{Sumi}, T. and {Bennett}, D.~P. and {Bond}, I.~A. and {Abe}, F. and {Botzler}, C.~S. and {Fukui}, A. and {Furusawa}, K. and {Itow}, Y. and {Ling}, C.~H. and {Masuda}, K. and {Matsubara}, Y. and {Muraki}, Y. and {Ohnishi}, K. and {Rattenbury}, N. and {Saito}, To. and {Sullivan}, D.~J. and {Suzuki}, D. and {Sweatman}, W.~L. and {Tristram}, P.~J. and {Wada}, K. and {Yock}, P.~C.~M. and {MOA Collaboratoin}, The},
        title = "{The Microlensing Event Rate and Optical Depth toward the Galactic Bulge from MOA-II}",
      journal = {\apj},
         year = 2013,
        month = dec,
       volume = {778},
       number = {2},
          eid = {150},
        pages = {150},
          doi = {10.1088/0004-637X/778/2/150},
archivePrefix = {arXiv},
       eprint = {1305.0186},
 primaryClass = {astro-ph.GA},
       adsurl = {https://ui.adsabs.harvard.edu/abs/2013ApJ...778..150S}
}

@ARTICLE{Mroz2019,
       author = {{Mr{\'o}z}, Przemek and {Udalski}, Andrzej and {Skowron}, Jan and {Szyma{\'n}ski}, Micha{\l} K. and {Soszy{\'n}ski}, Igor and {Wyrzykowski}, {\L}ukasz and {Pietrukowicz}, Pawe{\l} and {Koz{\l}owski}, Szymon and {Poleski}, Rados{\l}aw and {Ulaczyk}, Krzysztof and {Rybicki}, Krzysztof and {Iwanek}, Patryk},
        title = "{Microlensing Optical Depth and Event Rate toward the Galactic Bulge from 8 yr of OGLE-IV Observations}",
      journal = {\apjs},
         year = 2019,
        month = oct,
       volume = {244},
       number = {2},
          eid = {29},
        pages = {29},
          doi = {10.3847/1538-4365/ab426b},
archivePrefix = {arXiv},
       eprint = {1906.02210},
 primaryClass = {astro-ph.SR},
       adsurl = {https://ui.adsabs.harvard.edu/abs/2019ApJS..244...29M}
}

@ARTICLE{Specht2020,
       author = {{Specht}, David and {Kerins}, Eamonn and {Awiphan}, Supachai and {Robin}, Annie C.},
        title = "{MaB{\ensuremath{\mu}}lS-2: high-precision microlensing modelling for the large-scale survey era}",
      journal = {\mnras},
         year = 2020,
        month = oct,
       volume = {498},
       number = {2},
        pages = {2196-2218},
          doi = {10.1093/mnras/staa2375},
archivePrefix = {arXiv},
       eprint = {2005.14668},
 primaryClass = {astro-ph.EP},
       adsurl = {https://ui.adsabs.harvard.edu/abs/2020MNRAS.498.2196S}
}

@ARTICLE{Paczynski1995,
       author = {{Paczynski}, B.},
        title = "{The Masses of Nearby Dwarfs Can Be Determined with Gravitational Microlensing}",
      journal = {\actaa},
         year = 1995,
        month = apr,
       volume = {45},
        pages = {345-348},
          doi = {10.48550/arXiv.astro-ph/9504099},
archivePrefix = {arXiv},
       eprint = {astro-ph/9504099},
 primaryClass = {astro-ph},
       adsurl = {https://ui.adsabs.harvard.edu/abs/1995AcA....45..345P}
}

@ARTICLE{Mroz2020ztf,
       author = {{Mr{\'o}z}, Przemek and {Street}, R.~A. and {Bachelet}, E. and {Ofek}, E.~O. and {Bellm}, E.~C. and {Dekany}, R. and {Duev}, D.~A. and {Gal-Yam}, A. and {Graham}, M.~J. and {Masci}, F.~J. and {Porter}, M. and {Rusholme}, B. and {Smith}, R.~M. and {Soumagnac}, M.~T. and {Zolkower}, J.},
        title = "{Gravitational Microlensing Events from the First Year of the Northern Galactic Plane Survey by the Zwicky Transient Facility}",
      journal = {Research Notes of the American Astronomical Society},
         year = 2020,
        month = jan,
       volume = {4},
       number = {1},
          eid = {13},
        pages = {13},
          doi = {10.3847/2515-5172/ab7021},
archivePrefix = {arXiv},
       eprint = {2001.09584},
 primaryClass = {astro-ph.SR},
       adsurl = {https://ui.adsabs.harvard.edu/abs/2020RNAAS...4...13M}
}

@ARTICLE{Navarro2017,
       author = {{Navarro}, Mar{\'\i}a Gabriela and {Minniti}, Dante and
         {Contreras Ramos}, Rodrigo},
        title = "{VVV Survey Microlensing Events in the Galactic Center Region}",
      journal = {\apjl},
         year = 2017,
        month = dec,
       volume = {851},
       number = {1},
          eid = {L13},
        pages = {L13},
          doi = {10.3847/2041-8213/aa9b29},
archivePrefix = {arXiv},
       eprint = {1712.07667},
 primaryClass = {astro-ph.SR},
       adsurl = {https://ui.adsabs.harvard.edu/abs/2017ApJ...851L..13N}
}

\begin{appendix}
\onecolumn
\section{A complete visualization of modelling results for an example event}
\label{sec:full_visualization}
\begin{figure*}[htb!]
	\includegraphics[width=1.0\columnwidth]{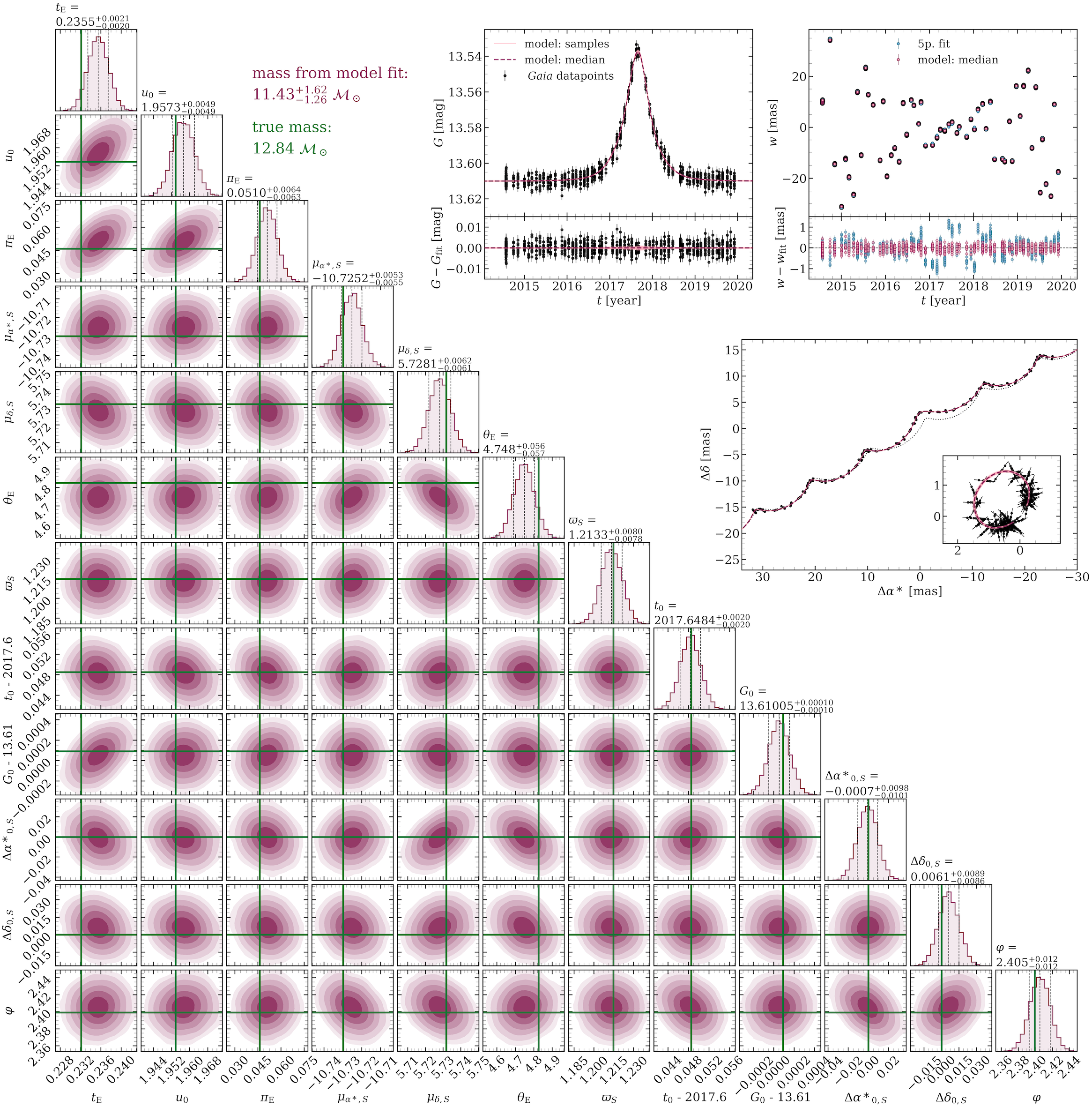}
    \caption{Full modelling results for an example well-modelled event, where the lens is a 12.8 $M_\odot$ BH. \textit{Main:} Cornerplot of the fitted parameters. {Levels of purple shading are separated by 0.5, 1, 1.5, 2, 2.5 and 3}$\sigma$ contours of posterior sample distributions. Green vertical and horizontal lines show true input parameter values for comparison. Added text compares the inferred mass from samples compared to the true mass. \textit{Inset, top left:} Photometric time-series data (black) with 300 random posterior samples (light pink) and a model assuming median parameter values from all samples (purple, dashed). The posterior samples are only visible in some sections of the plot, as they overlap closely with the median model. The bottom subplot shows residuals from the median model. \textit{Inset, top right:} Astrometric time-series data (black) with corresponding values from a median model (pink) and the {five-parameter} fit (blue). In the top subplot, the three point types largely overlap, as the absolute $w$ values are significantly larger than the astrometric lensing signal. The bottom subplot shows residuals from both fits, demonstrating a clear improvement with the addition of the astrometric lensing effect into the model. \textit{Inset, bottom right:} The reconstructed 2D on-sky tracks; colour coding as in the photometric plot. Datapoints with corresponding error bars are placed in the 2D space at an across-scan value, calculated using the median model, and an along-scan value, calculated using the median model and then shifted by (data -- model) residuals. The inset shows just the astrometric shift (subtracting the unlensed source motion).}
    \label{fig:corner}
\end{figure*}
\end{appendix}
\end{document}